\documentclass[superscriptaddress,showpacs,prd,floatfix]{revtex4}
\usepackage{times}
\usepackage[reqno,tbtags]{amsmath}
\usepackage{epsfig}
\usepackage{color} 
\usepackage{graphicx,psfrag}
\allowdisplaybreaks
\newcommand{\nl}{\nonumber \\}
\def \litwo {{\rm{Li_2}}}
\def \litri {{\rm{Li_3}}}
\def \lin   {{\rm{Li_n}}}

\newcommand{\bq}{\begin{equation}}
\newcommand{\eq}{\end{equation}}
\newcommand{\ba}{\begin{eqnarray}}
\newcommand{\ea}{\end{eqnarray}}
\begin{document}
\vspace{1cm}
\begin{flushleft}
{\tt 
\noindent
DESY 08-101 \hfill 
\\
PITHA 08-18
\\
}
\end{flushleft}

\bigskip

\title{%
Virtual  Hadronic and Heavy-Fermion 
  ${\cal O}(\alpha^2)$ Corrections
  to Bhabha Scattering 
}
\author{Stefano Actis}
\affiliation{Institut f\"ur Theoretische Physik E, RWTH Aachen University,      
            D-52056 Aachen, Germany}
\email{actis@physik.rwth-aachen.de,mczakon@yahoo.com,gluza@us.edu.pl,Tord.Riemann@desy.de}
\author{Micha{\l} Czakon}
\affiliation{Institut f\"ur Theoretische Physik und Astrophysik, Universit\"at W\"urzburg, 
Am Hubland, D-97074 W\"urzburg, Germany}
\affiliation{Institute  of Physics, University of
Silesia, Uniwersytecka 4, PL-40-007 Katowice, Poland }
\author{Janusz Gluza}
\affiliation{Institute  of Physics, University of
Silesia, Uniwersytecka 4, PL-40-007 Katowice, Poland }
\author{Tord Riemann}
\affiliation{DESY, Platanenallee 6, 15738 Zeuthen, Germany}
\begin{abstract}
  Effects of vacuum polarization by hadronic and heavy-fermion insertions were the last unknown two-loop QED 
  corrections to high-energy Bhabha scattering and have been first announced in \cite{Actis:2007fs}.  
  Here we describe the corrections in detail and explore their numerical influence.
  The hadronic contributions to the virtual ${\cal O}(\alpha^2)$ QED corrections 
  to the Bhabha-scattering cross-section are evaluated using dispersion
  relations and computing the convolution of hadronic data with 
  perturbatively calculated kernel functions.
  The technique of dispersion integrals is also employed to derive the virtual ${\cal O}(\alpha^2)$ corrections
  generated by muon-, tau- and top-quark loops in the small 
  electron-mass limit for arbitrary values of the internal-fermion masses.
  At a meson factory with 1 GeV  center-of-mass energy the complete effect of hadronic and heavy-fermion
  corrections amounts to less than 0.5 per mille and reaches, at 10 GeV, 
  up to about 2 per mille.
  At the $Z$ resonance it amounts to 2.3 per mille at 3 degrees; overall, hadronic corrections
  are less than 4 per mille.
  For ILC energies (500 GeV or above), the combined effect of hadrons and heavy fermions becomes 6 per mille  at 
  3 degrees; hadrons contribute less than 20 per mille in the whole angular region.
\end{abstract}


\maketitle
\section{\label{sec-introduction} INTRODUCTION}
Elastic $e^{+}e^{-}$ scattering, or Bhabha scattering, 
\begin{equation}\label{bhabhamomenta}
  e^-\,(p_1) \, +   \,
  e^+\,(p_2) \, \to \,
  e^-\,(p_3) \, +   \, e^+\,(p_4)\, 
,
\end{equation}
was one of the first scattering processes that were observed and predicted in quantum mechanics \cite{Bhabha:1936xx}.
It has a unique and clean experimental signature.
The accuracy of theoretical predictions profits from its purely leptonic external particle content and from the extremely small electron mass.
The first complete one-loop prediction in the Standard Model was \cite{Consoli:1979xw},
the first $O(\alpha)$ predictions in the Standard Model with account of hard bremsstrahlung were
determined in \cite{%
Consoli:1982ib,Caffo:1984jb,%
Bohm:1984yt,%
Tobimatsu:1985pp,Tobimatsu:1985vd,
Bohm:1986fg%
},
the effects from hadronic vacuum polarization were first studied in
\cite{Berends:1987jm},
and the leading NNLO corrections from the top quark in \cite{Bardin:1990xe}.
The complete electroweak two-loop corrections are available in form of few form factors \cite{Awramik:2003rn,Awramik:2006uz}, but they are not implemented for Bhabha scattering so far. 
During the years, a rich literature on the subject arose, both concerning QED Monte Carlo results and virtual electroweak corrections; see 
\cite{%
Berends:1976zn,%
Berends:1983fs,%
Berends:1984ge,
Greco:1986dc,%
Kuroda:1987yi,Karlen:1987vk,%
Aversa:1990ek,Fujimoto:1990tb,%
Caffo:1991cg,Cacciari:1991rm,Cacciari:1991qy,Beenakker:1991es,Beenakker:1991mb,Aversa:1991rw,Riemann:1991ga,%
Fadin:1992uem,Bjoerkevoll:1992cu,%
Bardin:1992jc,%
Montagna:1993py,Caffo:1993hc,Fujimoto:1993qh,%
Caffo:1994dm,Caffo:1994fy,Fadin:1994xe,%
Bjoerkevoll:1992uu,%
Field:1995dk,%
Cacciari:1995fq,%
Jadach:1995nk,%
Arbuzov:1995qd,Arbuzov:1995vi,Arbuzov:1995vj,Arbuzov:1995ix,%
Caffo:1996vi,Caffo:1996mi,Arbuzov:1996jj,Arbuzov:1996su,Arbuzov:1996qb,Arbuzov:1996zp,%
Jadach:1996md,Jadach:1996is,Jadach:1996gu,Jadach:1996hy,%
Beenakker:1997fi,Caffo:1997yy,%
Arbuzov:1997pj,Merenkov:1997zm,%
Arbuzov:1998du,%
Montagna:1998vb,Arbuzov:1998ax,%
Bardin:1999yd,Arbuzov:1999db,Placzek:1999xc,Jadach:1999tr,%
Montagna:1999tf,%
CarloniCalame:1999aw,Antonelli:1999pe,%
CarloniCalame:2000pz,%
Battaglia:2001dg,CarloniCalame:2001ny,Karlen:2001hw,%
Ward:2002qq,%
Jadach:2003zr,%
Arbuzov:2004wp,Fleischer:2004ah,Gluza:2004tq,Lorca:2004dk,%
Arbuzov:2005pt,Arbuzov:2005ma,%
Arbuzov:2006mu,Balossini:2006sd,Balossini:2006wc,Fleischer:2006ht%
}, and also the references therein.

Quite recently, an experimental precision at the per mille level or beyond seems feasible both at meson factories and in the ILC (and GigaZ) project \cite{moenig:sfb2005,denig:sfb2005,trentadue:sfb2005,jadach:sfb2005a,%
Balossini:2007zz,Balossini:2008ht}.
As a reaction to that, a program of systematic evaluation of the complete next-to-next-to leading order (NNLO) contributions was emerging
\cite{Smirnov:2001cm,Bern:2000ie,Glover:2001ev,%
Bonciani:2003ai,Bonciani:2003cj,Bonciani:2003te,%
Bonciani:2004qt,Czakon:2004tg,Czakon:2004wm,Heinrich:2004iq,%
penin:2005kf,Penin:2005eh,%
Bonciani:2005im,Czakon:2005gi,%
Bonciani:2006qu,Czakon:2006pa,%
Mitov:2006xs,Actis:2006dj,%
Becher:2007cu,%
Actis:2007fs,Actis:2007gi,Actis:2007pn,Actis:2007pn2,Bonciani:2007eh,Fleischer:2007ph,%
Bonciani:2008ep%
}. 

In this article, we extensively describe the evaluation of the last building block of QED two-loop corrections, namely the corrections from heavy fermions and hadronic vacuum polarization.
Note that the latter result
has been confirmed very recently in~\cite{Kuhn:2008zs} (upon using the same parametrisation of 
the vacuum polarization, the agreement between the two studies is perfect, 5 digits for the ${\cal O}(\alpha^4)$
NNLO terms).
Both for reasons of completeness and in order to ensure easy comparisons,
we will also include in the discussion the $N_f=1$ corrections which consist of purely photonic corrections and electron loop insertions, the soft bremsstrahlung and soft electron pair emission corrections.    
All the two-loop contributions are calculated in our numerical Fortran package  \texttt{bhbhnnlohf.F} and will be made available at the webpage \cite{webPage:2007xx}.

The organization of the paper is as follows. 
In Section~\ref{sec:1} we introduce notations and the Born cross-section.
Section~\ref{sec:pi-vac} collects the known facts on pure vacuum-polarization corrections as they will be used,
and Section~\ref{sec-vac} the pure self-energy corrections to the cross-section. 
Section~\ref{sec-irred-vert} contains the irreducible vertex corrections and 
Section~\ref{sec-IR} the various infrared divergent corrections, including reducible corrections, soft-photon emission and the most complicated ones from the irreducible two-loop box diagrams.
The three kernel functions for the latter have been evaluated for the first time.
Section~\ref{sec:NumericalResults} contains a discussion of numerical effects at a variety of energies, typical of meson factories, LEP, ILC.
In the Summary we will also point to potential further research.
Appendices \ref{app:pho} to  \ref{app-polylog} are devoted to technical details of fermionic vacuum polarization, one-loop master integrals, soft real bremsstrahlung, real pair emission, the evaluation of the hadronic cross-section ratio $R_{\rm had}$,
and on our evaluation of complex polylogarithms.
Some Mathematica files of potential public interest and the Fortran package are available at the webpage \cite{webPage:2007xx}.
\section{\label{sec:1}THE BORN CROSS-SECTION}
The QED tree-level differential Bhabha-scattering 
cross section with respect to the solid angle $\Omega$, 
in the kinematic region $m_e^2 \ll s,|t|,|u|$, is:
\begin{eqnarray}\label{born}
 \frac{d\sigma_0}{d\Omega} &=&
\frac{\alpha^2}{2s} \left\lbrace  \frac{v_1(s,t)}{s^2} + 2 \frac{v_2(s,t)}{st}+\frac{v_1(t,s)}{t^2}\right\rbrace\nonumber
\\
&=& \frac{\alpha^2}{s} \left(\frac{s}{t}+1+\frac{t}{s} \right)^2 .
\end{eqnarray}
Here, $\alpha$ is the fine-structure constant~\cite{Eidelman:2004},
\begin{eqnarray}
 \label{alpha}
\alpha &=& 1/137.035 999 679(94),
\end{eqnarray}
 and 
\begin{eqnarray}\label{v1v2}
 v_1(x,y) &=&x^2 + 2 y^2 + 2 x y ,
\\
v_2(x,y) &=&(x + y)^2 .
\end{eqnarray}
The cross-section  depends on the Mandelstam invariants $s$, $t$ and $u$, which are related 
to $E$, the incoming-particle energy in the center-of-mass frame, 
and $\theta$, the scattering angle:
\begin{eqnarray}\label{Mandelstam}
s &=& \left( p_1+p_2 \right)^2 =\,  4 \, E^2, \nonumber\\
t &=& \left( p_1-p_3 \right)^2 =\, -\,4 \, E^2 
      \,\sin^2 \left(\frac{\theta}{2}\right),\nonumber\\
u &=& \left( p_1-p_4 \right)^2 =\, -\,4 \, E^2  
      \,\cos^2 \left(\frac{\theta}{2}\right),
\end{eqnarray}
where 
\begin{eqnarray}\label{stu}
s\,+\,t\,+\,u\,=\,0. 
\end{eqnarray}


For the numerical estimates at higher energies, it is reasonable to normalize the higher order corrections to the complete electroweak effective Born cross-section:
\begin{eqnarray}\label{ewborn}
 \frac{d \sigma_{ew}}{d \Omega} &=& \frac{\alpha^2}{4s}\left(T_s+T_{st}+T_t \right) ,
\end{eqnarray}
with
\begin{eqnarray}
 T_s &=& (1+\cos^2\theta ) \left[ 1 +  2 \text{Re} \chi(s) \left( v^2\right) + |\chi(s)|^2\left(1+v^2 \right)^2 \right] 
 + 2 \cos\theta \left[  2 \text{Re} \chi(s) + |\chi(s)|^2\left(4v^2 \right) \right] ,
\\
T_{st} &=&
-2\frac{(1+\cos\theta)^2}{(1-\cos\theta)}  \left\lbrace 1 +    [ \chi(t)+ \text{Re}\chi(s)]  \left( 1+v^2\right) +
 \chi(t) \text{Re}\chi(s)\left[(1+v^2)^2+4v^2\right]  \right\rbrace  ,
\\
T_t &=& 2~\frac{(1+\cos\theta)^2}{(1-\cos\theta)^2}  \left\lbrace  1 +  2  \chi(t) \left( 1+v^2\right) + \chi(t)^2\left[(1+v^2)^2 +4v^2 \right] \right\rbrace 
\nonumber\\
&& + ~\frac{8}{(1-\cos\theta)^2} \left[ 1-  \chi(t) \left(1-v^2\right)\right]^2 .
\end{eqnarray}
We choose the following conventions:
\begin{eqnarray}
 v&=& 1-4 s_w^2,
\\
\chi(s) &=& \frac{G_{F}}{\sqrt{2}} \frac{M_Z^2}{8\pi \alpha} \frac{s}{s-M_Z^2+i M_Z \Gamma_Z},
\\
\chi(t) &=& \frac{G_{F}}{\sqrt{2}} \frac{M_Z^2}{8\pi \alpha} \frac{t}{t-M_Z^2}.
\end{eqnarray}
Among the quantities $\alpha, G_{F}, s_w^2, M_Z$ there are only three independent, and  $\Gamma_Z$ is predicted by the theory as well.
The phrasing \emph{effective Born cross-section} means here that we use, besides $\alpha$ (introduced in (\ref{alpha})),  the following input variables:
\begin{eqnarray}
 s_w^2 &=& 0.23,
\\
M_Z &=& 91.188 \mathrm{~ GeV},
\\
\Gamma_Z &=& 2.495  \mathrm{~ GeV},
\\
G_{F} &=& 1.16637 \times 10^{-5} \mathrm{~ GeV}^{-2}.
\end{eqnarray}
The values are, in a strict sense, related in the Standard Model, and may be determined e.g. by using the package ZFITTER \cite{Bardin:1999yd,Arbuzov:2005ma}.
Here, we took them from \cite{Eidelman:2004}.

\begin{figure}[t]
 \centering
\includegraphics[scale=0.8]{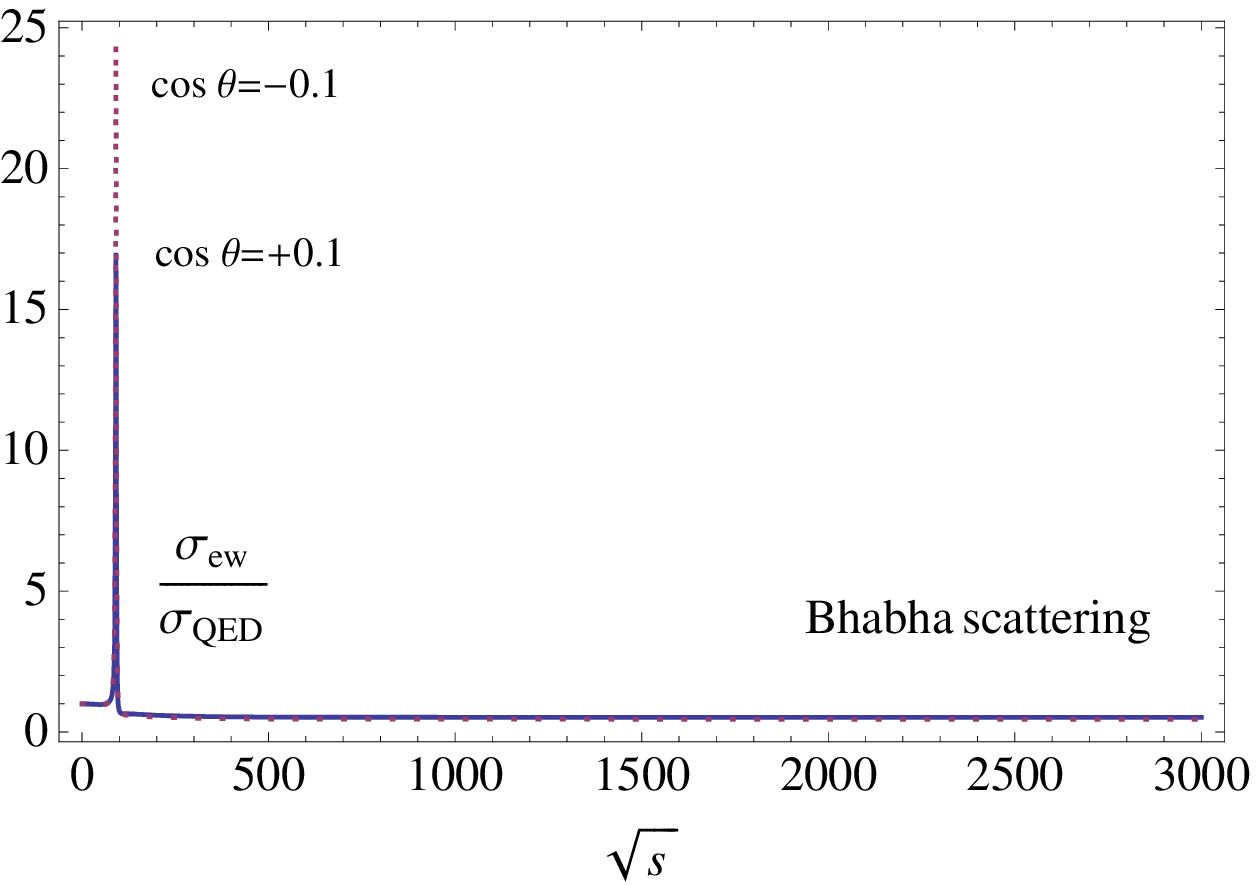}

\includegraphics[scale=0.8]{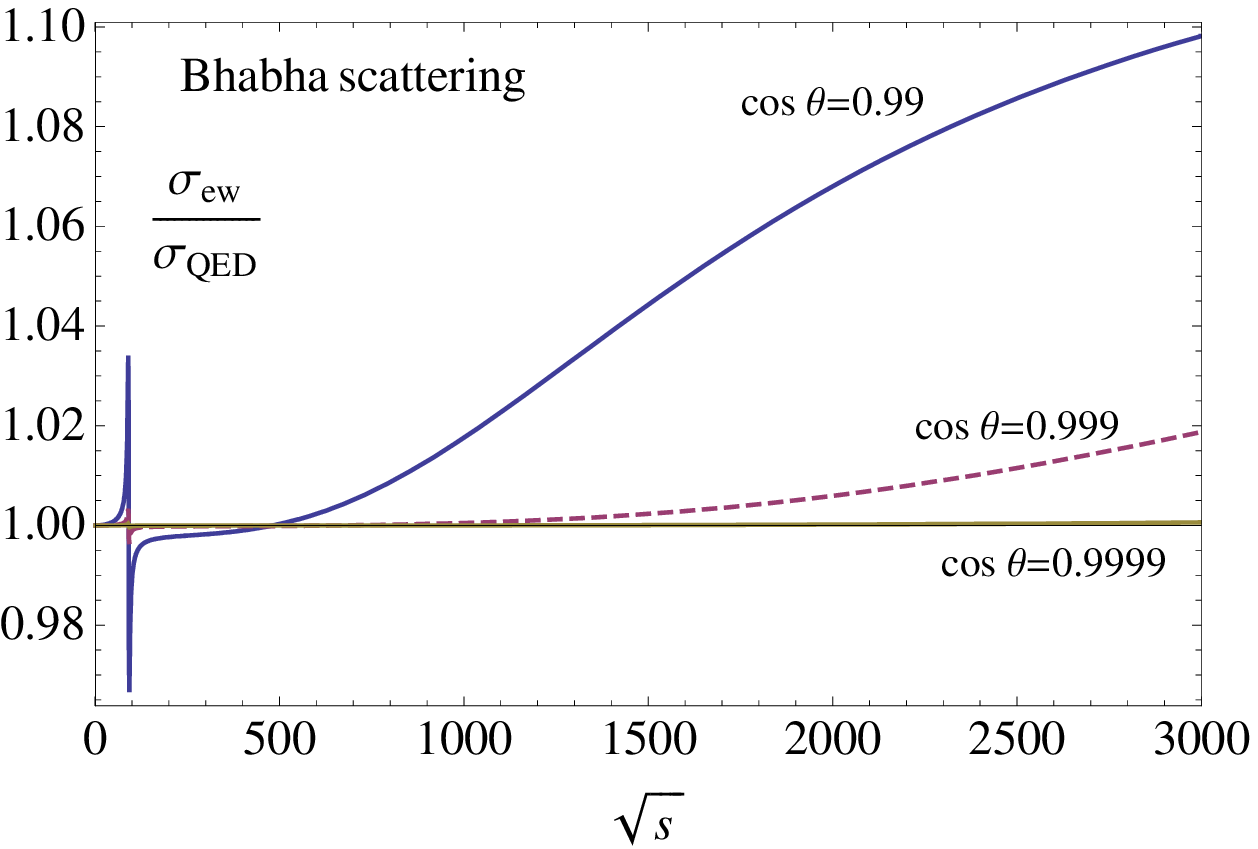}
 \caption[Ratio of electroweak to QED Bhabha scattering cross-section at large 
and small angles  as a function of $\sqrt{s}$.]
{\em Ratio of electroweak to QED Bhabha scattering cross-section at large angles (up)
and small angles (down) as a function of $\sqrt{s}$.}
 \label{fig:rat-SA2}
\end{figure}

\begin{figure}[t]
 \centering
\includegraphics[scale=0.8]{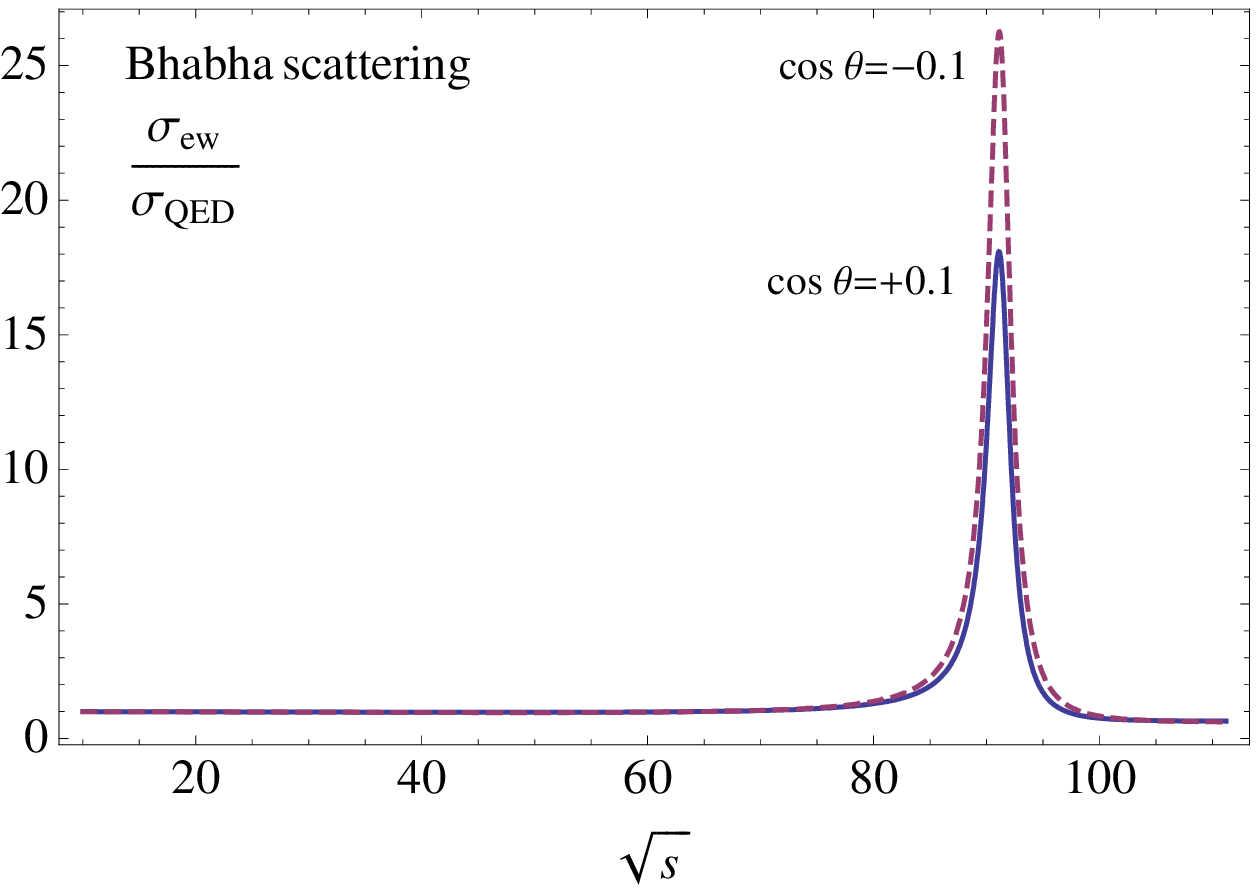}

\includegraphics[scale=0.8]{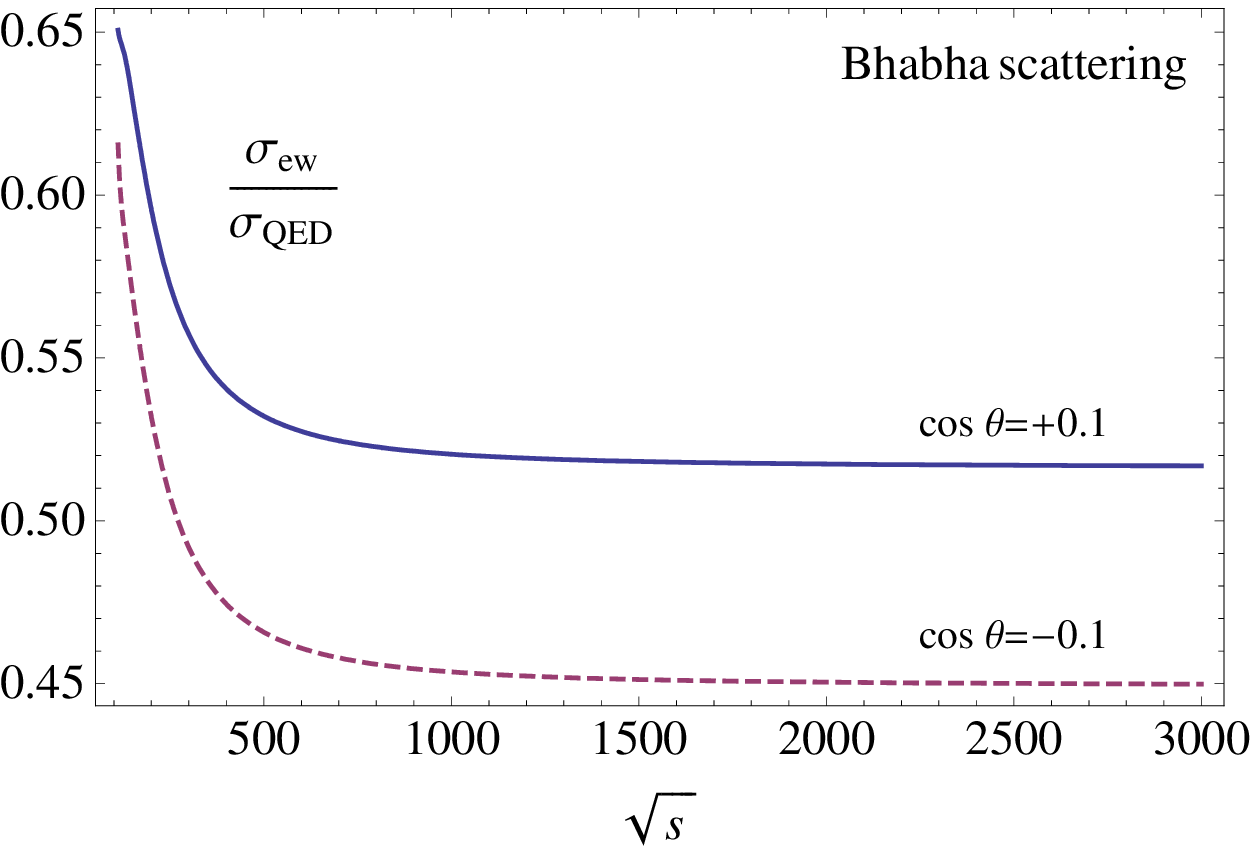}
 \caption[Ratio of electroweak to QED Bhabha scattering cross-section at large angles in the energy ranges of LEP 1  and ILC.]
{\em Ratio of electroweak to QED Bhabha scattering cross-section at large angles in the energy ranges of LEP1/GigaZ (up) and ILC (down).}
 \label{fig:rat-LA}
\end{figure}

We may now estimate the relevance of the $Z$-boson exchange to Bhabha scattering in different kinematic regions of interest.
It is minor at smallest energies where $s,|t|<<M_Z^2$, because there  $\chi(x) \sim x/M_Z^2<<1, x=s,t$.
The strength of the $Z$ exchange amplitude, relative to the photon exchange, becomes at large $s,|t|$ asymptotically:
\begin{eqnarray}
 \frac{G_{F}}{\sqrt{2}} \frac{M_Z^2}{8\pi \alpha}&=& 0.3739.
\end{eqnarray}
The other scale of relevance here is the ratio of photon propagators in the $s$- and $t$-channels:
\begin{eqnarray}
 \frac{s}{t} &=& -~\frac{2}{1-\cos\theta}.
\end{eqnarray}
In fact, at meson factory energies, the electroweak Born cross-section agrees with the QED prediction within few per mille,  and at LEP2 or the ILC within better than 50 \%,
 while at LEP1 or at GigaZ the ratio may become bigger than 25; this happens of course only for large scattering angles. 
At small angles, the corrections may safely be normalized to the QED Born cross-section everywhere.
The gross features are illustrated in  Figure~\ref{fig:rat-SA2} for large and small angle Bhabha scattering.
For large angles, we show the cross-section ratio separately for LEP1/GigaZ and the ILC in Figure~\ref{fig:rat-LA}.
We conclude that only for large angles at LEP 1 energies it is better to relate the corrections from  higher order contributions to the weak Born prediction, while for all other kinematics one may use the simple QED Born cross-section.

\section{\label{sec:pi-vac}THE VACUUM POLARIZATION}
Higher-order fermionic corrections to the Bhabha-scattering cross 
section can be obtained inserting the renormalized irreducible photon
vacuum-polarization function, $\Pi$, in the appropriate virtual-photon propagator,
\begin{equation}
\label{1stReplace}
\frac{g_{\mu\nu}}{q^2+i\, \delta} \, \to \,
\frac{g_{\mu\alpha}}{q^2+i\, \delta} \,
\left( q^2\, g^{\alpha\beta} - q^\alpha\, q^\beta \right) \,\Pi(q^2)\,
\frac{g_{\beta\nu}}{q^2+i\, \delta}.
\end{equation}
Here $q$ is the momentum carried by the virtual photon,
$\delta\to 0_+$.
The vacuum polarization
 $\Pi$ can be represented by the once-subtracted dispersion integral \cite{Cabibbo:1961sz}:
\begin{equation}
\label{DispInt}
\Pi(q^2) =
- \frac{q^2}{\pi} \, 
  \int_{4 M^2}^{\infty} \, d z \, 
  \frac{\text{Im} \, \Pi(z)}{z} \, 
  \frac{1}{q^2-z+i\, \delta},
\end{equation}
where the appropriate production threshold for the intermediate state in $\Pi$ is located at $q^2=4M^2$.
We leave as understood the subtraction at $q^2=0$ for the renormalized photon
self-energy.

Contributions to $\Pi$ arising from leptons and the top quark can be 
computed directly in perturbation theory, setting $M=m_f$
in Eq.~\eqref{DispInt}, where $m_f$ is the mass of the fermion
appearing in the loop, and inserting the imaginary part of the analytic 
result for $\Pi$.

We have at one-loop accuracy:
\begin{eqnarray}
\label{Im}
\text{Im}\, \Pi_f(z) =
&&-~ \left(\frac{\alpha}{\pi}\right)\, F_\epsilon\, 
\left(\frac{m_e^2}{m_f^2}\right)^\epsilon\,Q_f^2\, C_f\,
\theta\left( z-4\,m_f^2\right)\,
\frac{\pi}{3} \,
\Bigl\{\,
\frac{\beta_f(z)}{2} \, \Bigl[\,3-\beta_f^2(z)\,\Bigr]
\nonumber\\
&&+~ \epsilon\,\beta_f(z)\,
\Bigl[\, 3
      + \frac{3}{2}
      L_{\beta_f}(z)\,
      - \, \frac{4}{3}\, \beta_f^2(z)
      - \frac{\beta_f^2(z)}{2}\,
      L_{\beta_f}(z)\, 
\Bigr] \, \Bigr\}
+{\cal O}(\alpha^2),
\end{eqnarray}
where $Q_f$ is the electric charge, $Q_f=-1$ for leptons,
$Q_f= 2\slash 3$ for up-type quarks and $Q_f=-1\slash 3$ for down-type
quarks, and $C_f$ is the color factor, $C_f=1$ for
leptons and $C_f=3$ for quarks. In addition, we have introduced
the $\theta$ function, $\theta(x)=1$ for $x \geq 0$ and $\theta(x)=0$ for $x<0$,
and the threshold factor,
\begin{eqnarray}
\beta_f(z)&=&\sqrt{1-4\,\frac{m_f^2}{z}},
\\
L_{\beta_f}(z)&=&\ln\left(\frac{1-\beta_f^2(z)}{4\,\beta_f^2(z)}\right).
\end{eqnarray}
The overall regularization-dependent factor reads as
\begin{equation}
\label{norm}
F_\epsilon = 
\left( \frac{ m_e^2\, \pi\, e^{\gamma_E} }{\mu^2} \right)^{-\epsilon},
\end{equation}
where $\mu$ is the 't Hooft mass unit and $\gamma_E$ is the 
Euler-Mascheroni constant. 

The inclusion of the ${\cal O}(\epsilon)$ terms in Eq.~\eqref{Im} deserves a comment. 
These terms might play a role when combining $\text{Im}\, \Pi_f$ with a pole term of another one-loop insertion in a reducible two-loop Feynman diagram.
The Bhabha-scattering cross section we are going to consider is an 
infrared-finite quantity, provided one takes into account the real emission of soft photons. 
Therefore, when summing up all contributions, the result does not
show any pole in the $\epsilon$ plane and all radiative corrections,
including the one-loop photon self-energy, can be evaluated
at ${\cal O}(\epsilon^0)$. 
However, we retain the higher $\epsilon$ order
in Eq.~\eqref{Im} for comparing partial results with those of ~\cite{Actis:2007gi}.

In contrast to leptons and the top quark, light-quark contributions get 
modified by low-energy strong-interaction effects, which cannot be computed using
perturbative QCD. 
However, these contributions can be evaluated
using the optical theorem~\cite{Cutkosky:1960sp}. 
After relating $\text{Im} \, \Pi_{\rm had}$ to the 
hadronic cross-section ratio $R_{\rm had}$~\cite{Cabibbo:1961sz},
\begin{eqnarray}
\label{Rhad0}
\text{Im} \, \Pi_{\rm had}(z)&=& 
- \frac{\alpha}{3} \, R_{\rm had}(z),
\\
\label{Rhad}
R_{\rm had}(z) &=& 
\frac{\sigma(\{e^+e^-\to\gamma^\star\to \text{hadrons}\};z)}
     {(4 \pi \alpha^2)\slash (3z)},
\end{eqnarray}
$\text{Im}~\Pi_{\rm had}$ can be obtained from the experimental data for $R_{\rm had}$
in the low-energy region and around hadronic resonances, and the perturbative-QCD
prediction in the remaining regions.
The lower integration boundary is given by $M=m_\pi$, where $m_\pi$ is the pion mass.
For self-energy corrections to Bhabha scattering at one-loop order this was first employed in \cite{Berends:1976zn}.
Two-loop applications, similar to our study, are the evaluation of the hadronic vertex correction \cite{Kniehl:1988id} and of two-loop hadronic corrections to the lifetime of the muon \cite{vanRitbergen:1998hn}.
The latter study faces quite similar technical problems to those met here, like the infrared divergency of single contributions and the existence of several scales.

\begin{figure}[bt]
\begin{center}
\includegraphics[scale=0.65]{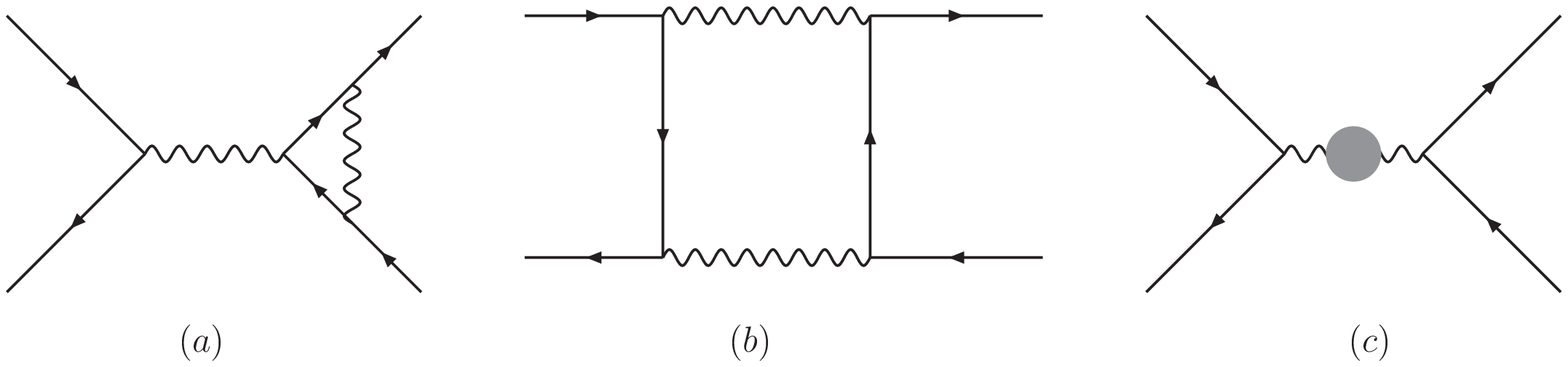}
\end{center}
\caption[The one-loop topologies for Bhabha scattering.]
{\em 
The one-loop topologies for Bhabha scattering.
The gray circle in (c) denotes the vacuum polarization under consideration, which may be understood to include fermionic and hadronic one- and two-loop irreducible self-energy corrections.}
\label{OneLoop}
\end{figure}

\begin{figure}[tbhp]
\begin{center}
\hfill
\includegraphics[scale=0.45]{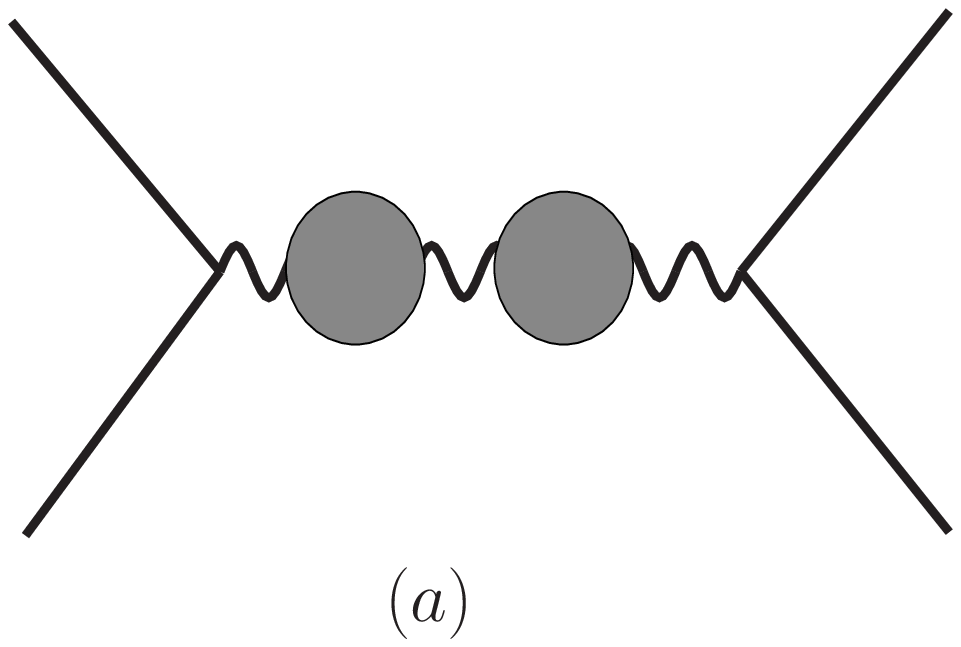}\hfill
\includegraphics[scale=0.45]{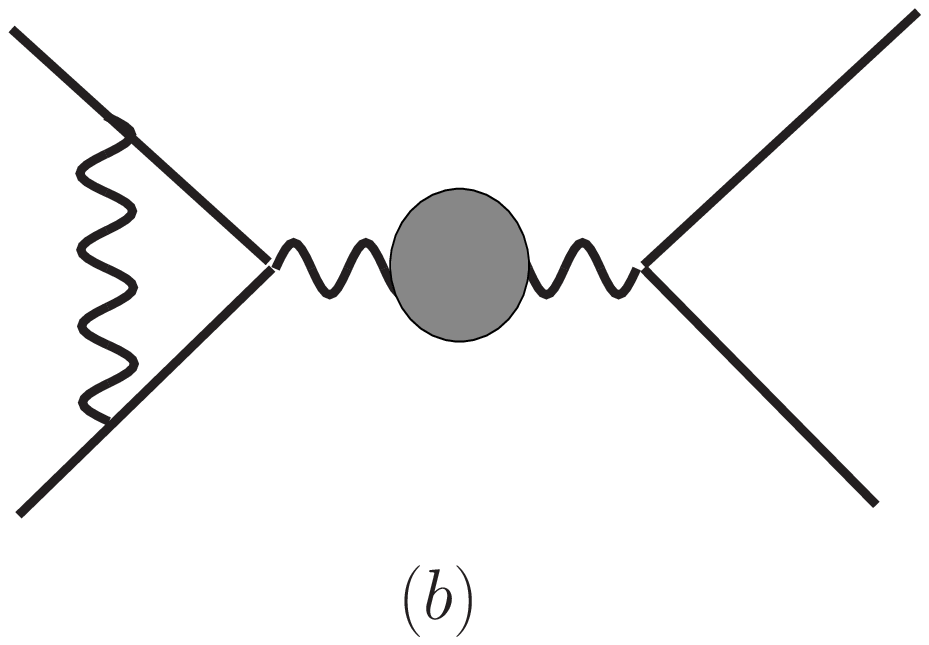}\hfill
\includegraphics[scale=0.45]{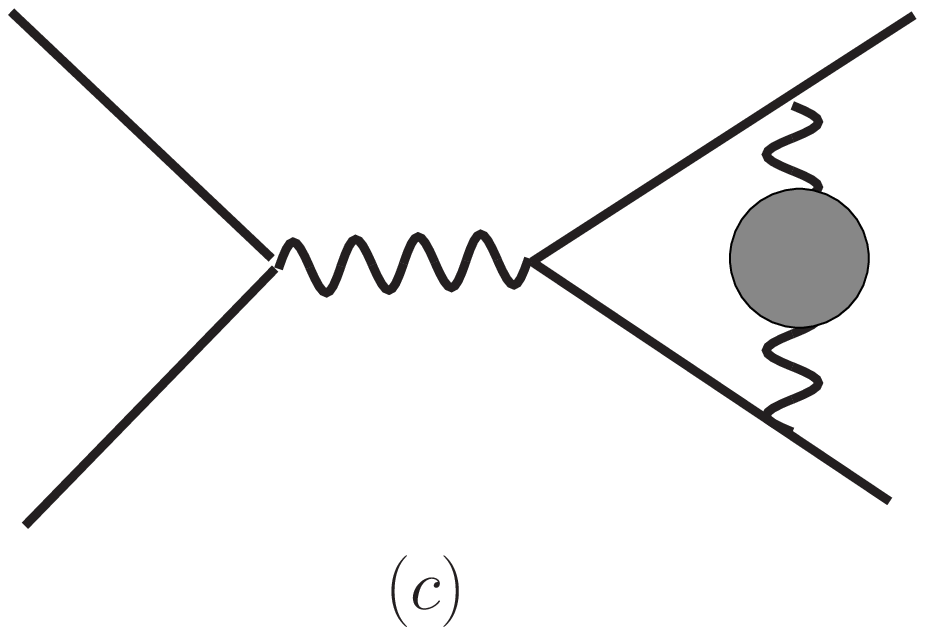}\hfill
\includegraphics[scale=0.45]{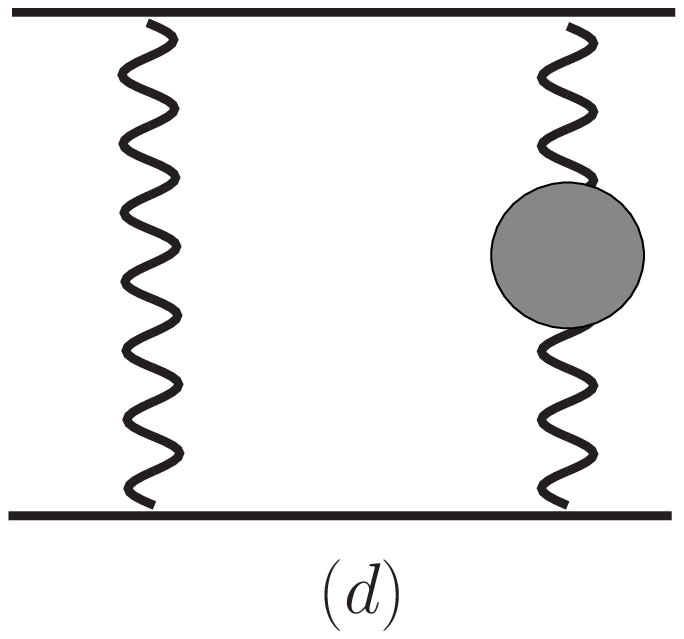}\hfill
\end{center}
\caption[Two-loop topologies for Bhabha scattering with vacuum polarization insertions.]
{\em 
 Two-loop topologies for Bhabha scattering with vacuum polarization insertions: 
reducible self-energy (a) and vertex (b) corrections as well as 
 irreducible vertex (c) and box (d) corrections;
for the irreducible self-energy corrections see Fig.~\ref{OneLoop}(c).}
\label{fig1}
\end{figure}

For the fermionic and hadronic corrections to Bhabha scattering at one-loop accuracy, there is only the \emph{self-energy diagram} shown in Fig.~\ref{OneLoop}(c).
The two-loop 
\emph{irreducible}  self-energy contributions have the topology shown in Fig.~\ref{OneLoop}(c).
One has additionally the four classes of \emph{two-loop diagrams} shown in Fig.~\ref{fig1} 
The \emph{reducible} self-energy (Figure~\ref{fig1}(a)) and vertex (Figure~\ref{fig1}(b)) topologies are much easier to evaluate than the \emph{irreducible} vertex (Figure~\ref{fig1}(c)) and box (Figure~\ref{fig1}(d)) topologies.
In fact, only the two-loop boxes were unknown until quite recently.  

The two-loop corrections have to be added with the loop-by-loop contributions (the interferences of the topologies of Fig.~\ref{OneLoop}) and with the soft photon corrections.
All these terms will be discussed in the following sections.

To summarize this section,
the  hadronic and heavy-fermion corrections to the Bhabha-scattering cross section
can be obtained by replacing appropriately the photon propagator by a massive 
propagator, whose effective mass $z$ is subsequently integrated over. 
Inserting ~\eqref{DispInt} and ~\eqref{Rhad0} in ~\eqref{1stReplace} we get:
\begin{equation}
\label{PropReplace}
\frac{g_{\mu\nu}}{q^2+i\, \delta} \,
\to\,
\frac{\alpha}{3 \pi}\,
\int_{4 M^2}^{\infty} \, d z \,
\frac{R(z)}{z}\, 
\frac{1}{q^2-z+i\, \delta}\,
\left(\,
g_{\mu\nu} - \frac{q_\mu\, q_\nu}{q^2+i\, \delta}\,
\right).
\end{equation}
In the following, we will call  the massive propagator function in (\ref{PropReplace}) the self-energy kernel function:
\begin{eqnarray}\label{kse}
 K_{\rm SE}(q^2; z) &=& \frac{1}{q^2-z+i\, \delta} .
\end{eqnarray}
The weight function
 $R(z)$ is given by the sum of the non-perturbative light-quark component
of Eq.~\eqref{Rhad} and the perturbative result of Eq.~\eqref{Im},
valid for leptons, $f=e,\mu,\tau$, and the top quark, $f=t$:
\begin{eqnarray}
\label{eq:finalR}
R(z)&=& R_{\rm had}^{(5)}(z) - \frac{3}{\alpha}\,\sum_{f=e,\mu,\tau,t} \,\text{Im}\,\Pi_f(z)
\nonumber\\
    &=& R_{\rm had}^{(5)}(z) +\sum_{f=e,\mu,\tau,t}\,R_f(z;m_f),
\\
\label{rzmf}
R_f(z;m_f)&=&
 Q_f^2\, C_f \,\left(1+2\, \frac{m_f^2}{z}\right)\,
        \sqrt{1-4\,\frac{m_f^2}{z}}.
\end{eqnarray}
Compared to (\ref{Im}), we omit here the terms of order $O(\epsilon)$.
The function $R_{\rm had}^{(5)}(z)$ will be discussed in Appendix \ref{app:rhad}.

Corrections related to electron insertions ($f = e$) will  be discussed separately.
For pure self-energy insertions  (see Appendix \ref{app:pho}), we may consider the electron mass as being small and neglect terms of order $O(m_e^2/x)$, $x=s,|t|,|u|$.
At the expense of that, even the three-loop corrections are known \cite{Steinhauser:1998rq}.
For two-loop irreducible vertex and box corrections, we may either consider $m_e$ being finite and treat a two-scale problem ($s/m_e^2, t/m_e^2$), or  we may assume also here $m_e^2 << s,|t|,|u|$.
Instead, for the diagrams with self-energy insertions of other fermions $f$, we will assume $m_e^2 << m_f^2, s,|t|,|u|$, but we will make no additional assumption on $m_f^2$.
\section{\label{sec-vac}PURE SELF-ENERGY CORRECTIONS}
The pure vacuum polarization contributions  to Bhabha scattering form a gauge invariant subset of diagrams.
So, their numerics may be discussed separately.
They can be readily obtained from the 
tree-level result ~\eqref{born} by introducing appropriately a 
running fine-structure constant $\alpha(x)$, where $x=s,t$,
\begin{eqnarray}\label{sigalfrun} 
\frac{ d\sigma_{\alpha\, {\rm run.}} }{ d\Omega } &=&
\frac{1}{2s}\,
\Bigl[\, |\alpha(s)|^2\frac{v_1(s,t)}{s^2} 
+ 2\alpha(t)\,\text{Re}\, \alpha(s) \frac{v_2(s,t)}{s\, t}
+\alpha^2(t) \frac{v_1(t,s)}{t^2}
\Bigr] + {\cal O}(m_e^2),
\end{eqnarray}
and where the running of $\alpha$ is defined as
\begin{equation}\label{alfar2} 
\alpha(x)\,=\,
\frac{\alpha}{1-\Delta\alpha(x)}.
\end{equation}
Here $\Delta \alpha$ is given by the sum of the non-perturbative
light-quark contribution $\Delta \alpha_{\rm had}^{(5)}$~\cite{Eidelman:1995ny}
(see Refs.~\cite{Burkhardt:2005se,Jegerlehner:2006ju,Hagiwara:2006jt}
and references therein for recent developments),
a perturbative electron-loop component evaluated in the small
electron-mass limit, $\Pi_e$, and a fermion-loop term computed exactly,
$\Pi_f$, with $f=\mu,\tau,t$,
\begin{eqnarray}
\label{dispA}
\Delta \alpha(x)&=& \Delta \alpha_{\rm had}^{(5)}(x) + \Pi_e(x) + \sum_{f=\mu,\tau,t} \Pi_f(x),
\\
\label{dispAlpha}
\Delta \alpha_{\rm had}^{(5)}(x)\,&=&\, \frac{\alpha}{\pi}\,\frac{x}{3} \, \int_{4 m_\pi^2}^{\infty}\, dz\, 
\frac{R_{\rm had}^{(5)}(z)}{z}\, 
 K_{\rm SE}(x; z) ,
\end{eqnarray}
with the self-energy kernel function $K_{\rm SE}(x; z)$ (\ref{kse}).

For $x<4 m_\pi^2$, Eq.~\eqref{dispAlpha} is well defined.
For $x> 4 m_\pi^2$, the  real and imaginary parts are  after a subtraction:
\begin{eqnarray}
\label{dispSE}
\text{Re} \left[ \Delta \alpha_{\rm had}^{(5)}(x) \right] \,&=&\, 
\frac{\alpha}{\pi}\,\frac{x}{3} \, \int_{4 m_\pi^2}^{\infty}\, dz\,
\frac{\left[ R_{\rm had}^{(5)}(z)-R_{\rm had}^{(5)}(x) \right] }{z \, (x-z)} + \frac{\alpha}{3 \pi}
R_{\rm had}^{(5)}(x) \log \left[ \frac{x}{4 m_\pi^2}-1 \right], \\
\text{Im} \left[ \Delta \alpha_{\rm had}^{(5)}(x) \right] \,&=&\, -\frac{\alpha}{3}\,R_{\rm had}^{(5)}(x).
 \end{eqnarray}
The $\text{Im} \left[ \Delta \alpha_{\rm had}^{(5)}(x) \right]$ coincides with Eq.~\eqref{Rhad0}.
Expressions for the perturbative contributions to the photon
vacuum-polarization function, $\Pi_f$ and $\Pi_e$, are available in QED exactly up to 
two loops~\cite{Kallen:1955fb} and in the small electron-mass limit up to 
three loops~\cite{Steinhauser:1998rq}.
For convenience, their explicit expressions are collected in Appendix~\ref{app:pho}.
For our analysis, we use the exact results of Eqs.~\eqref{oneloop} and \eqref{twoloops} for 
fermion loops ($f\neq e$),
and the high-energy expressions of Eqs.~\eqref{oneloopE}, \eqref{twoloopsE} and \eqref{threeloopsE} 
for electron loops.

In Tables~\ref{table:dalpha2} and ~\ref{table:dalpha1}
we show numerical values for the various
components of $\Delta \alpha$ of Eq.~\eqref{dispA} for space-like and time-like values 
of $x$ ($t$- and $s$-channel). 
Note that $\Delta \alpha$ develops an
imaginary part in the $s$-channel above the two-particle production threshold
(see Table~\ref{table:dalpha2}). 
Besides the Fortran package \texttt{hadr5.f} for hadronic contributions \cite{Jegerlehner-hadr5n:2003aa}, 
we employed the Mathematica package HPL~\cite{Maitre:2005uu,Maitre:2007kp}
and, as a cross check,  our Fortran routines (see Appendices \ref{app:pho} and \ref{app-polylog}).

\begin{table}[ht]\centering
\setlength{\arraycolsep}{\tabcolsep}
\renewcommand\arraystretch{1.1}
\begin{tabular}{|r|r|r|r|r|r|r|}
\hline 
$\sqrt{s}$ [GeV] & 1 & 10 & $M_Z$  & 500 \\
\hline 
\hline
1 loop  $e$ & 104.462 -- 24.3245 ~i & 
              140.119 -- 24.3245 ~i &
              174.347 -- 24.3245 ~i & 
              200.698 -- 24.3245 ~i \\
\hline 
$\mu$       &  21.352 -- 24.3060 ~i  & 
               57.551 -- 24.3245 ~i  &
               91.784 -- 24.3245 ~i  &
              118.136 -- 24.3245 ~i  \\
\hline 
$\tau$      & -- 0.508                & 
                12.194 -- 24.1724 ~i  &
                48.060 -- 24.3245 ~i  & 
                74.429 -- 24.3245 ~i \\
\hline 
$t$         &   $<10^{-3}$              &  
              -- 0.007                 &  
              -- 0.595                 &  
              -- 5.180 -- 29.0633 ~i   \\
\hline 
\hline
2 loops  $e$ & 0.258 -- 0.0424 ~i   &
               0.320 -- 0.0424 ~i   & 
               0.380 -- 0.0424 ~i   &
               0.426 -- 0.0424 ~i   \\
\hline 
$\mu$        & 0.123 -- 0.0487 ~i   &
               0.177 -- 0.0424 ~i   &
               0.236 -- 0.0424 ~i   & 
               0.282 -- 0.0424 ~i  \\
\hline 
$\tau$       &  -- 0.005  & 
               0.118 -- 0.0626 ~i  & 
               0.160 -- 0.0426 ~i  & 
               0.206 -- 0.0424 ~i  \\
\hline 
$t$          & $<10^{-3}$  &
               $<10^{-3}$  &  
                 -- 0.002  & 
             0.061 -- 0.0876 ~i \\
\hline
\hline 
3 loops  $e$ & 0.001   -- 0.0005 ~i    &
               0.002   -- 0.0006 ~i    &
               0.003   -- 0.0008 ~i    & 
               0.004   -- 0.0009 ~i    \\
\hline 
\hline
hadrons & -- 74.420  -- 37.9089 ~i& 
            138.850  -- 97.4106 ~i  & 
            276.213  -- 97.2980 ~i  & 
            370.744  -- 97.2980 ~i  \\

\hline 
\hline
SUM  & 51.263  -- 86.6310~i  & 
       349.324 -- 170.3800~i  & 
       590.586 -- 170.3997~i  & 
       759.806 -- 199.5505~i \\
\hline 
\end{tabular}
\caption[]{
Contributions to $\Delta \alpha$ in units of 
  $10^{-4}$ in the $s$-channel (see Eq.~\eqref{dispA}). The real part of the hadronic contributions 
  is obtained with help of the subroutine 
  \texttt{hadr5.f}~\cite{Jegerlehner-hadr5n:2003aa}, the imaginary
  part follows from the Burkhardt parametrization~\cite{Burkhardt:1981jk}.}
\label{table:dalpha2}
\end{table}

\begin{table}[ht]
\centering
\setlength{\arraycolsep}{\tabcolsep}
\renewcommand\arraystretch{1.1}
\begin{tabular}{|r|r|r|r|r|}
\hline
$\theta$ [$^\circ$] \, $\vert$ \, $\sqrt{s}$ [GeV] & $\theta=20$\, $\vert$ \, 1 & $\theta=20$\, $\vert$ \, 10 & 
$\theta=3$\, $\vert$ \, $M_Z$  & $\theta=3$\, $\vert$ \, 500 \\
\hline 
\hline
1 loop $e$ & 77.3512   &  113.008 & 117.935   & 144.286  \\
\hline 
     $\mu$ &  3.3069   &  30.614  &  35.463   &  61.727  \\
\hline 
    $\tau$ &  0.0148  &   1.346  &  2.365   &  18.804  \\
\hline 
       $t$ & $<10^{-4}$ &   $<10^{-3}$   &   $<10^{-3}$   &   0.012  \\
\hline
\hline 
2 loops $e$ & 0.2109    & 0.273     & 0.282   & 0.327  \\
\hline
      $\mu$ & 0.0260   &  0.126   & 0.136   & 0.184  \\
\hline 
     $\tau$ & 0.0001   & 0.011    & 0.019    & 0.097  \\
\hline 
        $t$ & $<10^{-4}$ & $<10^{-3}$ & $<10^{-3}$  &  $<10^{-3}$  \\
\hline
\hline 
3 loops $e$ & 0.0006 & 0.001  & 0.001  & 0.002  \\
\hline
\hline 
hadrons     &  2.6072      &  57.830       & 71.643        & 162.280        \\
\hline
\hline 
SUM     & 83.5177       &   203.209      & 227.844        & 387.719        \\
\hline
\hline
$\theta=90^\circ$\, $\vert$ \, $\sqrt{s}$ [GeV] & 1 & 10 & $M_Z$  & 500 \\
\hline 
\hline
1 loop $e$ & 99.0951  & 134.752 & 168.980 & 195.331 \\
\hline 
     $\mu$ & 17.4725  & 52.200 & 86.418   & 112.769  \\
\hline 
    $\tau$ & 0.2412  & 10.841   & 42.746   & 69.064  \\
\hline 
       $t$ & $< 10^{-4}$ & 0.003    & 0.284    & 6.208  \\
\hline
\hline 
2 loops $e$ & 0.2487  & 0.311 & 0.370 & 0.416 \\
\hline
      $\mu$ & 0.0924  & 0.167    & 0.227  & 0.273 \\
\hline 
     $\tau$ & 0.0021   & 0.068  & 0.150  & 0.196 \\
\hline 
        $t$ & $<10^{-4}$  & $< 10^{-3}$  & 0.001  & 0.021 \\
\hline
\hline 
3 loops $e$ & 0.0009 & 0.002 & 0.003 & 0.003  \\
\hline
\hline 
hadrons &  25.0834       & 127.219        & 256.279        &  362.375       \\
\hline 
\hline 
SUM     & 142.2363       &   492.396      & 555.458        & 746.656        \\
\hline
\end{tabular}
\caption[]{
Contributions to $\Delta \alpha$ in units of $10^{-4}$ in the 
  $t$-channel for three values of the scattering angle, $\theta=3^\circ$,
  $\theta=20^\circ$ and $\theta=90^\circ$, $t=-s\, \sin^2(\theta\slash 2)$.
  See the caption of Tab.~\ref{table:dalpha2} for further details.}
\label{table:dalpha1}
\end{table}

\begin{figure}[tbhp]
\begin{center}
\includegraphics[scale=0.55]{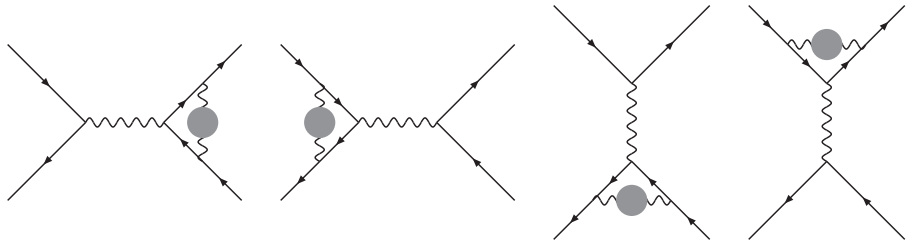}
\end{center}
\caption[Hadronic and fermionic irreducible vertex diagrams]
{\em Hadronic and fermionic irreducible vertex diagrams. 
The gray circles mark the corresponding one-loop insertions.
}
\label{Vertices2loopsIR}
\end{figure}

\section{\label{sec-irred-vert}IRREDUCIBLE VERTEX CORRECTIONS}
Hadronic and heavy-fermion irreducible vertex corrections are obtained through the 
interference of the diagrams of Figure~\ref{Vertices2loopsIR} with the 
tree-level amplitude. 
The contributions from the irreducible vertices are gauge invariant by themselves.
Their contribution to the ${\cal O}(\alpha^2)$ differential
cross section is given by
\begin{eqnarray}\label{sig-irr-vert}
\frac{ d\sigma_{\rm vert} }{ d\Omega } \!=\!
4 \left( \frac{\alpha}{\pi} \right)^2
 \left(\frac{  \, \alpha^2 }{2 s}\right) 
 \Bigl\{\, 
  \frac{v_1(s,t)}{s^2}\,  \text{Re} \, V_2(s) 
 \!+\!\frac{v_1(t,s)}{t^2}\,    V_2(t)
\!+\! \frac{v_2(s,t)}{s\,t} \,  \Bigl[\, \text{Re} \, V_2(s) + V_2(t) \,\Bigr]
 \,\Bigl\} + {\cal O}(m_e^2).
\end{eqnarray}
Here $V_2$ summarizes all two-loop fermionic corrections to the QED Dirac form factor,
whose computation can be traced back to the  seminal work of 
Refs.~\cite{Barbieri:1972as} and~\cite{Barbieri:1972hn}.
The full result can be organized as
\begin{eqnarray}
\label{fullVertex}
V_2(x)&=& V_{2e}(x) + V_{2{\rm rest}}(x),
\end{eqnarray}
where $V_{2e}$ denotes the electron-loop component.
Closed analytical expressions in the case of electron loops at finite $m_e$ can
be found in Ref.~\cite{Bonciani:2003ai}. 
In the high-energy limit,
compact expressions are available thanks to Ref.~\cite{Burgers:1985qg}:
\begin{eqnarray}
\label{eq:V2}
V_{2e}(x)\!= \!\frac{1}{36} \ln^3\Bigl(-\frac{m_e^2}{x}\Bigr)
+ \frac{19}{72}\, \ln^2\Bigl(-\frac{m_e^2}{x}\Bigr)
+\frac{1}{6}\,\left( \frac{265}{36}+\zeta_2 \right)\, \ln\Bigl(-\frac{m_e^2}{x}\Bigr) +~
\frac{1}{4}\left( \frac{383}{27} - \zeta_2 \right)+{\cal O}(m_e^2).
\end{eqnarray}
After a combination with soft real electron pair emission contributions (\ref{sp}), the leading logarithmic 
contributions  $\ln^3(s/m_e^2)$ get cancelled in (\ref{sig-irr-vert}). 

Heavy-fermion and hadronic contributions, instead, can be evaluated as in 
Ref.~\cite{Kniehl:1988id} through the dispersion integral
\begin{equation}
\label{eq:dispVer}
V_{2{\rm rest}}(x) = \int_{4 M^2}^{\infty} \, dz \, \frac{R(z)}{z} \, K_V(x+i\delta;z),
\end{equation}
where $R$ is given in Eq.~\eqref{eq:finalR} and the two-loop irreducible vertex kernel function $K_V$, in the limit of a vanishing electron mass, reads as
\begin{eqnarray}\label{eq:kernelV}
K_V(x;z) = \frac{1}{3} \, \Bigl\{\,
 - \, \frac{7}{8} 
 - \, \frac{z}{2\,x} 
 + \Bigl(\, \frac{3}{4} + \frac{z}{2\,x} \,\Bigr) \, 
   \ln\left(-\frac{x}{z}\right)
 - \, \frac{1}{2} \, \Bigl(\, 1 + \frac{z}{x}\, \Bigr)^2 \,
      \Bigl[\, \zeta_2 - \text{Li}_2\, 
      \left( 1 + \frac{x}{z} \right)\, \Bigr]\,
\Bigr\}.\nonumber\\
\end{eqnarray}
Here $\text{Li}_2(x)$ is the usual dilogarithm and $\zeta_2=\litwo(1) = \pi^2\slash 6$.
The kernel is at the upper integration boundary of the order $O(1/z)$, the integrand of order $O(1/z^2)$  
so that the dispersion integral is finite there:
\begin{eqnarray}\label{vertexbigz}
 K_V(x;z) & \approx& \frac{1}{3} \, \Bigl\{\,
\frac{11}{36} u -\frac{1}{6}u\ln(-u) + \left( -\frac{13}{288} + \frac{1}{24}\ln(-u)\right) u^2
\nl
&&+~ \left(\frac{47}{3600} -\frac{1}{60}\ln(-u)\right) u^3
\Bigr\} \mathrm{~~for~} u=\frac{x}{z}\to 0.
\end{eqnarray}
At the lower integration bound, the integrand becomes for small $z/x$:
\begin{eqnarray}\label{vertexsmz}
  K_V(x;z) \approx \frac{1}{3} \, \Bigl\{\,
-\frac{7}{8} -\zeta(2) +\frac{3}{4} \ln(-u) - \frac{1}{4} \ln^2(-u) - \left[1+2\zeta(2)+\frac{1}{2}\ln^2(-u) \right]\frac{1}{u} 
\Bigr\} \mathrm{~~for~} u=\frac{x}{z}\to \infty.
\nl
\end{eqnarray}
This asymptotic behavior yields at most terms of the order of  $\ln^3(x/M^2)$ if $M^2<<x$.

An interesting question is the identification of mass logarithms in case of fermion insertions.
Let us rewrite:
\begin{eqnarray}\label{fullVertexrest}
V_{2{\rm rest}}(x)&=& V_{2{\rm had}}^{(5)}(x) + \sum_{f=\mu,\tau,t} Q_f^2\, C_f\, V_{2f}(x),
\end{eqnarray}
where
$V_{2{\rm had}}^{(5)}$ denotes the non-perturbative light-quark term
and $V_{2f}$ the perturbative contribution of a fermion of flavor $f\neq e$.
Potentially large logarithms arise from parts of the integrand for the  $z$ integration which are singular at the lower integration bound, $z\to 4M^2$, when allowing thereby $M^2$ to become small.
For fermions, one has to analyze
 $R_f(z) K_V(x;z)/z$
in that limit.

The corresponding analytical integrations may be performed easily after applying the transformation
\begin{eqnarray}\label{rkv2}
 z &=& \frac{4m_f^2}{1-u^2},
\end{eqnarray} 
thereby getting rid of the square root function in $R_f(z)$: 
\begin{eqnarray}\label{rkv3}
 R_f(z)&=& C_f Q_f^2 \frac{u}{2}(3-u^2).
\end{eqnarray}
After that transformation, the dispersion integral becomes: 
\begin{eqnarray}\label{rkv4}
 V_{2f}(x) 
&=& \int_0^1 du \left[-2+u^2+\frac{1}{1-u}
+\frac{1}{1+u}  \right] K_V\left( x+i \delta; \frac{4m_f^2}{1-u^2}\right).   
\end{eqnarray}
 From the vertex kernel function $K_V(x;z)$, we have additionally  dependences on
 $\ln(-x/z)$ and on $\litwo(1+x/z)$.
Although after the  variable change (\ref{rkv2}) the arguments of logarithm and dilogarithm become non-linear, all the integrals may be taken trivially, and we will not go into further details.
The result 
contains $\litri$ and powers of logarithms $\ln^n(x/m_f^2)$ with $n\leq 3$.
In fact, one will rediscover in the kinematically interesting ultra-relativistic case the formula known from
 ~\cite{Burgers:1985qg} and e.g. also from ~\cite{Actis:2007gi}:
\begin{eqnarray}\label{eq:V3}
V_{2f}(x) &=&\frac{1}{36}\, \ln^3\Bigl(-\frac{m_f^2}{x}\Bigr)
+ \frac{19}{72}\, \ln^2\Bigl(-\frac{m_f^2}{x}\Bigr)
+\frac{1}{6}\,\left( \frac{265}{36}+\zeta_2 \right)\, \ln\Bigl(-\frac{m_f^2}{x}\Bigr) 
\nl
&&+~
\frac{1}{6}\left( \frac{3355}{216} +\frac{19}{6}\,\zeta_2 - 2\,\zeta_3 \right)
+{\cal O}(m_f^2).
\end{eqnarray}
The same soft- real pair cancellation mechanism as described for electrons works also for heavy fermions,
and the leading logarithmic powers $\ln^3(s/m_f^2)$ will get cancelled in the cross-section.
This is of physical relevance if the soft pair emissions remain unobserved.
In our numerical studies, we will, conventionally,  include the soft electron pair emission cross-section, but not that for heavy fermions or hadrons.
For further details see Section \ref{sec:realpairs}, and some numerical results were presented in
\cite{Actis:2008sk}, where we used the parameterization \cite{Burkhardt:1981jk} with flag setting $\texttt{IPAR} =1$.

We just mention that the transformation (\ref{rkv2}), when applied to  
the simple one-loop self-energy kernel (\ref{kse}), 
\begin{eqnarray}
 K_{\rm SE}(x;z) &=& \frac{1}{x-z} ~~=~~ \frac{1}{x}\left[ 1+ \frac{4m_f^2/x}{1-u^2-4m_f^2/x}\right] ,
\end{eqnarray}
gives a rational integrand for the $u$-integration, and one gets as a result a function at most linear in $\ln(s/m_f^2)$. For the explicit expressions see Equations (\ref{oneloop}) (constant term in $\epsilon$) and  (\ref{oneloopE}).
\section{\label{sec-IR}INFRARED-DIVERGENT CORRECTIONS}
There are various origins of heavy-fermion or hadronic infrared divergent cross-section contributions of order $O(\alpha^4)$:
\begin{itemize}
\item Factorisable diagrams with one-loop vertex or box insertions 
\item Irreducible two-loop box diagrams
\item soft real photon corrections  
\end{itemize}
The sum of these corrections is gauge-invariant and infrared finite.

\begin{figure}[bt]
\begin{center}
\includegraphics[scale=0.55]{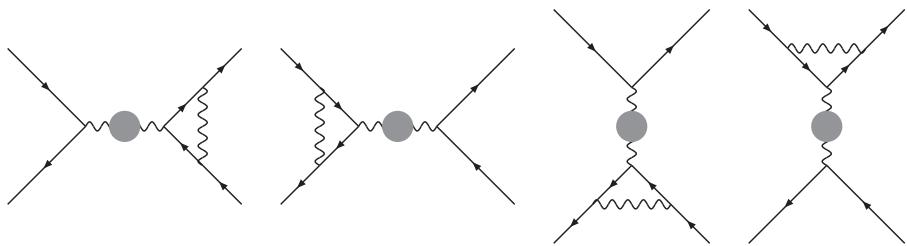}
\end{center}
\caption[Hadronic and fermionic reducible vertex diagrams.]
{\em Hadronic and fermionic reducible vertex diagrams. 
The gray circles mark the corresponding one-loop insertions.
}
\label{Vertices2loopsRED}
\end{figure}

We will consider five classes of contributions:
\begin{itemize}
 \item[(a)]  
Interference of  Born diagrams with reducible [vertex+self-energy] corrections of Fig.~\ref{Vertices2loopsRED};
\item[(b)]  Interference of one-loop vertex and self-energy diagrams, both  of Fig.~\ref{OneLoop};
\item[(c)]  Interference of one-loop  box and self-energy diagrams, both  of Fig.~\ref{OneLoop};
\item[(d)] Interference of real soft photon emission diagrams, one of them with a self-energy insertion;   
\item[(box)] Interference of   Born diagrams with two-loop box diagrams of Figure~\ref{Boxes2loops}.
\end{itemize}

For ease of notation, in the following we collect the 
overall dependence on $\alpha$ and rewrite the factorizing contributions of
class $i$, $i=a,\ldots,d$:
\begin{equation}
\label{labell}
  \frac{ d          \sigma_{\rm fact.}^i  }{ d\Omega } = 
       \left( \frac{ \alpha }{ \pi } \right)^2 \, 
              \frac{ \alpha^2 }{s}             \,
  \frac{ d\overline{\sigma}_{\rm fact.}^i }{ d\Omega },
\end{equation}
and analogously for the two-loop boxes.
In addition, we define 
\begin{eqnarray}
\hat{s}= \frac{s}{m_e^2},
\\
r=-\frac{t}{s},
\end{eqnarray}
and 
 introduce short-hand notations for those kinematic factors which 
appear more than once in the following formulas:
\begin{eqnarray}
A_r&=& -\frac{v_1(s,t)}{s^2}-\frac{v_2(s,t)}{st}
~=~
\frac{1}{r}\Bigl[\left(1-r\right)^3-r^3\Bigr],
\nonumber\\
B_r&=& \frac{v_1(t,s)}{t^2}+\frac{v_2(s,t)}{st}
~=~
\frac{1}{r^2}\Bigl[2\,\left(1-r\right)^2+r\,\left(1+r-r^2\right)\Bigr],
\nonumber\\
C_r&=& \frac{v_1(s,t)}{s^2}
~=~
\left(1-r\right)^2+r^2,
\nonumber\\
D_r&=& -\frac{v_2(s,t)}{st}
~=~
\frac{1}{r} \left(1-r\right)^2,
\nonumber\\
E_r&=& 3 \frac{v_1(t,s)}{t^2}+\frac{v_2(s,t)}{st}
~=~
\frac{1}{r^2}\Bigl[ 6\,\left(1-r\right)^2+r\,\left(5-r-r^2\right)\Bigr],
\nonumber\\
F_r&=& \frac{1}{2} \frac{v_1(t,s)}{t^2}+\frac{1}{4}\frac{v_2(s,t)}{st}
~=~
\frac{1}{4\,r^2} \Bigl[4\,\left(1-r\right)^2+r\,\left(3-r^2\right)\Bigr],
\nonumber\\
G_r&=& \frac{v_1(t,s)}{t^2}
~=~\frac{1}{r^2}\left(1+(1-r)^2\right).
\end{eqnarray}

\subsection{\label{sec-fact}Factorisable corrections with vertex or box insertions}
The
infrared-divergent factorisable heavy fermion and hadronic corrections for $m_e^2 << M^2,s,|t|,|u|$ can be readily 
obtained from Ref.~\cite{Actis:2007gi} by replacing the photon 
vacuum-polarization function in the $s$- or $t$-channel with the dispersion integral
\begin{eqnarray}
\label{disppi}
\Pi(x) \,~=~  \Delta\alpha(x)      \,  &=&\, \frac{\alpha}{\pi} ~I(x),
\\\label{disp}
I(x)\,&=&\, \frac{x}{3} \, \int_{4 M^2}^{\infty}\, \frac{dz}{z}\, 
\frac{R(z)}{x-z+i \,\delta},\qquad x=s,t,
\end{eqnarray}
where $\Delta\alpha(x)$ is given in ~\eqref{dispA} and $R$ in ~\eqref{eq:finalR}.

We begin with the reducible vertex corrections (a).
 From Eq.~(3.8) of Ref.~\cite{Actis:2007gi} we derive:
\begin{eqnarray}
\label{IR1}
\frac{d\overline{\sigma}_{\rm fact.}^a}{d\Omega} &=&
\frac{F_\epsilon}{\epsilon}\, \Bigl\{\,
    A_r \, \Bigl[\, \Bigl(1-\ln(\hat{s})\Bigr) \,
                  \text{Re}\,I(s) - \pi\,\text{Im}\,I(s)\, \Bigr]
  + B_r \, \Bigl[\, \ln(\hat{s})
+ \ln(r) - 1 \Bigr]\,I(t) \Bigr\}\nonumber\\
&&+~ \frac{1}{2}\, \Bigl\{ A_r\, 
  \Bigl[ \ln^2(\hat{s}) - 8\,\zeta_2 \Bigr]
           - \Bigl(A_r
- 2\,C_r\Bigr) \, \ln(\hat{s})
 + 2\,\Bigl( A_r - C_r\Bigr)\, \Bigr\}\, \text{Re}\,I(s)\nonumber\\
&&+~ \frac{1}{2}\, \Bigl[
    2\,A_r\, \ln(\hat{s})
  - \, A_r + 2\, C_r \Bigr] \,\pi\, \text{Im}\,I(s)
- \frac{1}{2}\, \Bigl\{ 
  B_r\, 
  \Bigl[ \ln^2(\hat{s}) + \ln^2(r)\Bigr]\nonumber\\
  &&-~ \Bigl[ E_r
- 2\, B_r\,\ln(r) \Bigr] 
    \ln(\hat{s}) -
  E_r\,\ln(r)
- 2\,B_r\, \zeta_2 + 8\,F_r\,
\Bigr\} \, I(t)
,
\end{eqnarray}
where the normalization factor $F_\epsilon$ is given  in Eq.~\eqref{norm}.
It appears here in the combination
\begin{eqnarray}
 \frac{F_\epsilon}{\epsilon} = \frac{1}{\epsilon} - \ln\left(\frac{m_e^2}{\mu^2} \right)
-\ln(\pi)  - \gamma_E + \mathcal{O}(\epsilon).
 \end{eqnarray}
In strict analogy, the interference of the one-loop vertex diagrams 
of Figure~\ref{OneLoop}~(a), with the vacuum-polarization diagrams 
of  Figure~\ref{OneLoop}~(c) can be extracted from Eq.~(3.26) of 
Ref.~\cite{Actis:2007gi}:
\begin{eqnarray}
\label{IR2}
\frac{ d\overline{\sigma}_{\rm fact.}^b }{ d\Omega } &=& 
\frac{F_\epsilon}{\epsilon}
\Bigl\{ \Bigl[  A_r\, \Bigl( 1- \ln(\hat{s}) \Bigr)- D_r\, \ln(r) \Bigr]\,
        \text{Re}\,I(s)\, 
 - C_r\, \pi\, \text{Im}\, I(s) \nonumber\\
&&+~ \Bigl[ B_r\, \Bigl(\, \ln(\hat{s}) - 1 \Bigr)
    + G_r \, \ln(r)\Bigr]\, I(t)\, \Bigr\}
+ \frac{1}{2}\,\Bigl\{
  \,A_r\, \ln^2(\hat{s}) \nonumber\\
  &&-~  2\, \Bigl[ \Bigl(1-4\,r\Bigr)\,D_r -4\, r^2 \Bigr]\, \zeta_2
- \Bigl[ A_r - 2\, C_r 
 - 2\, D_r \ln(r) \Bigr]
\ln(\hat{s}) \nonumber\\
&&+~ D_r\, \ln^2(r)
- \Bigl( D_r-2 \Bigr)\,\ln(r)
+ 2\, \Bigl[\Bigl(1-2\,r\Bigr)\,D_r-2\,r^2 \Bigr] 
\Bigr\} \, \text{Re}\,I(s)\nonumber\\
&&+~
\frac{1}{2} \Bigl\{2\, C_r\, \ln(\hat{s}) 
- \Bigl[C_r - 4\, r\Bigl(1-r\Bigr)\, \Bigr] \Bigr\}
 \, \pi\, \text{Im}\,I(s)
- \frac{1}{2}
\Bigl\{ B_r\, \ln^2(\hat{s})\nonumber\\
&&-~   \Bigl[ E_r
- \frac{2}{r^2}\, \Bigl( 1+r\,D_r \Bigr)\, \ln(r)\Bigr]\,
\ln(\hat{s})
+ \frac{1}{r^2} \Bigl( 1+r\,D_r\Bigr)\,\ln^2(r)
- \frac{1}{r^2}\,\times\nonumber\\
&&\times ~ \Bigl[ 6\,\Bigl(1-r\Bigr)+r^2 \Bigr]\,\ln(r)
- \frac{2}{r^2}\Bigl[ r\,\Bigl(1-4\,r\Bigr)\,D_r +1\Bigr]\, \zeta_2\
+ 8\, F_r \Bigr\}\,I(t).\nonumber\\
\end{eqnarray}
Finally, the contributions from
the  one-loop box diagrams of Figure~\ref{OneLoop}~(b)
may be derived from Eq.~(3.28) of Ref.~\cite{Actis:2007gi}:
\begin{eqnarray}
\label{IR3}
\frac{d\overline{\sigma}_{\rm fact.}^c}{d\Omega} &=&
\frac{ F_\epsilon }{ \epsilon }\,
\Bigl\{\,  \Bigl[\, C_r\,\ln(r) + A_r\,\ln(1-r)\, \Bigr]\,\text{Re}\,I(s)
       + D_r\, \pi\,\text{Im}\,I(s) 
       -  \Bigl[\,D_r\,\ln(r)\nonumber\\
& +& B_r\,\ln(1-r)\, \Bigr]\, I(t)\, \Bigr\}
- \Bigl\{\, \Bigl[\, C_r\,\ln(r) + A_r\,\ln(1-r)\, \Bigr]\, \ln(\hat{s})
       + \ln(r) \nonumber\\ &&+~ \frac{1}{2}\,\Bigl(2\, D_r+r\, \Bigr)\,\ln(1-r)
       +\frac{3}{4}\,\Bigl( 1-r\Bigr)\,\ln^2(r)
       + \frac{1}{4}\,\Bigl( 1-2\,r \Bigr) \,\ln^2(1-r) \nonumber\\
       &&+~ D_r\,\ln(r)\,\ln(1-r)\, \Bigr\}\,\text{Re}\, I(s) 
- \Bigl\{
   D_r\, \ln(\hat{s})
   + \frac{1}{2\,r}\,\Bigl[ D_r\,r\,\Bigl(1-r\Bigr)
   +1
   -3\, r^3 \Bigr]\,\times \nonumber \\ 
   &&\times ~ \ln(r)
   + \frac{1}{2}\Bigl[ 3\, \Bigl(1-2\,r\Bigr)+ 4\,r^2 \Bigr]\ln(1-r)
   +\frac{1}{2\,r} \, \Bigl(r\,D_r+1
    + 2\,r^2 \Bigr) \Bigr\}\, \pi\, \text{Im}\, I(s)\nonumber\\
&&+~ \Bigl\{\,  \Bigl[\, D_r\, \ln(r)
+ B_r\, \ln(1-r)\,\Bigr]\, \ln(\hat{s})
+  \frac{1}{2\,r}\Bigl( C_r+2 \Bigr)\,\ln(r)
- \frac{1}{2\,r}\Bigl[ r\,D_r\nonumber\\
& +&2\,\Bigl(1-r\Bigr)+r^2 \Bigr]\,\ln(1-r)
 +  \frac{1}{4\,r}\Bigl( 5-4\,r\Bigr)\, \ln^2(r)
 + \frac{1}{4\,r} \Bigl( 2-r\Bigr)\, \ln^2(1-r)\nonumber\\
 &&+~ \frac{1}{2\,r^2}\Bigl[ 2\,r\,D_r + 2\,\Bigl(1-r\Bigr)+r^2 \Bigr] \ln(r)\ln(1-r)
 + \frac{3}{2\,r}\, \Bigl( 2-r\Bigr)\,\zeta_2
\Bigr\}\, I(t)
.\nonumber\\
\end{eqnarray}

All three types of corrections are infrared divergent.
The vertex diagrams contribute leading electron mass singularities of the order $\ln^2(s/m_e^2)$, while  for the factorisable box diagrams the leading order is $\ln(s/m_e^2)$.
In addition, the self-energy insertions $I(x)$ yield a dependence on $\ln(s/m_f^2)$, in case $m_f^2$ is small compared to $s$.
This may be most easily seen from the $\epsilon$-independent terms in (\ref{oneloop}).
So, we collect here at most terms of the order $\ln^2(s/m_e^2) \ln(s/m_f^2)$.
\subsection{\label{sec-realsoft}Soft real photon emission}
In order to obtain an infrared-finite quantity, we take into 
account 
the interferences of diagrams with
real emission of soft photons from the external legs, where one of the diagram has a  
vacuum-polarization insertion.
The anatomy of these real corrections is exemplified in Appendix~\ref{app-soft}, where the soft photon factor is shown both for non-vanishing electron mass $m_e$ and in the ultra-relativistic approximation.
 The result may be also read off 
from Eq.~(4.4) of Ref.~\cite{Actis:2007gi} and reads as
\begin{eqnarray}
\label{IR4}
\frac{ d\overline{\sigma}_{\rm fact.}^d}{ d\Omega } &=& 
\frac{ d\overline{\sigma}_{\rm fact.}^{d,1} }{ d\Omega }+
\ln\left(\frac{2\,\omega}{\sqrt{s}}\right)\,
\frac{ d\overline{\sigma}_{\rm fact.}^{d,2} }{ d\Omega }
,
\end{eqnarray}
where $\omega$ is the maximum energy carried by a soft photon in the final state.
We obtain
\begin{eqnarray}
\frac{ d\overline{\sigma}_{\rm fact.}^{d,1} }{ d\Omega }&=&
\frac{F_\epsilon}{\epsilon}\,2\,
\Bigl[ \ln(\hat{s})+\ln(r)-\ln(1-r)-1\Bigr]\,
\Bigl[ A_r\, \text{Re}\, I(s) - B_r\, I(t)\Bigr] \nonumber\\
&&-~2\, \Bigl\{ \Bigl[
\frac{1}{2}\,\ln^2(\hat{s}) + \ln(\hat{s})\,\Bigl( \ln(r)-\ln(1-r) \Bigr)
+\frac{1}{2}\ln^2(r) - \frac{1}{2}\ln^2(1-r)\nonumber\\
& -&\ln(r)\,\ln(1-r) -2\,\text{Li}_2(r)-\zeta_2
\Bigr]\,\Bigl[A_r\, \text{Re}\, I(s) - B_r\, I(t) \Bigr]\nonumber\\
&&+~D_r\,\Bigl[
\ln(\hat{s})+\ln(r)
- \ln(1-r)-1
\Bigr] \Bigl[ \text{Re}\,I(s)+I(t)\Bigr]
 \Bigr\},\\
&&\nonumber\\
\frac{ d\overline{\sigma}_{\rm fact.}^{d,2} }{ d\Omega }&=&
-4\, \Bigl[\ln(\hat{s})+\ln(r)-\ln(1-r)-1 \Bigr]\,
\Bigl[A_r\, \text{Re}\,I(s) -B_r\, I(t)\Bigr].
\end{eqnarray}
Again, the infra-red divergency is contained in the factor $F_{\epsilon}/\epsilon$,
and the mass singularities are at most of the orders  $\ln(x/m_f^2), x=s,t$,  and $\ln^2(x/m_e^2)$ for the $\omega$-independent part and $\ln(x/m_e^2)$ for the $\omega$-dependent part.
\subsection{\label{sec:3}Two-loop irreducible box corrections}
\begin{figure}[t]
\begin{center}
\includegraphics[scale=0.65]{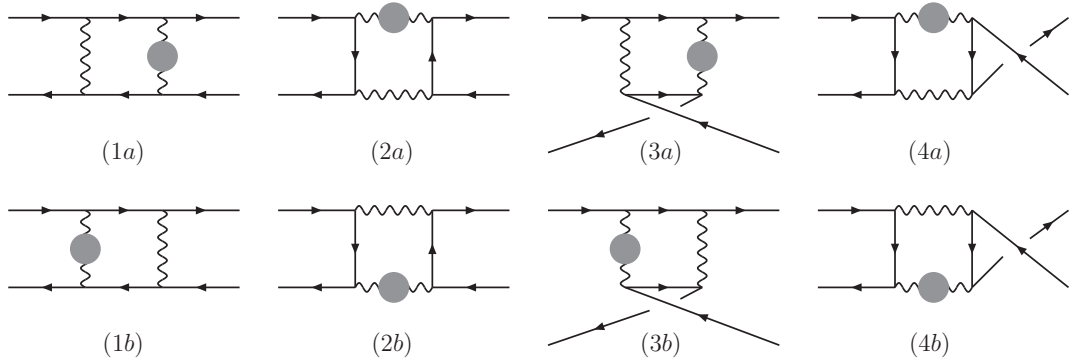}
\end{center}
\caption[Irreducible box diagrams.]
{\em Irreducible box diagrams. 
The gray circle denotes the hadronic or fermionic insertions.}
\label{Boxes2loops}
\end{figure}

 From the technical point of view, the two-loop irreducible box corrections of this section, represented by the  three box kernel functions, are the main result of the article. 
Their contributions  to the Bhabha-scattering cross
section arise from the interference of the diagrams of 
Figure~\ref{Boxes2loops} with the tree-level amplitude and can be 
written as
\begin{eqnarray}
\label{boxs}
  \frac{d\sigma_{\rm box}}{d\Omega} 
&=&
  \left(\frac{\alpha}{\pi}\right)^2 \, \frac{\alpha^2}{s}\,
   \frac{d\overline{\sigma}_{\rm box}}{d\Omega} \nonumber
\\
&=&
  \left(\frac{\alpha}{\pi}\right)^2 \, \frac{\alpha^2}{4\,s} \,
  2\, \left( \frac{\text{Re}\, A_s}{s}  +
         \frac{\text{Re}\, A_t}{t} \,  \right) .
\end{eqnarray}
Here the functions $A_s$ and $A_t$ contain the interferences of box 
diagrams with the $s$-channel and $t$-channel tree-level diagrams 
and can be represented through three independent form factors, 
evaluated with different kinematic arguments:
\begin{eqnarray}
\label{boxes}
  A_s &=& B_A(s,t) + B_B(t,s) + B_C(u,t) - B_B(u,s), 
 \\
  A_t &=& B_B(s,t) + B_A(t,s) - B_B(u,t) + B_C(u,s).
\end{eqnarray}
In addition, note that in Eq.~\eqref{boxs} we have collected an overall
factor $1\slash 4$, coming from the sum over the spins, and a factor
$2$, taking into account the fact that the contributions generated by
the diagrams $(1a)$, $(2a)$, $(3a)$ and $(4a)$
are equivalent to those of diagrams  $(1b)$, $(2b)$, $(3b)$ and $(4b)$
of Figure~\ref{Boxes2loops}. Finally, the correspondence among the form factors 
of Eq.~\eqref{boxes} and the diagrams of Figure~\ref{Boxes2loops} reads as follows:
\begin{eqnarray}\label{boxcombi}
{\rm diag.\, 1}\, &\times&\, {\rm tree_s}\, \Rightarrow\, B_A(s,t),\qquad
{\rm diag.\, 1}\,  \times\, {\rm tree_t}\, \Rightarrow\, B_B(s,t),\nonumber\\
{\rm diag.\, 2}\, &\times&\, {\rm tree_s}\, \Rightarrow\, B_B(t,s),\qquad
{\rm diag.\, 2}\,  \times\, {\rm tree_t}\, \Rightarrow\, B_A(t,s),\nonumber\\
{\rm diag.\, 3}\, &\times&\, {\rm tree_s}\, \Rightarrow\, B_C(u,t),\qquad
{\rm diag.\, 3}\,  \times\, {\rm tree_t}\, \Rightarrow\, -B_B(u,t),\nonumber\\
{\rm diag.\, 4}\, &\times&\, {\rm tree_s}\, \Rightarrow\, -B_B(u,s),\quad
{\rm diag.\, 4}\,  \times\, {\rm tree_t}\, \Rightarrow\, B_C(u,s).
\end{eqnarray}

We evaluate the three form factors $B_i$ using dispersion relations and computing 
thereby the convolution of the hadronic or fermionic cross-section ratio $R$ with three 
kernel functions $K_i$,
\begin{equation}
  \label{Bfc}
  B_i(x,y) = \int_{4M^2}^{\infty} \, dz \, \frac{R(z)}{z} \, 
  K_i(x ,y;z),
\end{equation}
where $R$ has been introduced in Eq.~\eqref{eq:finalR}, and the kernel function are to be calculated.
For positive $x$ or $y$, one has to replace $x \to x+i\delta$ or $y \to y+i\delta$.

The self-energy insertion is represented by a dispersion relation, thus replacing the one-loop photon propagator by a massive effective propagator as in Eq.~\eqref{PropReplace}. 
This procedure reduces the evaluation of the two-loop diagrams to one-loop complexity with a subsequent dispersion integration.
Employing standard techniques, together with the Mathematica packages \texttt{AMBRE} \cite{Gluza:2007rt} and \texttt{MB} \cite{Czakon:2005rk}, for a reduction of one-loop integrals to scalar master integrals, the kernel functions have  been finally expressed by eight one-loop master integrals $M^{(j)}(x,y;z)$,
\begin{equation}
\label{deco}
K_i(x,y;z) =  F_\epsilon \, \sum_{j=1}^8 \, C_i^{(j)}(x,y;z)\, M^{(j)}(x,y;z),
\end{equation}
where $F_\epsilon$ is the usual normalization factor of 
Eq.~\eqref{norm}, and $C_i^{(j)}$ are rational functions of the kinematic 
invariants, of the space-time dimension $d$, and of the two masses $m_e, z$. 
The master integrals $M^{(j)}$ are shown in Figure \ref{1loopmasters} and analytical expressions for them can be found in Appendix~\ref{app-masters}. 
Due to their length, we do not reproduce here the explicit (exact in $m_e$ and $d$ dimensions) right hand side of (\ref{deco}), but refer for them to the Mathematica file 
at the webpage \cite{webPage:2007xx}.

\begin{figure}[t]
\begin{center}
\includegraphics[scale=0.5]{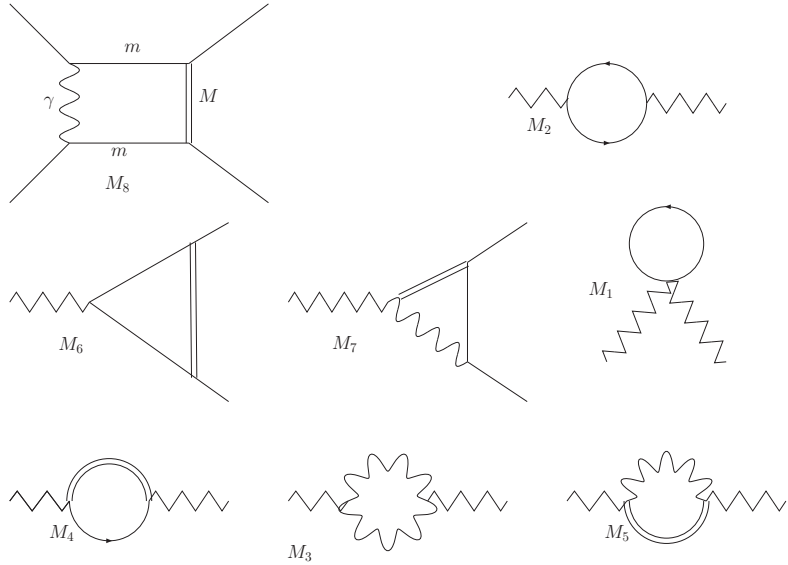}
\end{center}
\caption[The one-loop master integrals.]
{\em The one-loop master integrals with an additional mass scale $M=\sqrt{z}$ for the dispersive two-loop box evaluation.}
\label{1loopmasters}
\end{figure}

In the small electron-mass limit we obtain the two-loop box kernel functions:
\begin{eqnarray}
K_A(x,y;z) &=&
\label{KA}
\frac{1}{3\,(y-z)} \Bigl\{
-   2 \frac{F_\epsilon}{\epsilon} \left(x+y\right)^2 \ln\Bigl(-\frac{m_e^2}{x}\Bigr)
+   4\, \zeta_2 \Bigl[ z^2-z\Bigl(\frac{x^2}{y}+y\Bigr)\nonumber\\
                       &&+~2x\left(x+y\right)+y^2 \Bigr]
+ 2\,\Bigl[z\,\left(x+y\right)+x^2\Bigr] \, \ln\Bigl(-\frac{m_e^2}{x}\Bigr)
+   \Bigl[z^2+2\,z\,x \nonumber\\
    &&-~y\,\left(2\,x+y\right)\Bigr] \, \ln^2\Bigl(-\frac{m_e^2}{x}\Bigr)
+   \Bigl[ 2\,z^2\,\Bigl(\frac{x}{y}+1\Bigr)
           -z\,\Bigl(\frac{x^2}{y}+6\,x+5\,y\Bigr)\nonumber\\
           &&+~x\,\left(x+4\,y\right)+3\,y^2 \Bigr] \,\ln\Bigl(-\frac{m_e^2}{y}\Bigr) 
+  \Bigl[ z^2-2\,z\,\Bigl(\frac{x^2}{y}+x+y\Bigr)
          \nonumber\\
          &&+~2\,x\,\left(x+y\right)+y^2 \Bigr] \, \ln^2\Bigl(-\frac{m_e^2}{y}\Bigr) 
-   2\, \Bigl[ z^2+2\,z\,x+2\,x \, \left(x+y\right) \nonumber\\
               &&+~y^2\Bigr] \,\ln\Bigl(-\frac{m_e^2}{x}\Bigr)\,\ln\Bigl(-\frac{m_e^2}{y}\Bigr) 
+ \Bigl[ 2\,z^2\,\Bigl(\frac{x}{y}+1\Bigr)-z\,\Bigl(
           \frac{x^2}{y}+4\,x+3\,y\Bigr)\nonumber\\
     &&+~ \left(x+y\right)^2 \Bigr] \,
    \ln\Bigl(\frac{z}{m_e^2}\Bigr)
+ \Bigl[ 2\,z\,\Bigl(\frac{x^2}{y}+2\,x+y\Bigr)
           -\left(x+y\right)^2 \Bigr] \,
    \ln^2\Bigl(\frac{z}{m_e^2}\Bigr)\nonumber\\
&&-~   2\,\left(x+y\right)^2\,\ln\Bigl(\frac{z}{m_e^2}\Bigr)\,\ln\Bigl(-\frac{m_e^2}{x}\Bigr) 
+ 2\, \Bigl[ z^2-2\,z\,\Bigl(\frac{x^2}{y}+x+y\Bigr)
              +2\,x\, (x\nonumber\\
    &&+~y )
     + y^2 \Bigr] \,
    \ln\Bigl(\frac{z}{m_e^2}\Bigr)\,
    \ln\Bigl(1-\frac{z}{y}\Bigr)
 - \Bigl[ 2\,z^2\,\Bigl(\frac{x}{y}+1\Bigr)-z\,
           \Bigl(\frac{x^2}{y}+6\,x+5\,y\Bigr)\nonumber\\
           &&-~\frac{y}{z}
           \left(x+y\right)^2
           +2\,x\,\left(x+3\,y\right)
           +4\,y^2\Bigr]\,
    \ln\Bigl(1-\frac{z}{y}\Bigr)
+   2 \, \Bigl[z^2+2\,z\,x\nonumber\\
    &&+~2\,x (x
    + y )+y^2\Bigr]\,
    \ln\Bigl(1-\frac{z}{y}\Bigr)\,\ln\Bigl(-\frac{m_e^2}{x}\Bigr) 
+ 4 \Bigl[ \frac{z^2}{2}-z \Bigl(\frac{x^2}{y}+x+y\Bigr)\nonumber\\
             &&+~x\left(x+y\right)
    + \frac{y^2}{2} \Bigr]
    \text{Li}_2\Bigl(\frac{z}{y}\Bigr)
+ 2 \left(x+z\right)^2 \text{Li}_2\Bigl( 1+\frac{z}{x}\Bigr)
\Bigr\}, \\
&& \nonumber \\
K_B(x,y;z) &=&
\label{KB}
\frac{1}{3\,\left(y-z\right)} \, \Bigl\{
-   4\, \frac{F_\epsilon}{\epsilon}\,
    \Bigl[x\,\left(x+y\right)+\frac{y^2}{2}\Bigr]\,\ln\Bigl(-\frac{m_e^2}{x}\Bigr)
+   4\,\zeta_2\,\Bigl[z^2\nonumber\\
    &&-~2\,z\,\Bigl(\frac{x^2}{y}+\frac{y}{2}\Bigr)
    +2\,x\,\left(2\,x+y\right) + y^2 \Bigr]
+   2 \Bigl[z\,\left(x+y\right)-x\,y\Bigr]
    \ln\Bigl(-\frac{m_e^2}{x}\Bigr)\nonumber\\
&&+~   \Bigl[z^2+2 z\,x-y\left(2\,x+y \right)\Bigr]
    \ln^2\Bigl(-\frac{m_e^2}{x}\Bigr)
+ \Bigl[2\,z^2\,\Bigl(\frac{x}{y}+1\Bigr)-z\,
    \Bigl(2\,\frac{x^2}{y}+6\,x\nonumber\\
    &&+~5\,y\Bigr)
    +y\,
    \left(4\,x+3\,y\right)+2x^2\Bigr]\,
    \ln\Bigl(-\frac{m_e^2}{y}\Bigr)
+  \Bigl[z^2
   -2\, z\, \Bigl(2\, \frac{x^2}{y}
    + x+y\Bigr)\nonumber\\
   &&+~2\,x\,\left(2\,x + y\right)+y^2 \Bigr]\,
    \ln^2\Bigl(-\frac{m_e^2}{y}\Bigr)
- 2\,\Bigl[z^2+2\,z\,x
    + 2\,x\,( 2\,x 
    +y )
    +y^2\Bigr]\,\times \nonumber\\
    &&\times ~
    \ln\Bigl(-\frac{m_e^2}{x}\Bigr)\,\ln\Bigl(-\frac{m_e^2}{y}\Bigr)
+   \Bigl[2\,z^2\,\Bigl(\frac{x}{y}+1\Bigr)-z\,
    \Bigl(2\,\frac{x^2}{y}
    +4\,x+3\,y\Bigr)\nonumber\\
    &&+~ 2\,x \left(x+y \right)
    +y^2\Bigr]\,
    \ln\Bigl(\frac{z}{m_e^2}\Bigr)
+   4 \Bigl[z \Bigl( \frac{x^2}{y}+ x+\frac{y}{2}\Bigr) 
    -\frac{x}{2}\left(x+y\right)
    -\frac{y^2}{4} \Bigr]\times\nonumber\\
    &&\times ~\ln^2\Bigl(\frac{z}{m_e^2}\Bigr)
-   4 \Bigl[ x\left(x+y\right)
    + \frac{y^2}{2} \Bigr]
    \ln\Bigl(\frac{z}{m_e^2}\Bigr) \ln\Bigl(-\frac{m_e^2}{x}\Bigr)
+   2 \Bigl[z^2
     - 4 z \Bigl(\frac{x^2}{y}\nonumber\\
    &&+~\frac{x}{2}+\frac{y}{2}\Bigr)
    +2 x \left(2 x+y\right)
    +y^2\Bigr]
    \ln\Bigl(\frac{z}{m_e^2}\Bigr)
    \ln\Bigl(1-\frac{z}{y}\Bigr)
-   \Bigl[2\,z^2\,\Bigl(\frac{x}{y}+1\Bigr)\nonumber\\
    &&-~ 2\,z\,\Bigl(\frac{x^2}{y}+3\,x+\frac{5}{2}\,y\Bigr)
    -2\,\frac{y}{z}\,\Bigl(x^2+x\,y+\frac{y^2}{2}\Bigr)
     + 2\,(2\,x^2+2\,y^2\nonumber\\
    &&+~3\,x\,y )\Bigr]\,
    \ln\Bigl(1-\frac{z}{y}\Bigr)
+   2\,\Bigl[z^2
    + 2\,z\,x
    +2\,x\,\left(2\,x+y\right)
    + y^2\Bigr]\,\times\nonumber\\
    &&\times ~\ln\Bigl(1-\frac{z}{y}\Bigr)\,
    \ln\Bigl(-\frac{m_e^2}{x}\Bigr)
 +  2\,\Bigl[z^2-2\,z\,\Bigl(2\,\frac{x^2}{y}+x+y\Bigr)
    + 2\,x \, \left(2\,x+y\right)\nonumber\\
    &&+~ y^2 \Bigr]\,
    \text{Li}_2\Bigl(\frac{z}{y}\Bigr)
+   2\,\left(z^2+2\,x\,z+2\,x^2\right)\,
    \text{Li}_2\Bigl( 1+\frac{z}{x}\Bigr)
\Bigr\},\\
&& \nonumber \\
K_C(x,y;z) &=&
\label{KC}
\frac{1}{3\,(y-z)} \, \Bigl\{
     2\,\frac{F_\epsilon}{\epsilon}\,x^2\,\ln\Bigl(-\frac{m_e^2}{x}\Bigr)
+ 4\,\zeta_2\,x^2\,\Bigl(\frac{z}{y}-2\Bigr)
- 2\,(x^2+y^2\nonumber\\
  &&+~x\,y )\,\ln\Bigl(-\frac{m_e^2}{x}\Bigr)
 + x^2\,\Bigl(\frac{z}{y}-1\Bigr)\,\ln\Bigl(-\frac{m_e^2}{y}\Bigr)
 + 2\,x^2\,\Bigl(\frac{z}{y}-1\Bigr)\,\ln^2\Bigl(-\frac{m_e^2}{y}\Bigr)\nonumber\\
 &&+~ 4\,x^2\,\ln\Bigl(-\frac{m_e^2}{x}\Bigr)\,\ln\Bigl(-\frac{m_e^2}{y}\Bigr)
+ x^2\,\Bigl(\frac{z}{y}-1\Bigr)\,\ln\Bigl(\frac{z}{m_e^2}\Bigr)
 - 2\,x^2\,\Bigl(\frac{z}{y}-\frac{1}{2}\Bigr)\,\times\nonumber\\
   &&\times ~  \ln^2\Bigl(\frac{z}{m_e^2}\Bigr)
+ 4\,x^2\,\Bigl(\frac{z}{y}-1\Bigr)\,\ln\Bigl(\frac{z}{m_e^2}\Bigr)\,
     \ln\Bigl(1-\frac{z}{y}\Bigr)
 + 2\,x^2\,\ln\Bigl(\frac{z}{m_e^2}\Bigr)\,\times\nonumber\\
  &&\times ~ \ln\Bigl(-\frac{m_e^2}{x}\Bigr)
 - x^2\,\Bigl(\frac{z}{y}+\frac{y}{z}-2\Bigr)\,
   \ln\Bigl(1-\frac{z}{y}\Bigr)
- 4\,x^2\,\ln\Bigl(1-\frac{z}{y}\Bigr)\,\ln\Bigl(-\frac{m_e^2}{x}\Bigr)
     \nonumber\\
 &&+~  4 \,x^2\,\Bigl(\frac{z}{y}-1\Bigr)\,
     \text{Li}_2\Bigl(\frac{z}{y}\Bigr)
- 2\,x^2\,\text{Li}_2\Bigl(1+\frac{z}{x}\Bigr)
\Bigr\}.
\end{eqnarray}
These kernel functions are reproduced in  Mathematica files at the webpage \cite{webPage:2007xx}
as functions $KA, KB, KC$ and 
$KAexp, KBexp, KCexp$.

The two-loop box kernel masters (\ref{KA}) to (\ref{KC}) are evaluated in the Feynman gauge; they are infrared divergent and contain collinear singularities in $m_e$.  
    
After inserting Eq.~\eqref{KA}, Eq.~\eqref{KB} and Eq.~\eqref{KC} 
in Eq.~\eqref{Bfc}, we derive the total contribution to the
cross section generated by box diagrams. Collecting powers of $\alpha$, 
we write
\begin{eqnarray}
\label{boxAssembl}
\frac{ d\overline{\sigma}_{\rm box} }{ d\Omega }&=&
\int_{4M^2}^{\infty}\, dz\, \frac{R(z)}{z}\, \frac{1}{t-z}\, I_1(z)\nonumber\\
&&+~
\text{Re}\,\int_{4M^2}^{\infty}\, dz\, \frac{R(z)}{z}\, \frac{1}{s-z+i\,\delta}\,
\Bigl[ \,I_2(z)\,+\, I_3(z)\, \ln\Bigl(1-\frac{z}{s+i\, \delta}\Bigr)\, \Bigr]\nonumber\\
&&+~\pi\, \text{Im}\, \int_{4M^2}^{\infty}\, dz\, \frac{R(z)}{z}\, 
\frac{1}{s-z+i\,\delta} \, I_3(z),
\end{eqnarray}
where the integrand functions are given by
\begin{eqnarray}\label{i1z}
I_1(z)&=&\frac{1}{3}\,\Bigl\{
-\Bigl[ \frac{F_\epsilon}{\epsilon}-\ln\Bigl(\frac{s}{m_e^2}\Bigr)+\ln\Bigl(\frac{z}{s}\Bigr)\Bigr]\,
 \ln\Bigl(-\frac{u}{s}\Bigr)\,
\Bigl[ \frac{v_1(t,s)}{t}+ \frac{v_2(s,t)}{s}  \Bigr]
\nonumber\\
&&-~\zeta_2\, \Bigl[ 2\,\frac{z^2}{t}
-4\,z\,\Bigl(1+\frac{s}{t}\Bigr)-\frac{t^2}{s}
-2\,\frac{s^2}{t}+s-t\Bigr]
-\Bigl[ z\,\frac{s}{t}-\frac{t^2}{s}-2\,\Bigl(s+t\Bigr)\Bigr]\,\times\nonumber\\
&&\times ~\ln\Bigl(1+\frac{t}{s}\Bigr)
-\frac{1}{2}\, \Bigl[ \frac{z^2}{t}-2\,z\,\Bigl(1+\frac{s}{t}\Bigr)
+2\,s+t \Bigr]\,
\ln^2\Bigl(1+\frac{t}{s}\Bigr)
+\Bigl[ z^2\,\Bigl( \frac{1}{s}
+2\,\frac{s}{t^2}\nonumber\\
&&+~ \frac{2}{t}\Bigr)
-z\,\Bigl( \frac{t}{s}+2\frac{s}{t}+2 \Bigr)\Bigr]
\ln\Bigl(\frac{z}{s}\Bigr)
-\Bigl[ z^2\,\Bigl(\frac{1}{s}+\frac{1}{t}\Bigr) +2\,z\,\Bigl(1+\frac{s}{t}\Bigr)
+s
+2\,\frac{s^2}{t}\Bigr]\,\times\nonumber\\
&&\times ~\ln\Bigl(\frac{z}{s}\Bigr)\,
\ln\Bigl(1+\frac{z}{s}\Bigr)
+ \Bigl[\frac{z^2}{s}
+4\,z\,\Bigl(1+\frac{s}{t}\Bigr)-\frac{t^2}{s}-4\,\Bigl(s+t\Bigr)\Bigr]\,
\Bigl[ \ln\Bigl(\frac{z}{s}\Bigr)\,\ln\Bigl(1-\frac{z}{t}\Bigr)\nonumber\\
&&+~\frac{1}{2}\,
\ln^2\Bigl(-\frac{t}{s}\Bigr)\Bigr]
- \Bigl[ z^2\,\Bigl(\frac{1}{s}+2\frac{s}{t^2}+ \frac{2}{t} \Bigr)
-2\, z\,\Bigl(\frac{t}{s}+2\,\frac{s}{t}
+2\Bigr)+\frac{t^2}{s}+2\,\Bigl( s+t\Bigr)\Bigr]\,\times\nonumber\\
&&\times ~ \Bigl[ \ln\Bigl(1-\frac{z}{t}\Bigr) + \ln\Bigl(-\frac{t}{s}\Bigr)\Bigr]
+\Bigl[ \frac{z^2}{t}-2\,z\,\Bigl(1+\frac{s}{t}\Bigr)+2\,\frac{t^2}{s}
+8\,s
+4\,\frac{s^2}{t}+7\,t \Bigr]\,\times\nonumber\\
&&\times ~ \Bigl[ \ln\Bigl(1-\frac{z}{t}\Bigr)
+ \ln\Bigl(-\frac{t}{s}\Bigr)\Bigr]\,\ln\Bigl(1+\frac{t}{s}\Bigr)
-\Bigl[
z^2\,\Bigl(\frac{1}{s}+\frac{1}{t}\Bigr) +2\,z\,\Bigl(1+\frac{s}{t}\Bigr)\nonumber\\
&&+~s + 2\,\frac{s^2}{t}
\Bigr]\, \text{Li}_2\,\Bigl(-\frac{z}{s}\Bigr)
+ \Bigl[\frac{z^2}{s}+4\,z\,\Bigl(1+\frac{s}{t}\Bigr)-\frac{t^2}{s}-4\,\Bigl(s+t\Bigr)\Bigr]\,
\text{Li}_2\,\Bigl(\frac{z}{t}\Bigr)\nonumber\\
&&-~ \Bigl[\,\frac{z^2}{t}
-2\,z\,\Bigl(1+\frac{s}{t}\Bigr)
+\frac{t^2}{s}+5\,s+2\,\frac{s^2}{t}
+4\,t\, \Bigr]\,\text{Li}_2\,\Bigl(1+\frac{z}{u}\Bigr)
\Bigr\},\\
&&\nonumber\\
I_2(z) &=& \frac{1}{3}\,\Bigl\{
-\Bigl[ \frac{F_\epsilon}{\epsilon} -\ln\Bigl(\frac{s}{m_e^2}\Bigr) + \ln\Bigl(\frac{z}{s}\Bigr)\Bigr]\,
   \ln\Bigl(\frac{u}{t}\Bigr)\,
  \Bigl[ \frac{v_1(s,t)}{s} + \frac{v_2(s,t)}{t}
\Bigr]
\nonumber\\
&&-~\Bigl[ z\,\frac{t}{s}-\frac{s^2}{t}-2\,\Bigl(s+t\Bigr)\Bigr]\,
\ln\Bigl(1+\frac{t}{s}\Bigr)
- \frac{1}{2}\, \Bigl[\,\frac{z^2}{s}-2\,z\,\Bigl( 1+\frac{t}{s}\Bigr)\nonumber\\ 
&&+~s+2\,t \Bigr]\,
\ln^2\Bigl(1+\frac{t}{s}\Bigr)
-z\,\Bigl( \frac{t}{s}+\frac{s}{t}+2 \Bigr)\,\ln\Bigl(-\frac{t}{s}\Bigr)
+\frac{1}{2}\,\Bigl[z^2\,\Bigl(\frac{1}{s}+\frac{1}{t}\Bigr)\nonumber\\
&&+~2\,z\,
\Bigl(1+\frac{t}{s}\Bigr)
-\frac{s^2}{t}-3\,s-2\,t\Bigr]\, \ln^2\Bigl(-\frac{t}{s}\Bigr)
+\Bigl[ z^2\,\Bigl(\frac{1}{t}+\frac{2}{s}+2\,\frac{t}{s^2}\Bigr)
-z\,\Bigl(\frac{s}{t}+2\nonumber\\
&&+~2\,\frac{t}{s}\Bigr)\Bigr]\,\ln\Bigl( \frac{z}{s} \Bigr)
-\Bigl[ \frac{z^2}{t}+4\,z\,\Bigl(1+\frac{t}{s}\Bigr)-\frac{s^2}{t}-4\,\Bigl(s+t\Bigr)\Bigr]\,
\text{Li}_2\,\Bigl(1-\frac{z}{s}\Bigr)\nonumber\\
&&+~\Bigl[ z^2\Bigl(\frac{1}{s}+\frac{1}{t}\Bigr)
+2\,z\,\Bigl(1+\frac{t}{s}\Bigr)+2\,\frac{t^2}{s}+t \Bigr]
\text{Li}_2\,\Bigl(1+\frac{z}{t}\Bigr)\nonumber\\
&&-~\Bigl[ \frac{z^2}{s}-2\,z\,\Bigl(1+\frac{t}{s}\Bigr)+\frac{s^2}{t}
+2\,\frac{t^2}{s}+4\,s
+5\,t\Bigr]\,\text{Li}_2\,\Bigl(1+\frac{z}{u}\Bigr)
\Bigr\},\\
&&\nonumber\\
\label{I3}
I_3(z)&=&\frac{1}{3}\,\Bigl\{\,
\Bigl[\, \frac{z^2}{s}-2\,z\,\Bigl(1+\frac{t}{s}\Bigr)+4\,\frac{t^2}{s}
+2\,\frac{s^2}{t}+7\,s+8\,t\Bigr]\,\ln\Bigl(1+\frac{t}{s}\Bigr)\nonumber\\
&&-~\Bigl[ z^2\,\Bigl(\frac{1}{s}+\frac{1}{t}\Bigr) +2\,z\,\Bigl(1+\frac{t}{s}\Bigr)
+4\,\frac{t^2}{s}+\frac{s^2}{t}+3\,s+4\,t \Bigr]\,\ln\Bigl(-\frac{t}{s}\Bigr)\nonumber\\
&&-~\Bigl[ z^2\,\Bigl(\frac{1}{t}+\frac{2}{s}+2\,\frac{t}{s^2}\Bigr)-2\,z\Bigl(2+\frac{s}{t}
+2\,\frac{t}{s}\Bigr)
+\frac{s^2}{t}+2\,\Bigl(s+t\Bigr)\,
\Bigr]\,
\Bigr\}.
\end{eqnarray}
The functions $I_1(z)$ to $I_3(z)$ are reproduced  as functions $I_1, I_2, I_3$ in a Mathematica file 
at the webpage \cite{webPage:2007xx}.

Note that, after assembling all irreducible box diagrams, 
their total contribution is free of collinear divergencies in $m_e$ because $\ln(m_e^2)$ vanishes in
the combination
\begin{eqnarray}\label{me2free}
\frac{F_\epsilon}{\epsilon} -\ln\Bigl(\frac{s}{m_e^2}\Bigr) &=&
\frac{1}{\epsilon} - \gamma_{E} -\ln(\pi) -\ln\Bigl(\frac{s}{\mu^2}\Bigr) + 0(\epsilon).
\end{eqnarray}
This fact might be observed already for any sum of single pairs of direct and their related crossed box diagrams, which is gauge-independent and  free of collinear singularities \cite{Frenkel:1976bj}; from (\ref{boxcombi}) and Figure \ref{Boxes2loops} one selects e.g. the following ones:
\begin{eqnarray}\label{me2fre2}
 K_B(t,s;z) - K_B(u,s;z),\nonumber
\\
K_A(s,t;z) + K_C(u,t;z).
\end{eqnarray}
In the limit $m_f^2<<s,|t|,|u|$, the $z$-integration over the $I_i(z), i=1,2,$ develops mass singularities from the lower integration bound:
\begin{eqnarray}\label{mf2a2}
 \int\limits_{4M^2}^{\infty}dz \frac{R(z)}{z}K_{\rm SE}(y;z) \left[ A(x,y,z) + B(x,y)\ln\Bigl(\frac{z}{s}\Bigr)\right] 
\end{eqnarray}
where $A, B$ are regular for $z\to 0$.
It follows immediately that the irreducible box diagrams yield terms of the order of at most $\ln^2(s/m_f^2)$, because $A$ joins, after integration, terms with a behavior like  a one-loop self-energy, and $B$ joins terms with one order more in the logarithmic structure.
This has been discussed already in \cite{Actis:2007gi}.

The residual infrared-singular part of the box cross-section is:
\begin{eqnarray}\label{i1teps}
\frac{ d\overline{\sigma}_{\rm box}^{IR} }{ d\Omega }&=&
- \left[ \frac{F_\epsilon}{\epsilon}-\ln({\hat s})\right] \!
\left\lbrace 
 \ln\Bigl(-\frac{u}{s}\Bigr)
 \Bigl[ \frac{v_1(t,s)}{t^2}+\frac{v_2(s,t)}{st}\Bigr] 
I(t)\right. 
+ \left. 
 \ln\Bigl(\frac{u}{t}\Bigr)
 \Bigl[ \frac{v_1(s,t)}{s^2}+\frac{v_2(s,t)}{st}\Bigr] 
I(s)\!
\right\rbrace \! .
\end{eqnarray}
The function $I(t)$ (see Eq.~(\ref{disp})) stems from diagrams with a vacuum polarization insertion in the $t$-channel, and $I(s)$ from insertions in the $s$-channel.
One may wonder which of the other infrared divergent parts are needed to compensate the double-box divergency (in the gauge chosen here).
This may be exemplified by collecting all the IR-divergencies of the diagrams with a vacuum polarization insertion $I(t)$ in the $t$-channel; for the others, quite analogue arguments hold.
 From Sections~\ref{sec-fact} and ~\ref{sec-realsoft} we may extract such terms.
There are the following divergencies due to vertex diagrams:
\begin{eqnarray}\label{redateps}
 \frac{d\overline{\sigma}_{\rm fact.}^{a,IR}} {d\Omega} &=&
\left[ \frac{F_\epsilon}{\epsilon}-\ln({\hat s})\right]
\left[\ln({\hat s})-1+\ln\left( -\frac{t}{s}\right)  \right] \left( \frac{v_1}{t^2}+\frac{v_2}{st}\right) ~I(t)  ,
\\\label{redbteps}
\frac{d\overline{\sigma}_{\rm fact.}^{b,IR}} {d\Omega} &=&
\left[ \frac{F_\epsilon}{\epsilon}-\ln({\hat s})\right]
\left\lbrace 
\left[\ln({\hat s})-1+\ln\left( -\frac{t}{s}\right)  \right] ~ \frac{v_1}{t^2}
+  \left[\ln({\hat s})-1 \right] ~\frac{v_2}{st} 
\right\rbrace 
~I(t)  .
\end{eqnarray}
The reducible box diagrams are (in the curly brackets) free of electron mass singularities, also in the terms not shown here.
They depend also on $u$:
\begin{eqnarray}\label{redcteps}
 \frac{d\overline{\sigma}_{\rm fact.}^{c,IR}} {d\Omega} &=&
\left[\frac{F_\epsilon}{\epsilon}-\ln({\hat s})\right]
\left\lbrace 
\left[
-\ln\left( -\frac{u}{s}\right)  \right] ~ \frac{v_1}{t^2}
+  
\left[
-\ln\left( -\frac{u}{s}\right) - \ln\left( -\frac{t}{s}\right) 
\right] ~\frac{v_2}{st} 
\right\rbrace 
~I(t)  .
\end{eqnarray}
For the soft real terms, we refer to Appendix~\ref{app-soft} and may distinguish between initial and final state corrections (which are equal) and the initial-final state interference:
\begin{eqnarray}\label{reddteps}
\frac{d\overline{\sigma}_{\rm fact.}^{d,int,IR}} {d\Omega} &=&
\left[\frac{F_\epsilon}{\epsilon}-\ln({\hat s})\right]
\left[
2 \ln\left( -\frac{u}{s}\right)  
-2\ln\left( -\frac{t}{s}\right)
\right]~
\left( \frac{v_1}{t^2}+\frac{v_2}{st}\right) ~I(t)  ,
\\
\frac{d\overline{\sigma}_{\rm fact.}^{d,ini+fin,IR}} {d\Omega} &=&
\left[\frac{F_\epsilon}{\epsilon}-\ln({\hat s})\right]
\left[
-2\ln\left( {\hat s}\right) +2  
\right]~
\left( \frac{v_1}{t^2}+\frac{v_2}{st}\right) ~I(t) 
 .
\end{eqnarray}
It is now easy to see that the IR-divergency of the double box diagrams, , being proportional to $\ln(-u/s)$,  gets completely cancelled by the sum of the reducible box diagrams and the interference part of soft bremsstrahlung.
Although, the latter introduce to the sum an IR-divergency with $\ln(-t/s)$, and this gets cancelled the reducible vertex diagrams, thus introducing an IR-divergency with $\ln(s/m_e^2)$, which will be cancelled finally by the initial and final state soft corrections.
The lesson is: a sensible, infrared safe cross-section contains the complete sum of all the single IR-divergent diagrams, or no one of them.

Despite of that, an isolated treatment of the pure self energies or of the irreducible vertex corrections is possible. 

Finally, we just mention that the analytical integrations over $z$ may be performed
following the hints in Section \ref{sec-irred-vert}.

\subsection{\label{sec:IR-safeKernels}Kernel functions for the infrared safe sum}
We are now in a state to evaluate the net cross-section contribution from the various infrared divergent  terms of Sections~\ref{sec-fact} and \ref{sec:3}. 
We have seen that they have to be treated together.
The sum of the box contributions of Eq.~\eqref{boxAssembl} with all infrared-divergent
factorisable corrections, given in Eq.~\eqref{IR1}, Eq.~\eqref{IR2},
Eq.~\eqref{IR3} and Eq.~\eqref{IR4}, is infrared-finite and can be cast in the 
following form:
\begin{eqnarray}
\label{eqrest}
\frac{ d\overline{\sigma}_{\rm rest} }{ d\Omega }&=& 
\frac{ d\overline{\sigma}_{\rm box} }{ d\Omega } +
\sum_{i=a,b,c,d}\frac{ d\overline{\sigma}^i_{\rm fact.} }{ d\Omega } \nonumber \\
&=&
\int_{4M^2}^{\infty}\, dz\, \frac{R(z)}{z}\, \frac{1}{t-z}\, F_1(z)\nonumber\\
&&+~
\text{Re}\,\int_{4M^2}^{\infty}\, dz\, \frac{R(z)}{z}\, \frac{1}{s-z+i\,\delta}\,
\Bigl[ \,F_2(z)\,+\, F_3(z)\, \ln\Bigl(1-\frac{z}{s+i\, \delta}\Bigr)\, \Bigr]\nonumber\\
&&+~\pi\, \text{Im}\, \int_{4M^2}^{\infty}\, dz\, \frac{R(z)}{z}\, 
\frac{1}{s-z+i\,\delta} \, F_4(z).
\end{eqnarray}
The lower bound is $4M^2=4m_{\pi}^2$ for hadrons and $4M^2=4m_{f}^2$ for fermions $f$.
The auxiliary functions $F_i(z)$ are given by
\begin{eqnarray}
F_1(z)&=& \frac{1}{3}\,\Bigl\{\,
\Bigl[\, 3\,\Bigl( \frac{t^2}{s}+2\,\frac{s^2}{t} \Bigr)+9\,\Bigl(s+t\Bigr)\Bigr]\, 
\ln\Bigl(\frac{s}{m_e^2}\Bigr)
+\Bigl[-z^2\Bigl(\frac{1}{s}+\frac{2}{t}+2\,\frac{s}{t^2}\Bigr) \nonumber\\
 &&+~ z\,\Bigl( 4+4\,\frac{s}{t}+2\,\frac{t}{s}\Bigr)
+\frac{1}{2}\,\frac{t^2}{s}+6\,\frac{s^2}{t}+5\,s+4\,t\Bigr]\,
\ln\Bigl(-\frac{t}{s}\Bigr)
+ s\,\Bigl(-\frac{z}{t}+\frac{3}{2}\Bigr)\,\times\nonumber\\
&&\times ~\ln\Bigl(1+\frac{t}{s}\Bigr)
+\Bigl[\frac{1}{2}\,\frac{z^2}{s}+2\,z\,\Bigl(1+\frac{s}{t}\Bigr)-\frac{11}{4}\,s-2\,t \Bigr]\,
\ln^2\Bigl(-\frac{t}{s}\Bigr)
-\Bigl[\frac{1}{2}\,\frac{z^2}{t}\nonumber\\
&&-~z\,\Bigl(1+\frac{s}{t}\Bigr)+\frac{t^2}{s}
+2\,\frac{s^2}{t}+\frac{9}{2}\,s+\frac{15}{4}\,t \Bigr]\,\ln^2\Bigl(1+\frac{t}{s}\Bigr)
+\Bigl[\frac{z^2}{t}-2\,z\,\Bigl(1+\frac{s}{t}\Bigr)\nonumber\\
&&+~2\,\frac{s^2}{t}+5\,s+\frac{5}{2}\,t
\Bigr]\,\ln\Bigl(-\frac{t}{s}\Bigr)\,\ln\Bigl(1+\frac{t}{s}\Bigr)
-4\,\Bigl[\frac{t^2}{s}+2\,\frac{s^2}{t}+3\,\Bigl(s+t\Bigr) \Bigr]\,
\Bigl[1\nonumber\\
&&+~\text{Li}_2\Bigl(-\frac{t}{s}\Bigr)\Bigr]
-\Bigl[ 2\,\frac{z^2}{t}-4\,z\,\Bigl(1+\frac{s}{t}\Bigr)
-4\,\frac{t^2}{s}
-2\,\frac{s^2}{t}+s-\frac{11}{2}\,t\Bigr]\,\zeta_2\nonumber\\
&&-~
 \Bigl[ \frac{t^2}{s}+2\,\frac{s^2}{t}+3\,\Bigl(s+t\Bigr)\Bigr]\,\ln\Bigl(\frac{z}{s}\Bigr)\,
 \ln\Bigl(1+\frac{t}{s}\Bigr)
+\Bigl[ z^2\,\Bigl( \frac{1}{s}
+2\,\frac{s}{t^2}
+ \frac{2}{t}\Bigr)\nonumber\\
&&-~z\,\Bigl( \frac{t}{s}+2\frac{s}{t}+2 \Bigr)\Bigr]
\ln\Bigl(\frac{z}{s}\Bigr)
-\Bigl[ z^2\,\Bigl(\frac{1}{s}+\frac{1}{t}\Bigr) +2\,z\,\Bigl(1+\frac{s}{t}\Bigr)
+s
+2\,\frac{s^2}{t}\Bigr]\,\times\nonumber\\
&&\times ~\ln\Bigl(\frac{z}{s}\Bigr)\,
\ln\Bigl(1+\frac{z}{s}\Bigr)
+ \Bigl[\frac{z^2}{s}
+4\,z\,\Bigl(1+\frac{s}{t}\Bigr)-\frac{t^2}{s}-4\,\Bigl(s+t\Bigr)\Bigr]\,
\ln\Bigl(\frac{z}{s}\Bigr)\,\ln\Bigl(1-\frac{z}{t}\Bigr)\nonumber\\
&&-~ \Bigl[ z^2\,\Bigl(\frac{1}{s}+2\frac{s}{t^2}+ \frac{2}{t} \Bigr)
-2\, z\,\Bigl(\frac{t}{s}+2\,\frac{s}{t}
+2\Bigr)+\frac{t^2}{s}+2\,\Bigl( s+t\Bigr)\Bigr]\,
\ln\Bigl(1-\frac{z}{t}\Bigr)\nonumber\\
&&+~\Bigl[ \frac{z^2}{t}-2\,z\,\Bigl(1+\frac{s}{t}\Bigr)+2\,\frac{t^2}{s}
+8\,s
+4\,\frac{s^2}{t}+7\,t \Bigr]\,
\ln\Bigl(1-\frac{z}{t}\Bigr)\,\ln\Bigl(1+\frac{t}{s}\Bigr)\nonumber\\
&&-~\Bigl[
z^2\,\Bigl(\frac{1}{s}+\frac{1}{t}\Bigr) +2\,z\,\Bigl(1+\frac{s}{t}\Bigr)
+s + 2\,\frac{s^2}{t}
\Bigr]\, \text{Li}_2\,\Bigl(-\frac{z}{s}\Bigr)
+ \Bigl[\frac{z^2}{s}+4\,z\,\Bigl(1+\frac{s}{t}\Bigr)\nonumber\\
&&-~\frac{t^2}{s}-4\,\Bigl(s+t\Bigr)\Bigr]\,
\text{Li}_2\,\Bigl(\frac{z}{t}\Bigr)
- \Bigl[\,\frac{z^2}{t}
-2\,z\,\Bigl(1+\frac{s}{t}\Bigr)
+\frac{t^2}{s}+5\,s+2\,\frac{s^2}{t}\nonumber\\
&&+~4\,t\, \Bigr]\,\text{Li}_2\,\Bigl(1+\frac{z}{u}\Bigr)
\Bigr\}
+4\,\Bigl(\frac{1}{3}\,\frac{t^2}{s}+\frac{2}{3}\,\frac{s^2}{t}+s+t\Bigr)\,
\ln\Bigl(\frac{2\,\omega}{\sqrt{s}}\Bigr)\,\Bigl[
\ln\Bigl(\frac{s}{m_e^2}\Bigr)\nonumber\\
&&+~\ln\Bigl(-\frac{t}{s}\Bigr)
-\ln\Bigl(1+\frac{t}{s}\Bigr)-1\Bigr]
,\\
&&\nonumber\\
F_2(z)&=& \frac{1}{3}\,\Bigl\{
\,\Bigl[6\,\frac{t^2}{s}+3\,\frac{s^2}{t}+9\,\Bigl(s+t\Bigr)\Bigr]\,
\ln\Bigl(\frac{s}{m_e^2}\Bigr)
-\Bigl[z\Bigl(\frac{t}{s}+\frac{s}{t}+2\Bigr)-5 \,
\Bigl( s+\frac{t}{2}\nonumber\\
&&+~\frac{1}{2}\frac{s^2}{t}\Bigr)\Bigr]\,
\ln\Bigl(-\frac{t}{s}\Bigr)
-t\Bigl( \frac{z}{s}-\frac{3}{2}\Bigr)\,\ln\Bigl(1+\frac{t}{s}\Bigr)
+\Bigl[ \frac{z^2}{2}\,\Bigl(\frac{1}{s}+\frac{1}{t}\Bigr)
+z\,\Bigl(1+\frac{t}{s}\Bigr)\nonumber\\
&&+~2\,\frac{t^2}{s}-\frac{s}{4}+\frac{3}{4}t
\Bigr]\,\ln^2\Bigl(-\frac{t}{s}\Bigr)
- \Bigl[ \frac{z^2}{2\,s}-z\Bigl(1+\frac{t}{s}\Bigr)+2\,\frac{t^2}{s}
+\frac{s^2}{t}+\frac{15}{4}\,s\nonumber\\
&&+~\frac{9}{2}\,t\Bigr]\,
\ln^2\Bigl(1+\frac{t}{s}\Bigr)
-\Bigl(4\,\frac{t^2}{s}+\frac{s^2}{t}+4\,s+5\,t\Bigr)\,\ln\Bigl(-\frac{t}{s}\Bigr)\,
\ln\Bigl(1+\frac{t}{s}\Bigr)
\nonumber\\
&&-~4\, \Bigl[\, 2\,\frac{t^2}{s}+\frac{s^2}{t}+3\,\Bigl(s+t\Bigr)\,\Bigr]\,
\Bigl[ 1+\text{Li}_2\Bigl(-\frac{t}{s}\Bigr)\,\Bigr]
+  \Bigl( 12\,\frac{t^2}{s}+3\,\frac{s^2}{t}+12\,s+15\,t\Bigr)\,\zeta_2\nonumber\\
&&-~ \Bigl[ 2\, \frac{t^2}{s}+\frac{s^2}{t}
+3\,\Bigl(s+t\Bigr)\Bigr]\, \ln\Bigl(\frac{z}{s}\Bigr)\,
 \Bigl[ \ln\Bigl(1+\frac{t}{s}\Bigr) - \ln\Bigl(-\frac{t}{s}\Bigr)\Bigr]
+\Bigl[ z^2\,\Bigl(\frac{1}{t}\nonumber\\
&&+~\frac{2}{s}+2\,\frac{t}{s^2}\Bigr)
-z\,\Bigl(\frac{s}{t}+2
+2\,\frac{t}{s}\Bigr)\Bigr]\,\ln\Bigl( \frac{z}{s} \Bigr)
-\Bigl[ \frac{z^2}{t}+4\,z\,\Bigl(1+\frac{t}{s}\Bigr)-\frac{s^2}{t}\nonumber\\
&&-~4\,\Bigl(s+t\Bigr)\Bigr]\,
\text{Li}_2\,\Bigl(1-\frac{z}{s}\Bigr)
+\Bigl[ z^2\Bigl(\frac{1}{s}+\frac{1}{t}\Bigr)
+2\,z\,\Bigl(1+\frac{t}{s}\Bigr)
+2\,\frac{t^2}{s}
+t \Bigr]\times\nonumber\\
&&\times ~\text{Li}_2\,\Bigl(1+\frac{z}{t}\Bigr)
-\Bigl[ \frac{z^2}{s}-2\,z\,\Bigl(1+\frac{t}{s}\Bigr)+\frac{s^2}{t}
+2\,\frac{t^2}{s}+4\,s
+5\,t\Bigr]\,\text{Li}_2\,\Bigl(1+\frac{z}{u}\Bigr)
\Bigr\}\nonumber\\
&&+~4\,\Bigl(\frac{2}{3}\,\frac{t^2}{s}+\frac{1}{3}\,\frac{s^2}{t}+s+t\Bigr)\,
\ln\Bigl(\frac{2\,\omega}{\sqrt{s}}\Bigr)\,\Bigl[
\ln\Bigl(\frac{s}{m_e^2}\Bigr)
+\ln\Bigl(-\frac{t}{s}\Bigr)
-\ln\Bigl(1+\frac{t}{s}\Bigr)-1\Bigr]
,\nonumber\\
&&\\
F_3(z)&=& I_3(z),
\nonumber\\
&&\\
F_4(z)&=& \frac{1}{3}\,\Bigl\{\,
\Bigl[\, \frac{z^2}{s}-2\,z\,\Bigl(1+\frac{t}{s}\Bigr)+2\,\frac{t^2}{s}
+2\,\frac{s^2}{t}+\frac{11}{2}\,s+5\,t\Bigr]\,\ln\Bigl(1+\frac{t}{s}\Bigr)\nonumber\\
&&-~\Bigl[ z^2\,\Bigl(\frac{1}{s}+\frac{1}{t}\Bigr) +2\,z\,\Bigl(1+\frac{t}{s}\Bigr)
+2\,\frac{t^2}{s}+\frac{3}{2}\,s+\frac{5}{2}\,t \Bigr]\,\ln\Bigl(-\frac{t}{s}\Bigr)\nonumber\\
&&-~\Bigl[ z^2\,\Bigl(\frac{1}{t}+\frac{2}{s}+2\,\frac{t}{s^2}\Bigr)-2\,z\Bigl(2+\frac{s}{t}
+2\,\frac{t}{s}\Bigr)
-\frac{1}{2}\,\frac{s^2}{t}-s \,
\Bigr]\,
\Bigr\}.
\end{eqnarray}
The $I_3(z)$ is defined in (\ref{I3}).
 For $0<s<4\,M^2$ we can write
\begin{eqnarray}
\frac{ d\overline{\sigma}_{\rm rest} }{ d\Omega }&=&
\int_{4M^2}^{\infty}\, dz\, \frac{R(z)}{z}\, \Bigl\{\,
\frac{1}{t-z}\, F_1(z)
+ \frac{1}{s-z}\, \Bigl[F_2(z)+F_3(z)\,\ln\Bigl(\frac{z}{s}-1\Bigr)\Bigr]
\, \Bigr\}.
\end{eqnarray}
For $s>4\,M^2$, we have to perform some subtractions in order to make the formulas explicitly stable around $z = s$, and at the time retain the sufficiently fast vanishing of the integrand at $z\to \infty$:
\begin{eqnarray}
\frac{ d\overline{\sigma}_{\rm rest} }{ d\Omega }&=&
\int_{4M^2}^{\infty}\, dz\, \frac{R(z)}{z}\,
\frac{1}{t-z}\, F_1(z)
\\
&&+~
\int_{4M^2}^{\infty}\, dz\, \frac{1}{z\,\left(s-z\right)}\,
\Bigl\{
R(z)F_2(z)-R(s)F_2(s) + \left[ R(z)F_3(z)-R(s)F_3(s)\right] \ln\left| 1-\frac{z}{s}\right|
\Bigr\}
\nonumber\\
&&+~ \frac{R(s)}{s}\Bigl\{
F_2(s)\,\ln\Bigl(\frac{s}{4\,M^2}-1\Bigr)
- 6\, \zeta_2\,F_4(s)
+~F_3(s)\,\Bigl[
2\,\zeta_2
+~\frac{1}{2}\,\ln^2\Bigl(\frac{s}{4\,M^2}-1\Bigr)
+\text{Li}_2\Bigl(1-\frac{s}{4\,M^2}\Bigr)\,
\Bigr]
\Bigr\}.\nonumber
\end{eqnarray}
In the limit $m_f^2<<s,|t|,|u|$, the $z$-integration over the $F_i(z), i=1,2,$ develops mass singularities from the lower integration bound:
\begin{eqnarray}\label{mf2a}
 \int\limits_{4M^2} dz \frac{R(z)}{z}K_{\rm SE}(y;z) \left[ A(x,y,z) + B(x,y)\ln\Bigl(\frac{z}{s}\Bigr)
+ C(x,y)\ln\Bigl(\frac{s}{m_e^2}\Bigr)\right] 
\end{eqnarray}
where $A, B, C$ are regular for $z\to 0$.
It follows immediately that the sum of all infrared divergent  diagrams yield terms of the order of at most $\ln^2(s/m_f^2)$ and  $\ln(s/m_e^2)\ln(s/m_f^2)$, because $A$ joins, after integration, terms with a behavior like  a one-loop self-energy,  $B$ joins terms with one order more in $\ln(s/m_f^2)$ and $C$ 
goes together with at most  $\ln(s/m_e^2)\ln(s/m_f^2)$; there are no cubic logarithms here.
This has been discussed already in \cite{Actis:2007gi}.

Further, for the numerical evaluation, the functions $F_1$, $F_2$ and $F_3$ are
replaced for $z\to\infty$ by their asymptotic values:
\begin{eqnarray}
F_1(z)&\sim &
\Bigl[\, \frac{t^2}{s}+2\,\frac{s^2}{t} +3\,\Bigl(s+t\Bigr)\,\Bigr]\, 
\ln\Bigl(\frac{s}{m_e^2}\Bigr)
+\Bigl[\, \frac{1}{2}\frac{t^2}{s}+2\,\frac{s^2}{t}+\frac{7}{3}\,s+2\,t\,\Bigr]\,
\ln\Bigl(-\frac{t}{s}\Bigr)\nonumber\\
&&+~\frac{s}{2}\,\Bigl(\frac{1}{3}-\frac{s}{t}\Bigr)\,\ln\Bigl(1+\frac{t}{s}\Bigr)
-\frac{1}{2}\,\Bigl(\frac{s}{2}-\frac{1}{3}\,\frac{t^2}{s}\Bigr)\,\ln^2\Bigl(-\frac{t}{s}\Bigr)
-\frac{1}{3}\,\Bigl( \frac{1}{2}\,\frac{t^2}{s}+\frac{s^2}{t}+2\,s\nonumber\\
&&+~\frac{7}{4}\, t\Bigr)\,\ln^2\Bigl(1+\frac{t}{s}\Bigr)
-\Bigl[\frac{2}{3}\Bigl(\frac{t^2}{s}+\frac{s^2}{t}\Bigr)+s+\frac{3}{2}\,t\Bigr]\,
\ln\Bigl(-\frac{t}{s}\Bigr)\,\ln\Bigl(1+\frac{t}{s}\Bigr)\nonumber\\
&&-~4\,\Bigl[\frac{1}{3}\Bigl(\frac{t^2}{s}+2\,\frac{s^2}{t}\Bigr)+s+t\Bigr]\,
\text{Li}_2\Bigl(-\frac{t}{s}\Bigr)
+\Bigl[2\Bigl(\frac{t^2}{s}+\frac{s^2}{t}\Bigr) +3\,\Bigl(s+\frac{3}{2}\,t\Bigr) \Bigr]\,\zeta_2\nonumber\\
&&-~\Bigl[\frac{23}{12}\,\frac{t^2}{s}+\frac{8}{3}\,\frac{s^2}{t}+\frac{23}{4}\,\Bigl(s+t\Bigr)\,\Bigl]
-\frac{1}{2}\Bigl[\frac{t^2}{s}+3\,\Bigl(s+t\Bigr)\Bigr]\,\ln\Bigl(\frac{z}{s}\Bigr)
+4\,\Bigl(\frac{1}{3}\,\frac{t^2}{s}\nonumber\\
&&+~\frac{2}{3}\,\frac{s^2}{t}+s+t\Bigr)\,
\ln\Bigl(\frac{2\,\omega}{\sqrt{s}}\Bigr)\,\Bigl[
\ln\Bigl(\frac{s}{m_e^2}\Bigr)
+\ln\Bigl(-\frac{t}{s}\Bigr)
-\ln\Bigl(1+\frac{t}{s}\Bigr)-1\Bigr]+{\cal O}\Bigl(\frac{1}{z}\Bigr)
,
\\ 
F_s(z) &=& F_2(z)+F_3(z)\,\ln\Bigl(\frac{z}{s}-1\Bigr),
\\
F_s(z)&\sim&
\Bigl[2\,\frac{t^2}{s} + \frac{s^2}{t}+3\Bigl(s+t\Bigr)\Bigr]\,\ln\Bigl(\frac{s}{m_e^2}\Bigr)
+\Bigl(\frac{1}{2}\,\frac{t^2}{s}+\frac{s^2}{t}\nonumber\\
&&+~\frac{5}{2}\,s+2\,t\Bigr)\,\ln\Bigl(-\frac{t}{s}\Bigr)
+\frac{t}{2}\,\Bigl(\frac{1}{3}-\frac{t}{s}\Bigr)\,\ln\Bigl(1+\frac{t}{s}\Bigr)
+\frac{1}{3}\,\Bigl[\frac{t^2}{s}-\frac{1}{4}\,\Bigl(s-t\Bigr)\Bigr]\times\nonumber\\
&&\times ~ \ln^2\Bigl(-\frac{t}{s}\Bigr)
-\frac{1}{3}\,\Bigl(\frac{t^2}{s}+\frac{1}{2}\,\frac{s^2}{t}+\frac{7}{4}\,s+2\,t\Bigr)\,
\ln^2\Bigl(1+\frac{t}{s}\Bigr)
-\frac{1}{3}\,\Bigl(4\,\frac{t^2}{s}+\frac{s^2}{t}\nonumber\\
&&+~4\,s+5\,t\Bigr)\,
\ln\Bigl(-\frac{t}{s}\Bigr)\,\ln\Bigl(1+\frac{t}{s}\Bigr)
-4\,\Bigl(\frac{2}{3}\,\frac{t^2}{s}+\frac{1}{3}\,\frac{s^2}{t}+s+t\Bigr)\,
\text{Li}_2\Bigl(-\frac{t}{s}\Bigr)\nonumber\\
&&+~\Bigl(4\,\frac{t^2}{s}+\frac{s^2}{t}+4\,s+5\,t\Bigr)\,\zeta_2
-\Bigl[\frac{8}{3}\,\frac{t^2}{s}+\frac{23}{12}\,\frac{s^2}{t}+\frac{23}{4}\,\Bigl(
s+t\Bigr)\Bigr]\nonumber\\
&&-~\frac{1}{2}\,\Bigl[\frac{s^2}{t}+3\,\Bigl(s+t\Bigr)\Bigr]\,\ln\Bigl(\frac{z}{s}\Bigr)
+4\,\Bigl(\frac{2}{3}\,\frac{t^2}{s}+\frac{1}{3}\,\frac{s^2}{t}+s+t\Bigr)\,
\ln\Bigl(\frac{2\,\omega}{\sqrt{s}}\Bigr)\,\times\nonumber\\
&&\times ~ \Bigl[
\ln\Bigl(\frac{s}{m_e^2}\Bigr)
+\ln\Bigl(-\frac{t}{s}\Bigr)
-\ln\Bigl(1+\frac{t}{s}\Bigr)-1\Bigr]
+{\cal O}\Bigl(\frac{1}{z}\Bigr).
\end{eqnarray}


\section{\label{sec:NumericalResults}NUMERICAL RESULTS AT MESON FACTORIES, LEP/GigaZ, ILC}
We begin with
numerical results for Eq.~\eqref{eqrest}, multiplied by
the overall factor $(\alpha \slash \pi)^2\, \alpha^2\slash s$.
The expressions 
contain the contribution of irreducible two-loop boxes,
summed up with reducible two-loop vertex and loop-by-loop diagrams,  and combined with soft-photon
emission.
They are called here 'rest' from electrons, muons, tau-leptons, and from hadrons.
The top influence was also considered but comes out so marginal that we don't discuss it. 
The results are summarized in Table~\ref{nums1} and Table~\ref{nums2}
for small- and large-angle scattering and a variety of energy scales.
We do not discuss the isolated irreducible two-loop boxes because this would become more convention-dependent. 
Note further that in these tables the dependence on the maximal energy of the soft photons is switched off by setting $\omega= \sqrt{s}\slash 2$ (an analogous
  consideration holds for the soft pairs $e^+e^-$).
For comparison, the tables also contain entries with pure QED Born, 
QED Born with running coupling, and effective weak Born cross-sections, 
as well as  contributions from:
electron vertex insertions and soft $e^+e^-$ pairs (with a quite small sum of them); 
the sum of heavy fermion irreducible vertices.
The hadronic results
  have been obtained using the parametrization~\cite{Burkhardt:1981jk} with flag setting 
$\texttt{IPAR} = 0$ and implementing narrow resonances as described in
Appendix \ref{app:rhad}.

We see that the two-loop corrections from electron insertions (the so-called $N_f=1$ corrections) are the largest, and the second-largest ones are the hadronic corrections.
The tables also demonstrate that the approximation $m_f^2 << s,|t|,|u|$ as applied in e.g. \cite{Actis:2007gi} works well in the regions where this is expected.

\begin{table}[tb]\centering
\setlength{\arraycolsep}{\tabcolsep}
\renewcommand\arraystretch{1.1}
\begin{tabular}{|l|r|r|r|r|}
\hline 
$\theta$ $[^\circ]$\, $\vert$ \,$\sqrt{s}$ [GeV] & $\theta=20$\, $\vert$ \,  1& $\theta=20$\, $\vert$ \,  10& 
$\theta=3$\, $\vert$ \,  $M_Z$ &  $\theta=3$\, $\vert$ \,  500\\
\hline 
\hline
{\rm QED Born} 
 & 214.903  & 2.14903   & 53.0348   &  1.76398  \\ 
\hline 
\hline
{\rm weak Born}
 & 214.903  & 2.14930   & 53.0376   & 1.76390  \\ 
\hline 
\hline
{\rm QED Born, running} & 218.559 & 2.23814 & 55.5353 & 1.90910 \\ 
\hline 
\hline
{\rm vertices [$\mu$+$\tau$+hadr.]} 
 & -0.001086 & -0.00022513 & -0.007982 & -0.00129296 \\ 
\hline
\hline
{\rm vertices [$e$]}
 & -0.102787 &   -0.00325449 & -0.092546 & -0.00574577 \\
\hline 
{\rm soft pairs $e^+e^-$}
 & 0.130264 & 0.00403772 & 0.112763 & 0.00685890 \\ 
\hline 
\hline
{\rm rest:}\qquad\,\, $e$ 
 & 0.235562   & 0.00497834  & 0.135650  &0.00672652   \\
\hline 
 \qquad\qquad$\mu$ & 0.009518  & 0.00135040   & 0.040792 & 0.00287809  \\ 
       &-- 0.017214  & 0.00134282   & 0.040688  & 0.00287795  \\
\hline 
 \qquad\qquad$\tau$ &  0.000074   &   0.00005385  &  0.002706 & 0.00087639 \\ 
& $\times$ & $\times$  &-- 0.009610  & 0.00083969 \\
\hline
\qquad\qquad hadr. & 0.008642 & 0.00269490 &  0.087618 & 0.00810781 \\ 
\hline 
\end{tabular}
\caption[]{Numerical values for the differential cross section
  in nanobarns at scattering angles $\theta=20^\circ$ and
  $\theta=3^\circ$, in units of $10^2$. Concerning the finite remainder,
  containing irreducible box diagrams, we show for each fermion flavor 
  the result obtained through the dispersion-based approach (first line) and the one
  coming from the analytical expansion (second line), neglecting ${\cal O}(m_f^2 \slash x)$,
  where $x=s,|t|,|u|$. When $m_f^2>x$, the entry is suppressed.
  }
\label{nums1}
\end{table}

\begin{table}[ht]\centering
\setlength{\arraycolsep}{\tabcolsep}
\renewcommand\arraystretch{1.2}
\begin{tabular}{|l|r|r|r|r|}
\hline 
$\sqrt{s}$ [GeV] & 1& 10& $M_Z$& 500\\
\hline 
\hline
{\rm QED Born}  & 466537 & 4665.37  &56.1067  &1.86615  \\ 
\hline
\hline
{\rm weak Born}  &466526 & 4654.16  & 1238.7500 & 0.92890 \\ 
\hline
\hline
{\rm QED Born, running} & 480106 & 4984.83 & 62.9027 & 2.17957 \\ 
\hline 
\hline
{\rm vertices [$\mu$+$\tau$+hadr.] } & -16.351 & -2.0437 & -0.125208 & -0.0104275 \\ 
\hline
\hline
{\rm vertices [e]} & -477.620 &  -12.3010 & -0.298589 & -0.0155751 \\
\hline
\hline
{\rm soft pairs $e^+e^-$} & 648.275 & 16.0690 & 0.376531 & 0.0191990 \\ 
\hline 
\hline
{\rm rest:} \qquad\,\, $e$  & 807.476  & 14.5277   & 0.270575 & 0.0119285  \\
\hline 
\qquad \qquad $\mu$ & 160.197    &      6.0819  & 0.147046  &  0.0072579 \\ 
      & 152.890          & 6.0809        & 0.147046        & 0.0072579   \\
\hline 
\qquad \qquad $\tau$ & 2.383   & 1.3335 & 0.075268  &  0.0045713  \\ 
& $\times$ & 1.0739   & 0.075214  & 0.0045712  \\
\hline 
\qquad \qquad hadr. & 232.674  & 16.0670 & 0.469944 & 0.0246035  \\ 
\hline 
\end{tabular}
\caption[]{Numerical values for the differential cross section
  in nanobarns at a scattering angle $\theta=90^\circ$, in units of $10^{-4}$.
  See the caption of Table~\ref{nums1} for further details.}
\label{nums2}
\end{table}

\begin{figure}[tbh]
\begin{center}
\includegraphics[scale=0.5]{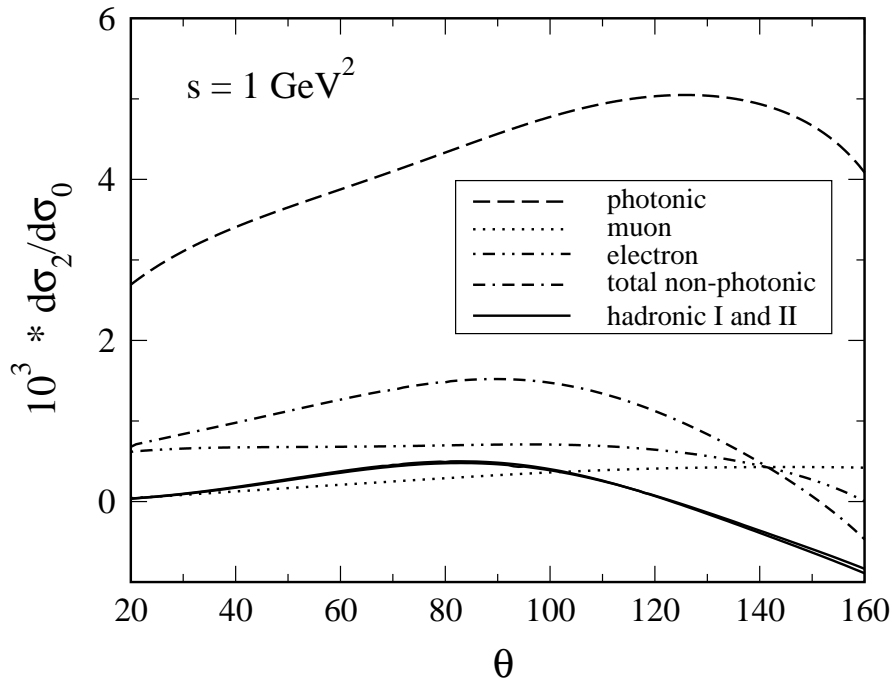}
\end{center}
\caption[]
{\em Two-loop corrections to Bhabha scattering at $\sqrt{s}=1$ GeV, normalized
to the QED tree-level cross section.}
\label{PRD1gev_noSE}
\end{figure}

\begin{figure}[tbhp]
\begin{center}
\includegraphics[scale=0.5]{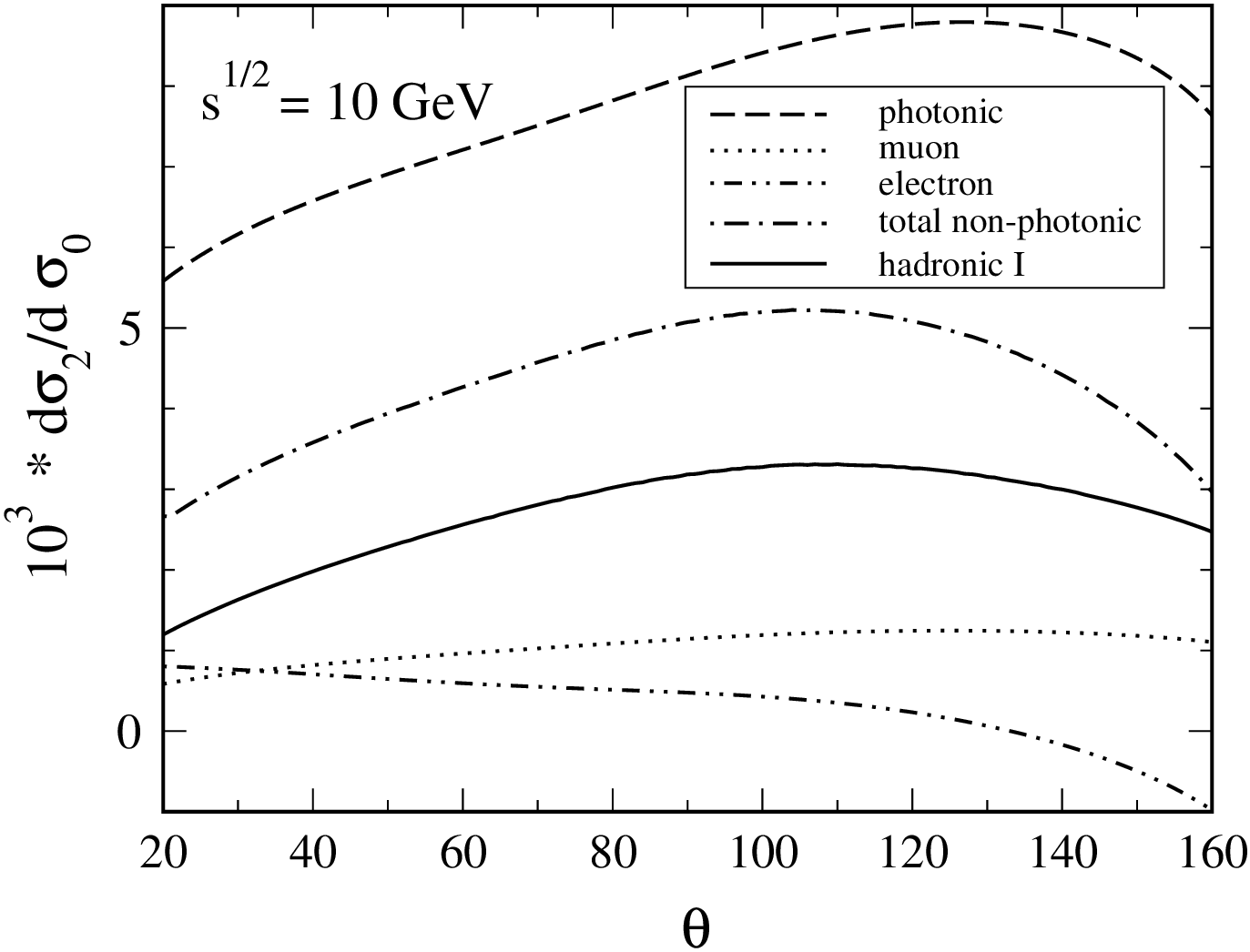}
\end{center}
\caption[]
{\em Two-loop corrections to Bhabha scattering at $\sqrt{s}=10$ GeV, normalized
to the QED tree-level cross section.}
\label{PRD10gev_noSE_hadI}
\end{figure}

\begin{figure}[tbhp]
\begin{center}
\includegraphics[scale=0.5]{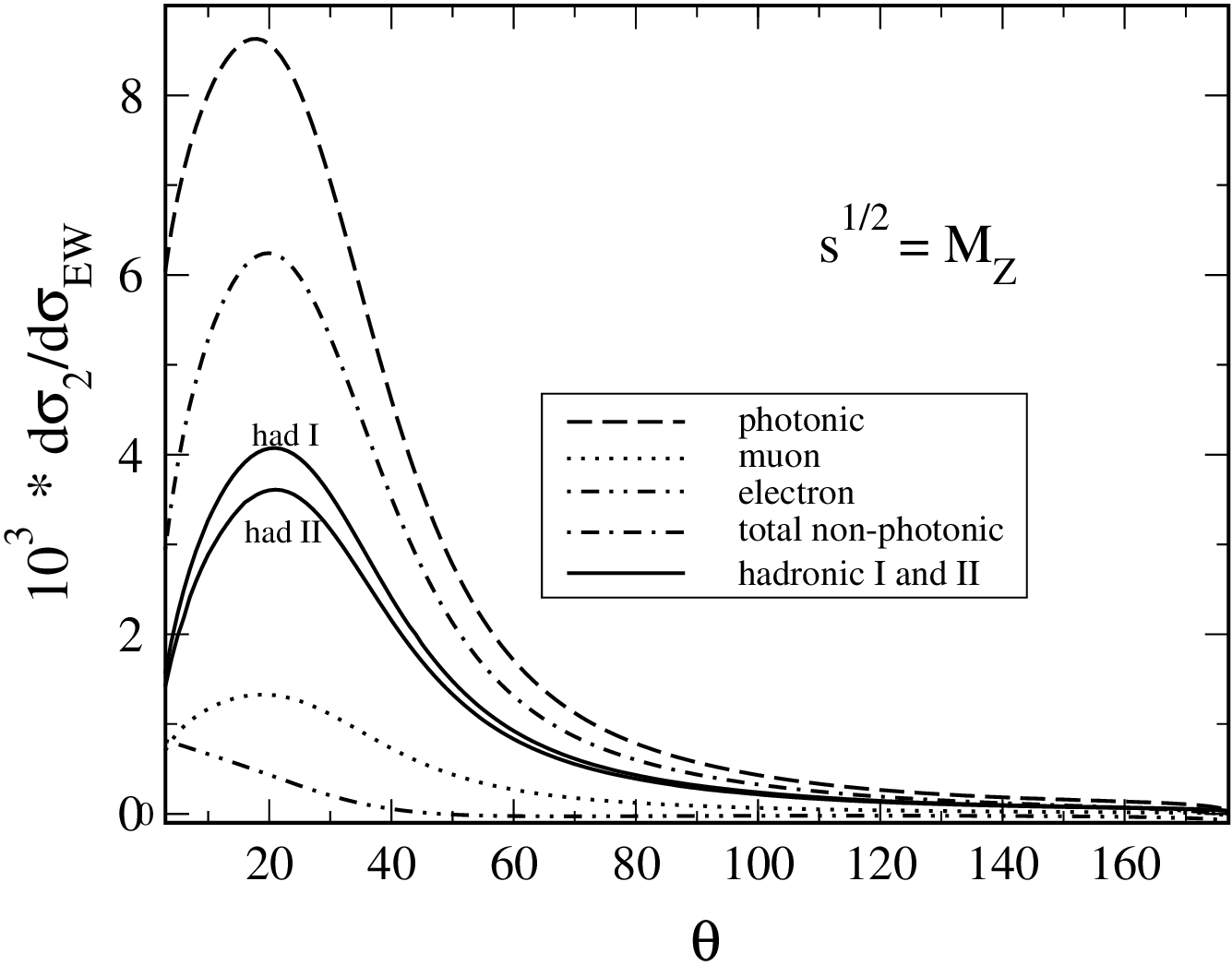}
\end{center}
\caption[]
{\em Two-loop corrections to Bhabha scattering at $\sqrt{s}=M_Z$, normalized
to the effective weak Born  cross section.}
\label{PRDMZgev_noSE}
\end{figure}

\begin{figure}[tbph]
\begin{center}
\includegraphics[scale=0.5]{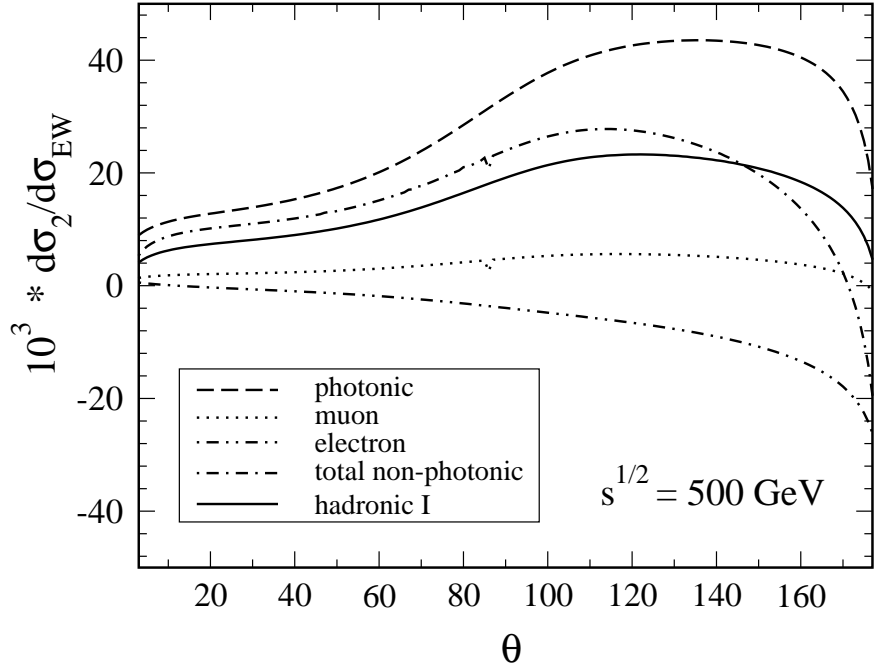}
\end{center}
\caption[]
{\em Two-loop corrections to Bhabha scattering at $\sqrt{s}=500$ GeV, normalized
to the effective weak Born  cross section.}
\label{PRD500gev_noSE_ewhadI}
\end{figure}

\begin{figure}[tbph]
\begin{center}
\includegraphics[scale=0.5]{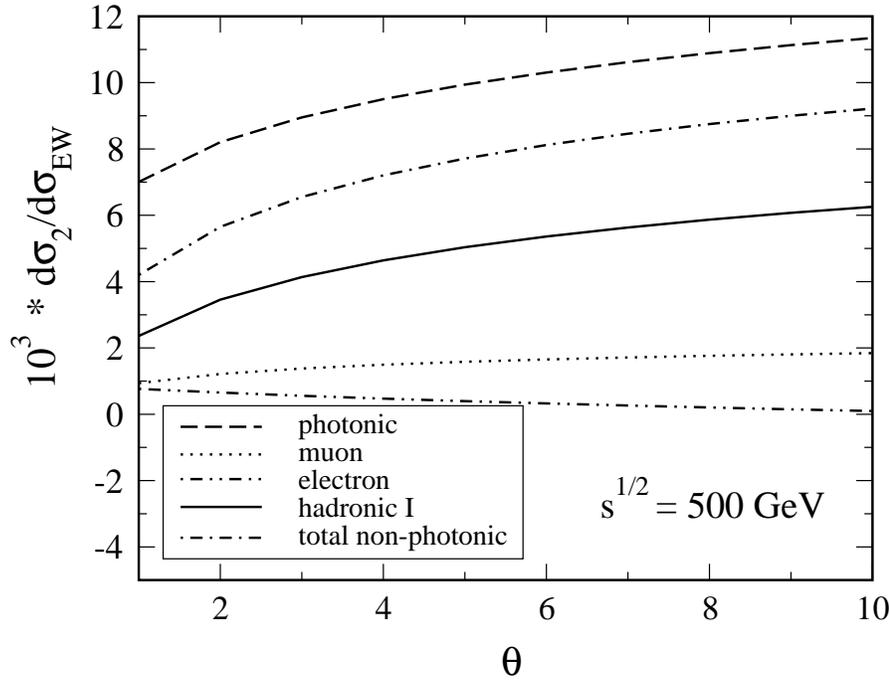}
\end{center}
\caption[]
{\em Same as in Fig.~\ref{PRD500gev_noSE_ewhadI}, for small angles.}
\label{PRD500gev_noSE_ew_small}
\end{figure}

\begin{figure}[tbph]
\begin{center}
\includegraphics[scale=0.5]{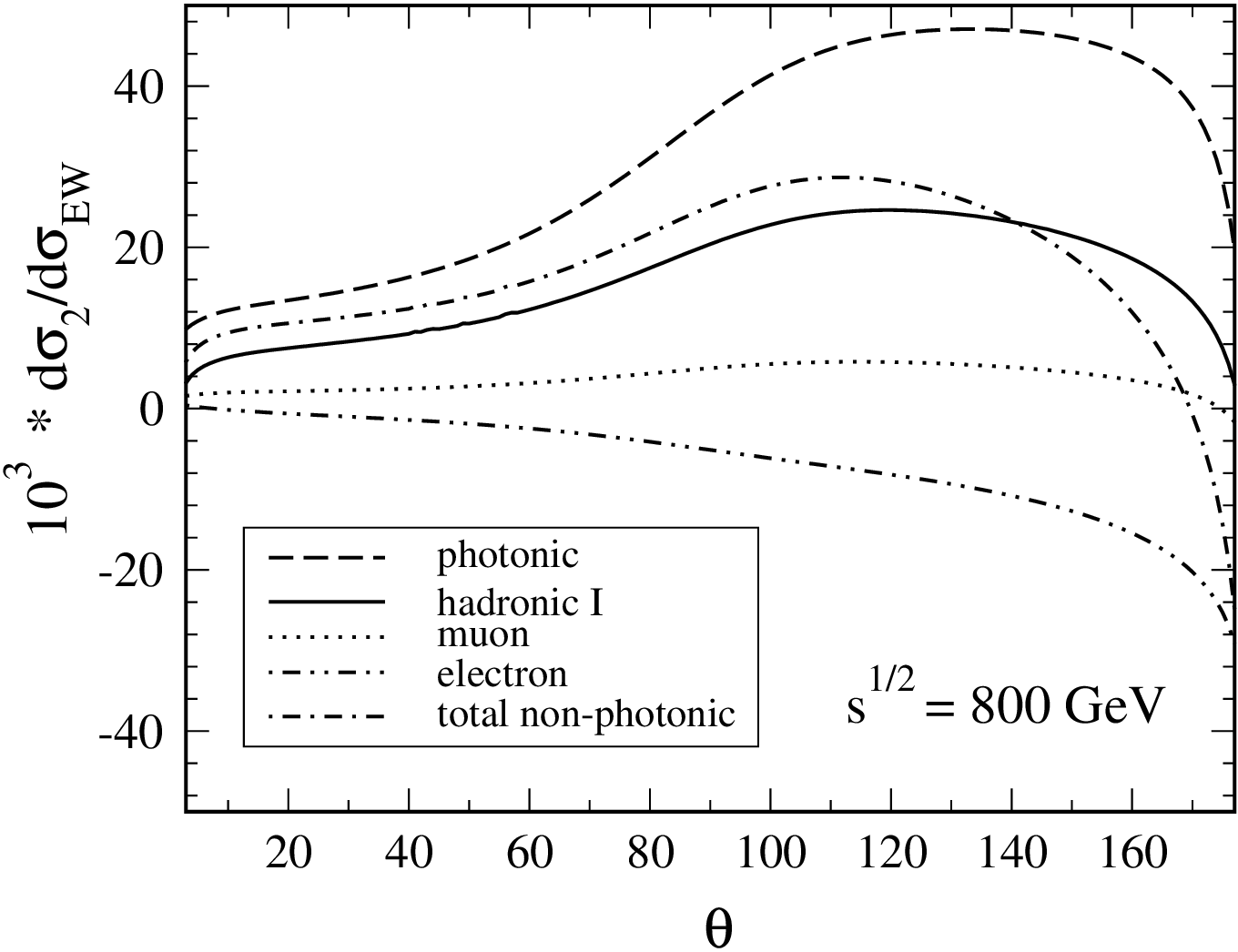}
\end{center}
\caption[]
{\em Two-loop corrections to Bhabha scattering at $\sqrt{s}=800$ GeV, normalized
to the effective weak Born  cross section.}
\label{PRD800gev_noSE_ew_hadI}
\end{figure}

A more detailed picture of the relevance of the fermionic and hadronic two-loop corrections may be got from figures \ref{PRD1gev_noSE} to \ref{PRD800gev_noSE_ew_hadI}, where we show the cross-section ratios
\begin{eqnarray}
 10^3~~\frac{d\sigma_{\rm NNLO}}{d\sigma_0},
\end{eqnarray}
where $d\sigma_0$ is the effective weak Born cross-section at $\sqrt{s}=M_Z, 500$ and $800$ GeV, and the QED  
Born cross-section elsewhere.
So, the figures show just the relative size of the corrections in per mille.
For a comparison, we show also the pure photonic corrections.
The $d\sigma_{\rm NNLO}$ is here the net sum of all the terms discussed arising from a fermion flavor ($e$ or $\mu$) or from the hadrons.
In case of electrons, we add also the real pair correction.
The total non-photonic term includes also the $\tau$ and top quark contributions.
For hadrons, we decided to use the parameterization $R_{\rm had,I}$ as given in \cite{Burkhardt:1981jk}
with parameter $\texttt{IPAR}=1$.
We applied also numerics with a combination  $R_{\rm had,II}$ of several adjusted pieces valid at different scales, as explained in Appendix \ref{app:rhad}.
In Figures \ref{PRD1gev_noSE} and \ref{PRDMZgev_noSE} it is seen that the predictions with  $R_{\rm had,I}$ and $R_{\rm had,II}$ are quite close to each other.
Because we did not get a stable numerics over all the parameter space with   $R_{\rm had,II}$, we decided not to use it for the final determination of the physical results until we have a better understanding of its behaviour.
 
At a meson factory with  $\sqrt{s} \approx 1$ GeV (Fig.~\ref{PRD1gev_noSE}) the heavy fermion effects are below 0.5 per mille and are thus certainly negligible.
At  $\sqrt{s} \approx 10$ GeV (Fig.~\ref{PRD10gev_noSE_hadI}), electron and hadron corrections amount to 2 to 5 per mille and might play some relevance.
At the higher energies, we have to consider small angles and large ones separately.
The hadronic corrections amount to up to 4 per mille at LEP1/GigaZ and 20 per mille at ILC energies at large angles, while at small angles they stay well below 5 per mille.
For $\sqrt{s} = 500$ GeV this is exemplified in Figure \ref{PRD500gev_noSE_ew_small}, and from the tables one may read exact values at $\theta = 3$ degrees:
for the infrared-finite remainder containing box diagrams, at LEP/GigaZ it is $\frac{d\sigma_{\rm 2}^{\rm had}}{d\sigma_0^{weak}} = 1.65$ per mille, and at $\sqrt{s} = 500$ GeV the corresponding value becomes 4.6 per mille.
Everywhere, the pure photonic corrections are the largest one, followed by the $N_f=1$ corrections.
This is, of course, due to the small electron mass producing large logarithmic mass effects and is extensively discussed in the literature.

\section{\label{sec:summa}SUMMARY}
The NNLO effects of heavy fermions and hadrons on the Bhabha cross-sections are accurately known now and the determination of QED two-loop corrections is completed. 
For each of the corrections there exist several independent calculations.
Quite recently, a second determination of the hadronic corrections in \cite{Kuhn:2008zs} fully confirmed our results as presented in \cite{Actis:2007pn,Actis:2007fs,Actis:2008sk} and at our webpage \cite{webPage:2006xx}.
We indeed checked, when preparing this longer write-up of our results, that, when using the same parameterization \cite{Burkhardt:1981jk}, all the digits shown in our Tables agree with those shown in \cite{Kuhn:2008zs} (see Tabs.~\ref{nums1}
and \ref{nums2}).
The numerical differences which were mentioned in \cite{Kuhn:2008zs} were due to a different choice of the parameter \texttt{IPAR} in \cite{Actis:2007pn} and \cite{Kuhn:2008zs}.

Summarizing the numerical discussion, it is quite obvious that for measurements aiming at an accuracy at the per mille level it is crucial to take the heavy fermion and hadron contributions into account. 
A detailed conclusion for a specific experiment evidently depends on the experimental set-ups and will deserve the use of a precise Monte-Carlo program.

Finally, we would like to mention that, in pure QED, not all of the contributions have been determined so far.
It would be quite interesting to know  also the influence from the so-called radiative loops.
This problem was treated in \cite{Melles:1996qa}, but so far without account of the radiative loop diagram, which include e.g. radiative boxes with the need of knowledge of five-point functions.
Also here, final conclusion will be made only with a precise Monte-Carlo program.

As a third field of future improvement we like to mention
the complete treatment of electroweak two-loop corrections to Bhabha scattering.
As already said there exists some literature on that subject.
The leading NNLO weak corrections due to top quarks have been determined long ago in \cite{Bardin:1990xe}. 
This was considered as a satisfactory approximation for LEP 1 and implemented e.g. in the packages  \texttt{ZFITTER} \cite{Arbuzov:2005ma} and in the program family \texttt{KORALZ} \cite{Jadach:1999tr}, \texttt{KKMC} \cite{Jadach:1999vf,Ward:2002qq}, \texttt{BHLUMI} \cite{Jadach:1996is}, \texttt{BHWIDE} \cite{Jadach:1995nk}; see also the workshop report \cite{Kobel:2000aw}.
An improvement of that might become necessary for large angle scattering at the ILC. 
This might be done similarly to the recent implementation of weak two-loop corrections for  muon pair production in \texttt{ZFITTER} v.6.42 \cite{Arbuzov:2005ma}, based on original work described in  \cite{Awramik:2003rn,Awramik:2006uz} and references therein.
 
--------------------------------------------------------
\begin{acknowledgments}
We would like to thank B.~Kniehl, H.~Burkhardt and T.~Teubner  for help concerning
$R_{\mathrm{had}}$ and A. Arbuzov, H. Czyz, S.-O. Moch, and K. M{\"o}nig for
discussions. 
Work supported by Sonderforschungsbereich/Transregio
SFB/TRR 9 of DFG ``Computergest\"utzte Theoretische Teilchenphysik", by the Sofja Kovalevskaja Programme of the Alexander von
Humboldt Foundation  sponsored by the German Federal Ministry of Education and Research, 
and by the European Community's Marie-Curie Research Training Networks
MRTN-CT-2006-035505 ``HEPTOOLS'' and MRTN-CT-2006-035482 ``FLAVIAnet''.

Feynman diagrams have been drawn with the packages 
{\sc Axodraw}~\cite{Vermaseren:1994je} and 
{\sc Jaxo\-draw}~\cite{Binosi:2003yf}.
\end{acknowledgments}


\appendix

\section{\label{app:pho}Analytic Results for the Fermionic Vacuum Polarization}
The contribution of a fermion of flavour $f$ to the irreducible 
renormalized photon vacuum-polarization function $\Pi$, introduced in 
Eq.~\eqref{1stReplace}, can be written in pure QED as
\begin{equation}
\label{Ksum}
\Pi_f(q^2)= \sum_{n=1}^{2}\, \left(\frac{\alpha}{\pi}\right)^n\,
F_\epsilon^n\,\left(\frac{m_e^2}{m_f^2}\right)^{n\epsilon}\,Q_f^{2n}\,C_f\,
\Pi_f^{(n)}(q^2) + {\cal O}(\alpha^3),
\end{equation}
where $Q_f$ is the electric-charge quantum number, $C_f$ is the color factor
and the normalization
factor $F_\epsilon$ is defined in Eq.~\eqref{norm}.

For our purposes we need both the $n=1$ and $n=2$ terms up to ${\cal O}(\epsilon^0)$.
However, since some components of the infrared-finite differential
cross section show single poles in the $\epsilon$ plane, we find
useful to consider also the ${\cal O}(\epsilon)$ part of the
one-loop photon self-energy for intermediate checks of the results. 

Both expressions can be written in a compact form introducing the variable
\begin{equation}\label{defofx}
x=\frac{\sqrt{-q^2+4\,m_f^2}-\sqrt{-q^2}}{\sqrt{-q^2+4\,m_f^2}+\sqrt{-q^2}}.
\end{equation}
The results can be found in Appendix~A of Ref.~\cite{Bonciani:2004gi}
and at the webpage \cite{webPage:2007xx}.
In the space-like region $-\infty<q^2<0$, it is $0<x<1$, and  one gets a real vacuum polarization:
\begin{eqnarray}
\label{oneloop}
\Pi_f^{(1)}(q^2)&=& -\frac{5}{9} +\frac{4}{3}\,\frac{x}{(1-x)^2}
+\frac{1}{3}\,\left[\, \frac{4}{(1-x)^3} - \frac{6}{(1-x)^2}+1\,
\right]\,\ln\left(x\right)\nonumber\\
&&+~\frac{\epsilon}{3}\,\Bigl\{\,
-\frac{28}{9}
+\frac{32}{3}\,\frac{x}{(1-x)^2}
+\frac{1}{3}\,\left[\, \frac{32}{(1-x)^3}-\frac{48}{(1-x)^2}+\frac{6}{1-x}
+ 5 \,\right]\,
\ln\left(x\right)\nonumber\\
&&-~ 2\,\left[ \frac{4}{(1-x)^3}
 - \frac{6}{(1-x)^2} + 1 \right]\,
\Bigl[\text{Li}_2(-x)+\ln(x)\,\ln(1+x)-\frac{1}{4}\,\ln^2(x)+\frac{\zeta_2}{2} \Bigr]\,
\Bigr\},\nonumber\\
&&\nonumber\\
&&\\
&&\nonumber\\
\label{twoloops}
\Pi_f^{(2)}(q^2)&=& -\frac{1}{6}\,\left[ \frac{5}{4}-13 \, \frac{x}{(1-x)^2} \right]
+\frac{1}{4}\,\left[\frac{12}{(1-x)^3}-\frac{18}{(1-x)^2}+\frac{4}{1-x}+ 1 \right]\,\ln(x)\nonumber\\
&&-~\frac{4}{3}\, \left[\frac{4}{(1-x)^3} -\frac{6}{(1-x)^2}+1\right]\,
\Bigl\{ \text{Li}_2(-x)+\frac{1}{2}\text{Li}_2(x)+\ln(x)\, \Bigl[ \ln(1+x)\nonumber\\
&&+~\frac{1}{2}\,\ln(1-x)\Bigr]\Bigr\}
-\frac{1}{6}\,\left[ \frac{7}{(1-x)^4} - \frac{26}{(1-x)^3} + \frac{23}{(1-x)^2}
+\frac{2}{1-x}
-6\right]\,
\ln^2(x)\nonumber\\
& +&\frac{1}{3}\,\left[\frac{4}{(1-x)^4}-\frac{8}{(1-x)^3}+\frac{4}{(1-x)^2}
-1\right]\,\Bigl\{ \ln^2(x)\Bigl[ \ln(1-x)
+2\,\ln(1+x)\Bigr]\nonumber\\
&&+~4\,\ln(x)\,\Bigl[ \text{Li}_2(x)+2\,\text{Li}_2(-x)\Bigr]
-6\,\Bigl[\text{Li}_3(x)
+2\,\text{Li}_3(-x)\Bigr]-3\,\zeta_3
\Bigr\}.\nonumber\\
\end{eqnarray}
For the time-like region,
we have to perform
an analytical continuation to $q^2>4\,m_f^2$ by setting $q^2\to q^2+i\,\delta$ in Eq. (\ref{defofx}).
Now, the conformal variable $x$ develops a small positive imaginary part and it is
$-1< \text{Re}  x <0$. 
In order to derive Im\,$\Pi$ of Eq.~\eqref{Im}, 
we may introduce an auxiliary variable $y$:
\begin{equation}
y=\frac{\sqrt{q^2}-\sqrt{q^2-4\,m_f^2}}{\sqrt{q^2}+\sqrt{q^2-4\,m_f^2}},
\end{equation}
and observe that $x=-y+i\delta$, with $y=0$ for $q^2\to\infty$
and $y=1$ for $q^2=4\,m_f^2$. 
With these conventions, it becomes evident for Eqs. (\ref{oneloop}) and (\ref{twoloops}) that $\text{Li}_2(\pm x)$, $\text{Li}_3(\pm x)$, and $\ln(1+x)$ 
stay well-defined, and one has to take care about $\ln(x)$:
\begin{eqnarray}
\ln(x)&\to&\ln(-y+i\,\delta)=\ln(y)+i \pi.
\end{eqnarray}
Of course, one may perform the evaluations with complex variables either.

The contribution of electron loops to the irreducible 
renormalized photon vacuum-polarization function $\Pi$
of Eq.~\eqref{1stReplace} in the small electron-mass limit
 is available in pure QED up to three loops,
\begin{equation}
\Pi_e(q^2)= \sum_{n=1}^{3}\, \left(\frac{\alpha}{\pi}\right)^n\,
\Pi_e^{(n)}(q^2) + {\cal O}(\alpha^4).
\end{equation}
The one- and two-loop contributions can be obtained by expanding
Eqs.~\eqref{oneloop} and \eqref{twoloops} and neglecting terms suppressed by
positive powers of the electron mass. 
The three-loop component, (we do not include  double-bubble diagrams with two different flavours),
can be found in Eqs.~(7) and (9) of Ref.~\cite{Steinhauser:1998rq}. 
The results for $q^2<0$ are:
\begin{eqnarray}
\label{oneloopE}
\Pi_e^{(1)}(q^2)&=& -\frac{5}{9}-\frac{1}{3}\,\ln\left(-\frac{m_e^2}{q^2}\right) +{\cal O}(m_e^2),\\
\label{twoloopsE}
\Pi_e^{(2)}(q^2)&=&  -\frac{5}{24}+\zeta_3-\frac{1}{4}\,\ln\left(-\frac{m_e^2}{q^2}\right) + {\cal O}(m_e^2),\\
\label{threeloopsE}
\Pi_e^{(3)}(q^2)&=& \frac{121}{192} - \Bigl[2\,\ln(2)-\frac{5}{4}\Bigr]\,\zeta_2 +\frac{99}{64}\,\zeta_3
-\frac{5}{2}\,\zeta_5 + \frac{1}{32}\,\ln\left(-\frac{m_e^2}{q^2}\right)\nonumber\\
&&+~\underbrace{\frac{307}{864} + \frac{2}{3}\,\zeta_2 -\frac{545}{576}\,\zeta_3
+\Bigl( \frac{11}{24}-\frac{\zeta_3}{3}\Bigr)\,\ln\left(-\frac{m_e^2}{q^2}\right)
+\frac{1}{24}\,\ln^2\left(-\frac{m_e^2}{q^2}\right)}_{\rm double\quad electron\quad bubble}+{\cal O}(m_e^2).\nonumber\\
\end{eqnarray}
The continuation to $q^2>0$ is again obtained by the replacement $q^2 \to q^2+i\delta$.

\section{\label{app-masters}Master Integrals for the Box Kernel Functions}
The three kernel functions for irreducible box diagrams of  Figure~\ref{Boxes2loops} may be found 
at webpage \cite{webPage:2007xx} with their exact dependences on $m_e$ and on  $\epsilon$.
They are expressed by eight master integrals, which were evaluated in the limit $m_e^2<<z,s,|t|,|u|$.
The master integrals of Eq.~\eqref{deco}, for $x=s$ and $y=t$, are evaluated to the power in  $\epsilon$ needed here:
\begin{eqnarray}\label{master1}
M^{(1)}&=& N 
\int \frac{d^D k}{\left( k^2-m_e^2 \right)}
\nonumber \\
&=& m_e^2\, \Bigl[ \,
          \frac{1}{\epsilon}+1+\epsilon\,\Bigl(\, 1+\frac{\zeta_2}{2}\, \Bigr)\,\Bigr],
\\\label{master2} 
M^{(2)}&=&N \int \frac{d^D k\,}{
        \left( k^2-m_e^2 \right) \,
               \left[ \left( k-p_1-p_2 \right)^2-m_e^2 \right] }
\nonumber \\
&=& \frac{1}{\epsilon} + 2 +\ln\Bigl(-\frac{m_e^2}{s}\Bigr) +\epsilon\,
           \Bigl[\, 4 -\frac{\zeta_2}{2}+2\, \ln\Bigl(-\frac{m_e^2}{s}\Bigr)
                 +\frac{1}{2}\, \ln^2\Bigl(-\frac{m_e^2}{s}\Bigr)\, \Bigr] +{\cal O}(m_e^2),
\\ \label{master3} 
M^{(3)}&=&N \int \frac{d^D k\,}{
       k^2\, \left(k-p_1+p_3\right)^2 }
\nonumber 
\\&=& \frac{1}{\epsilon} + 2 +\ln\Bigl(-\frac{m_e^2}{t}\Bigr) ,
\\ \label{master4} 
M^{(4)}&=&N \int \frac{d^D k}{
   \left(k^2-m_e^2\right)\,
              \left[\left(k-p_3\right)^2-z\right]},
\nonumber 
\\&=& {\cal O}(m_e^0),
\\ \label{master5} 
M^{(5)}&=&
N \int \frac{d^D k}{
     \left(k^2-z\right)\,\left(k-p_1+p_3\right)^2}
\\&=& \frac{1}{\epsilon} + 2 +\ln\Bigl(-\frac{m_e^2}{t}\Bigr)
           -\ln\Bigl(1-\frac{z}{t}\Bigr)-\frac{z}{t}\, 
           \Bigl[\, \ln\Bigl(-\frac{z}{t}\Bigr) - \ln\Bigl(1-\frac{z}{t}\Bigr)\,\Bigr],
\\ \label{master6} 
M^{(6)}&=&
N \int \frac{d^D k}{
        \left(k^2-z\right) \,
          \left[\left(k+p_3\right)^2-m_e^2\right]
      \left[\left(k+p_3-p_1-p_2\right)^2-m_e^2\right]}
\nonumber 
\\
&=& \frac{1}{s} \Bigl[ \zeta_2 + \frac{1}{2}\,\ln^2\Bigl(-\frac{z}{s}\Bigr) +
          \text{Li}_2\Bigl(1+\frac{z}{s}\Bigr)\, \Bigr] + {\cal O}(m_e^2),
\\ \label{master7} 
M^{(7)}&=&
N \int \frac{d^Dk\,}{\left(k^2-z\right)\,\left[\left(k+p_3\right)^2-m_e^2\right]\,
                     \left(k+p_3-p_1\right)^2}
\nonumber 
\\&=& \frac{1}{t}\,\Bigl\{ \zeta_2+\ln\Bigl(-\frac{z}{t}\Bigr)\,
          \Bigl[\,\ln\Bigl(-\frac{m_e^2}{t}\Bigr) - \frac{1}{2}\ln\Bigl(-\frac{z}{t}\Bigr)\, \Bigr]
\nonumber
\\
          && -~\ln\Bigl(1-\frac{z}{t}\Bigr)\,\Bigl[\ln\Bigl(-\frac{m_e^2}{t}\Bigr)
           -\ln\Bigl(-\frac{z}{t}\Bigr)\,\Bigl] + \,\text{Li}_2\, \Bigl(\frac{z}{t}\Bigr)\, \Bigr\}
           +{\cal O}(m_e^2),
\\ \label{master8} 
M^{(8)}&=&
N \int \frac{d^Dk}
{ \left(k^2-z\right)\left[\left(k+p_3\right)^2-m_e^2\right]
                     \left(k+p_3-p_1\right)^2 \left[\left(k+p_3-p_1-p_2\right)^2-m_e^2\right] }
\nonumber\\
&=& 
\frac{1}{s\,\left(t-z\right)}\,
\Bigl\{
\frac{1}{\epsilon}\Bigl[ \ln\Bigl(-\frac{m_e^2}{t}\Bigr)+\ln\Bigl(-\frac{z}{s}\Bigr)
                        -\ln\Bigl(-\frac{z}{t}\Bigr)\Bigr]-2\,\zeta_2
\nonumber\\
                        &&+~\ln\Bigl(-\frac{m_e^2}{t}\Bigr)\,\Bigl[
\frac{1}{2}\,\ln\Bigl(-\frac{m_e^2}{t}\Bigr)+\ln\Bigl(-\frac{z}{s}\Bigr) + \ln\Bigl(-\frac{z}{t}\Bigr)
-2\,\ln\Bigl( 1-\frac{z}{t} \Bigr)\Bigl]
\nonumber\\
&&-~\frac{3}{2}\,\ln^2\Bigl(-\frac{z}{t}\Bigr)
+\ln\Bigl(-\frac{z}{s}\Bigr)\,\ln\Bigl(-\frac{z}{t}\Bigr)
-2\,\ln\Bigl(1-\frac{z}{t}\Bigr)\,\Bigl[\, \ln\Bigl(-\frac{z}{s}\Bigr)-\ln\Bigl(-\frac{z}{t}\Bigr)\, 
\Bigr]
\nonumber\\
&&-~\text{Li}_2\,\Bigl(1+\frac{z}{s}\Bigr)\,
\Bigr\} + {\cal O}(m_e^2).
\end{eqnarray}
where $D=4-2\,\epsilon$ and
\begin{equation}
N= m_e^{2\epsilon}\, \frac{e^{\gamma_E\epsilon}}{i \pi^{2-\epsilon}}.
\end{equation}
For $M^{(1)}$ and $M^{(2)}$, results  are needed up to ${\cal O}(\epsilon)$,
since, after the reduction procedure, both coefficients $c_i^{(1)}$
and $c_i^{(2)}$, for $i=A,B,C$, include terms  ${\cal O}(\epsilon^{-1})$.
For all other basis integrals, ${\cal O}(\epsilon^0)$ results suffice. 
Note that for $M^{(1)}$ (tadpole), $M^{(3)}$ and $M^{(5)}$ (no dependence
on $m_e$, apart from the normalization factor $N$) results are exact. In other cases, the order of the
expansion in $m_e$ depends on the coefficients  $c_i^{(j)}$.
For example, we have $c_i^{(2)}={\cal O}(m_e^{-2})$, and we compute
$M^{(2)}$ up to ${\cal O}(m_e^0)$ (note the overall factor $m_e^2$
in Eq.~\eqref{boxs}). In contrast, we have $c_i^{(4)}={\cal O}(m_e^{0})$
and we do not need $M^{(4)}$ up to ${\cal O}(m_e^0)$.

\section{\label{app-soft}Soft Real Photon Emission}
The leading order contributions to the soft real photon corrections
\begin{equation}\label{soft1}
  e^-\,(p_1) \, +   \,
  e^+\,(p_2) \, \to \,
  e^-\,(p_3) \, +   \, e^+\,(p_4)\, +   \, \gamma\,(k)
\end{equation}
to the Bhabha cross section (\ref{born})
are 
 contained in the factor $F_{\rm soft}$:
\begin{equation}\label{npb786eq4.4}
\frac{d\sigma_{\gamma}^{LO}}{d\Omega} = \frac{d\sigma_0}{d\Omega} ~ \frac{\alpha}{\pi} ~ F_{\rm soft}(\omega,s,t,m_e^2),
\end{equation}
with $\omega$ being the upper limit of the energy of the non-observed soft photons:
\begin{equation}
 E_{\gamma} \in [0,\omega] .
\end{equation}
The $\omega$ has to be chosen as small as to guaranty that the emitted photon does not change the kinematics of the process (\ref{bhabhamomenta}).
The NLO radiative cross section with ${\cal O}(\alpha)$ vacuum polarization insertions is:

\begin{eqnarray}\label{npb786eq4.5}
 \frac{d\sigma_{\gamma}^{NLO}}{d\Omega} &=&
\frac{\alpha^2}{s} \left\lbrace  \frac{v_1(s,t)}{s^2} ~\text{Re} \Pi^{(1)}(s)
+ \frac{v_2(s,t)}{st}  ~\text{Re} \left[ \Pi^{(1)}(s)+ \Pi^{(1)}(t)\right] \right. 
\nonumber \\
&&~\left. +\frac{v_1(t,s)}{t^2}~\text{Re}  \Pi^{(1)}(t)
\right\rbrace ~\left(\frac{\alpha}{\pi}\right) ~F_{\rm soft}(\omega,s,t,m_e^2).
\end{eqnarray}
The result for the soft photon factor is split into initial and final state radiation and their interference:
\begin{equation}
F_{\rm soft}(\omega,s,t,m_e^2) = 
 \delta_{\rm ini} + \delta_{\rm int} + \delta_{\rm fin} ,
\end{equation}
where 
\begin{eqnarray}\label{delini}
\delta_{\rm ini} &=& (Q_1^2+Q_2^2) F_{11} + Q_1Q_2 F_{12} 
\\ \nonumber &=& 
~ ~ 2 F_{11} + F_{12},
\\\label{delint}
\delta_{\rm int} &=& (Q_1Q_3+Q_2Q_4) F_{13} + (Q_1Q_4+Q_2Q_3) F_{14}
\\ \nonumber &=& 
~ ~ 2 F_{13} + 2 F_{14},
\\\label{delfin}
\delta_{\rm fin} &=& (Q_3^2+Q_4^2) F_{33} + Q_3Q_4 F_{34} 
\\ \nonumber &=&
~ ~ 2 F_{33} + F_{34}.
\end{eqnarray}
Each of the terms  in Eqns. (\ref{delini}) to (\ref{delfin}) exhibits the radiating particles -- a factor $Q_iQ_j$ marks the emission of the photons from particles with momenta $p_i$ and $p_j$;
Of course, it is $Q_iQ_j=1$ here. 
Since the initial and final state particles have  equal masses, it is additionally:
\begin{eqnarray}
F_{33} &=& F_{11},
\\
F_{34} &=&  F_{12}.
\end{eqnarray}
So, it will be:
\begin{equation}
F_{\rm soft}(\omega,s,t,m_e^2) =  4 F_{11} + 2 F_{12} + 2 F_{13} + 2 F_{14}.
\end{equation}
The evaluation of $F_{\rm soft}$ follows standard textbook methods (see e.g. for details in Sec. (4.3) of \cite{Fleischer:2003kk}).
The exact result for the soft radiation functions is, for $d=4-2\epsilon$:
\begin{eqnarray}
 F_{11} &=&  \Delta_{\epsilon} +\frac{1}{2\beta}\log \left( \frac{1+\beta}{1-\beta}\right) ,
\\
F_{12} &=& \Delta_{\epsilon} \left[ -\frac{2(s-2m^2)}{s\beta}\log \left( \frac{1+\beta}{1-\beta}\right)\right] 
\\ \nonumber
&&+~\frac{2(s-2m^2)}{s\beta}\left[\litwo\left( \frac{2\beta}{\beta-1}\right) - \litwo\left(\frac{2\beta}{\beta+1} \right)\right], 
\\
F_{13} &=&  \Delta_{\epsilon}\left( -\frac{T}{\sqrt{\lambda_T}}\right) \ln\left( \frac{T+\sqrt{\lambda_T}}{T-\sqrt{\lambda_T}} \right)  + F_{13}^{\rm fin},
\\
F_{14} &=& - F_{13} \mathrm{~~with~~} (t \leftrightarrow u),
\end{eqnarray}
and 
\begin{eqnarray}
 F_{13}^{\rm fin} &=& \frac{(t-2m^2)}{t\beta_t}
\left[ \litwo\left( \frac{\beta-1/\beta_t}{1+\beta}\right)
-\litwo\left( \frac{\beta+1/\beta_t}{1+\beta}\right)
-\litwo\left(-\frac{\beta-1/\beta_t}{1-\beta} \right) \right. 
\\ \nonumber&&
\left. 
+~ \litwo\left(-\frac{\beta+1/\beta_t}{1-\beta} \right)\right] .
\end{eqnarray}
We use the abbreviations:
\begin{eqnarray}
\beta&=&\sqrt{1-4m^2/s},
\\
\beta_t&=& \sqrt{1-4m^2/t},
\\
 T &=&2m^2-t,
\\
\sqrt{\lambda_T}&=& \sqrt{T^2-4m^4},
\\
\beta_u&=& \sqrt{1-4m^2/u},
\\
U&=&2m^2-u,
\\
\sqrt{\lambda_U}&=& \sqrt{U^2-4m^4} .
\end{eqnarray}
Our kinematics fulfills here $s+t+u=4m^2$, and it is $T, U >0$.
If necessary, the logarithms and dilogarithms may be analytically continued with the replacement
\begin{eqnarray}
 s \to s+i\epsilon,
\end{eqnarray}
e.g.
\begin{eqnarray}
 \litwo\left( \frac{2\beta}{\beta-1}\right) &=&
- \litwo\left(\frac{\beta-1}{2\beta}\right)  -\litwo\left(1\right)-\frac{1}{2}\ln^2\left(\frac{2\beta}{1-\beta} \right). 
\end{eqnarray}
In the limit of small electron mass $m_e$, this simplifies considerably (${\hat s} = s/m_e^2$):
\begin{eqnarray}
 F_{11}&=&\Delta_{\epsilon}+\frac{1}{2}\ln\left({\hat s}\right),
\\
F_{12}&=&-2 \Delta_{\epsilon} \ln\left({\hat s}\right)  - \frac{1}{2} \ln\left({\hat s}\right)^2  - 2 \zeta_2,
\\
F_{13}&=&- 2 \Delta_{\epsilon} \ln\left(-\frac{t}{m_e^2}\right) -\frac{1}{2} \ln\left({\hat s}\right)^2   - 2  \zeta_2 - 
 \litwo \left( -\frac{u}{t} \right), 
\\
F_{14}&=& 2 \Delta_{\epsilon} \ln\left(-\frac{u}{m_e^2}\right) + \frac{1}{2} \ln\left({\hat s}\right)^2   + 2   \zeta_2 + 
 \litwo\left( -\frac{t}{u} \right) .
\end{eqnarray}
Finally, the divergent part is:
\begin{eqnarray}
 \Delta_{\epsilon} &=& \frac{1}{2}\left[ \frac{F_{\epsilon}}{\epsilon}-\ln\left({\hat s}\right)\right] -\ln\left( \frac{2\omega}{\sqrt{s}}\right) .
\end{eqnarray}
Taking all the terms together, we obtain:
\begin{eqnarray}
 F_{\rm soft}(\omega,s,t,m_e^2) &=&
\left[ 
\frac{F_{\epsilon}}{\epsilon} -\ln\left({\hat s}\right) - 2\ln\left(\dfrac{2\omega}{\sqrt{s}} \right)
\right] 
\left[-2\ln\left({\hat s}\right)+2 -2\ln\left( \frac{t}{u} \right) \right] \nonumber
\\
&&-~
\ln\left({\hat s}\right)^2 - 4\zeta_2 
+ 2 \ln\left({\hat s}\right) +2\litwo\left( -\frac{t}{u}\right) -  2 \litwo\left( -\frac{u}{t}\right) .
\end{eqnarray}
This expression agrees, of course, with e.g. Eq. (4.5) of \cite{Actis:2007gi}.

\section{\label{sec:realpairs}Real Fermion Pair or Hadron Emission}
The numerical influence of the virtual corrections gets modified by the non-observed emission of real pairs of electrons or other fermions, or of hadrons:
\begin{eqnarray}\label{sp}
 \frac{d\sigma^{\rm real}}{d\Omega}&=& \frac{d\sigma_0}{d\Omega}
\frac{\alpha^2}{\pi^2} \left[\delta^{e}+ \delta^{f}+\ \delta^{had}\right].
\end{eqnarray}
The real pairs or hadrons give non-singular contributions and depend, in the simplest configuration, on an energetic cut-off $D$ on the invariant mass of the non-observed pair or hadrons $E_{\rm real}$, and of course also on the production threshold $2M$.

There are two basically different situations.
In case $4M^2 << s,|t|,|u|$, one may additionally  choose  $ 2M < E_{\rm real} <  D E_{\rm beam}<< E_{\rm beam}$   (remember $E_{\rm beam}=\sqrt{s}/2$),  and observes a logarithmic dependence of the cross-sections on the two parameters $M, D$.
In the other case, assuming $M>> m_e$ but otherwise arbitrary, as it is done in the present study if not stated differently, the concept of soft pairs becomes senseless and one has to evaluate the pair and hadron emission cross-section numerically with MC methods.

For completeness and because of the numerical importance, we will include the soft pair emission contributions for electrons, which is by far the biggest one.
For this case, analytical expressions with logarithmic accuracy are known from \cite{Arbuzov:1995vj}:
\begin{eqnarray}
\label{spe}
 \delta_{\rm soft}^{e}&=& \frac{1}{3}\left[
\frac{1}{3}L_s^3+L_s^2\left(2\ln(D)-\frac{5}{3} \right) +L_s\left(4\ln^2(D)- \frac{20}{3}\ln(D)+A_s\right)  
\right. \nonumber \\ &&
+~\frac{1}{3}L_t^3+L_t^2\left(2\ln(D)-\frac{5}{3} \right) +L_t\left(4\ln^2(D)- \frac{20}{3}\ln(D)+A_t\right) 
\nonumber \\ &&
\left. 
-~\frac{1}{3}L_u^3-L_u^2\left(2\ln(D)-\frac{5}{3} \right) -L_u\left(4\ln^2(D)- \frac{20}{3}\ln(D)+A_u\right) 
 \right] ,
\end{eqnarray}
where 
\begin{eqnarray}\label{spe1}
 L_s &=&\ln\left(\frac{s}{m_e^2} \right),
\\\label{spe2}
 L_v &=&\ln\left(-\frac{v}{m_e^2} \right), ~~~~v=t,u,
\\ \label{spe3}
A_s&=& \frac{56}{9}-4\zeta_2.
\\\label{spe4}
A_v&=& A_s+2\litwo\left( \frac{1\pm\cos \theta}{2}\right) , ~~~~v=t,u.
\end{eqnarray}
The parameter $D$ has to fulfill:
\ba
2m_e << D E_{\rm beam} << E_{\rm beam}.
\ea
 From the sum of (\ref{sp}) and (\ref{sig-irr-vert}), the compensation of the leading mass singularities (contained here in the $L_s^3, L_t^3, L_u^3$ terms) in the cross-section becomes evident. 

 
\section{The Cross-Section Ratio $R_{\rm had}$}
\label{app:rhad}
The numerical values of the irreducible two-loop corrections  depend crucially on $R_{\rm had}(s)$ as defined in (\ref{Rhad}), while the reducible corrections may be evaluated with one of the publicly available  parameterizations of $\Pi(q^2)$ (see (\ref{DispInt})).
Unfortunately,
we did not find an actual, publicly available code for $R_{\rm had}(s)$ that
covers the complete integration region from the threshold at $s=4M_{\pi}^2$ to infinity.
In our short communication \cite{Actis:2007fs}, we used the Fortran routine of H. Burkhardt~\cite{Burkhardt:1981jk}.
This parameterization dates back to 1986 and was used for the numerics in \cite{Kniehl:1988id}, and it was available by contacting the author \cite{Burkhardt:1981jk}.
The Fortran file is made available at our website \cite{webPage:2007xx}.
It is to be expected that current hadronic data would not induce changes compared to the parametrization
of~\cite{Burkhardt:1981jk}  of more than about 10\%.
This would be tolerable in view  of the smallness of the  irreducible two-loop contributions in our analysis.
For the numerically much more sensitive reducible contributions, the running coupling $\alpha_{em}$ is needed, and 
implementations of that are publicly available, e.g. the Fortran package \texttt{hadr5.f} at \cite{Jegerlehner-hadr5n:2003aa}.

\begin{figure}[tbhp]
 \centering
\includegraphics[scale=0.5]{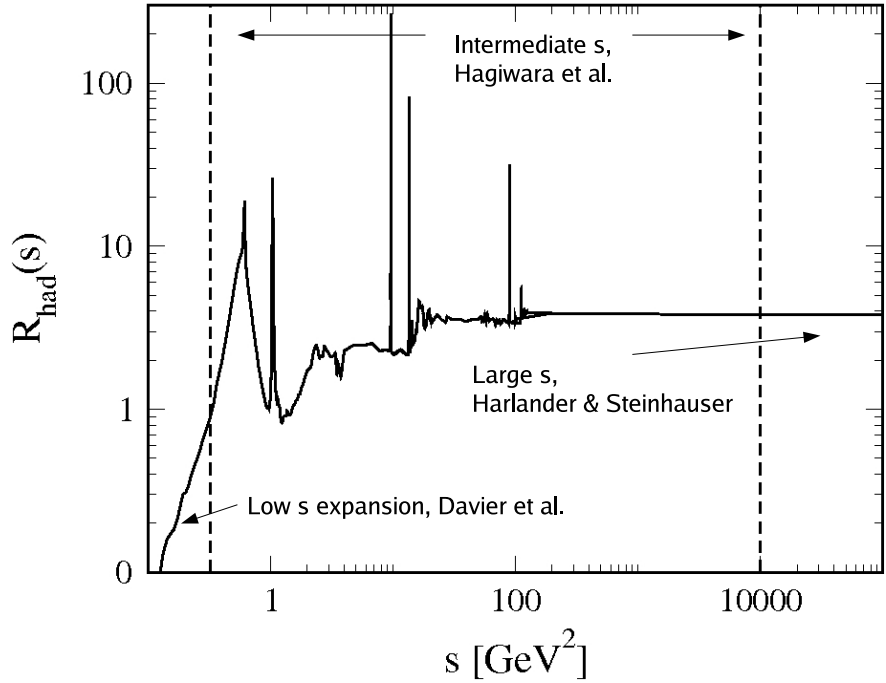}
 \caption[he implementation of $R_{\rm had}$ for irreducible two-loop corrections.]
{\em The implementation of $R_{\rm had}$ used for the numerical evaluation of irreducible two-loop corrections.}
 \label{fig:update.ps}
\end{figure}

For the present study, we improved our numerical basis for the evaluation of the irreducible vertex and box contributions by combining  packages for the evaluation of   $R_{\rm had}(s)$ in different kinematical regions:
\begin{itemize}
 \item[(A)] From threshold at $s=4m_\pi^2$ to $s=0.03$ GeV$^2$: We follow Section 8.1 of \cite{Davier:2002dy}: 
\begin{eqnarray}
 R_{\rm had}(s) = R_{\pi^+ \pi^-}(s) &=& \frac{1}{4}\left( 1-\frac{4m_{\pi}^2}{s}\right)^{3/2} |F_{\pi}(s)|^2,
\\ 
F_{\pi}(s) &=& 1 + 1.879 \left( \frac{s}{\mathrm{GeV}^2}\right) + 3.3  \left( \frac{s}{\mathrm{GeV}^2}\right)^2 -0.7 \left( \frac{s}{\mathrm{GeV}^2}\right)^3.
\end{eqnarray}
The above is based on a fit to  $e^+e^-$ data whose results are shown in Table 3 of  \cite{Davier:2002dy}; space-like data \cite{Amendolia:1986wj} are also taken into account. 
\item[(B)] From  $s=0.03$ GeV$^2$ to $s=10000$ GeV$^2$:
Use of 
subroutine \cite{rintpl:2008AA}.
%
\item[(C)] 	Above $s=10000$ GeV$^2$:
Use of subroutine \texttt{rhad.f v.1.00}, published in \cite{Harlander:2002ur}.
\end{itemize}
In Figure (\ref{fig:update.ps}) we show the  $R_{\rm had}$ resulting from our Fortran implementation for the regions (A) to (C) as described above.

\begin{figure}[tbhp] 
\vspace*{1.2cm}
 \centering
\includegraphics[scale=.5]{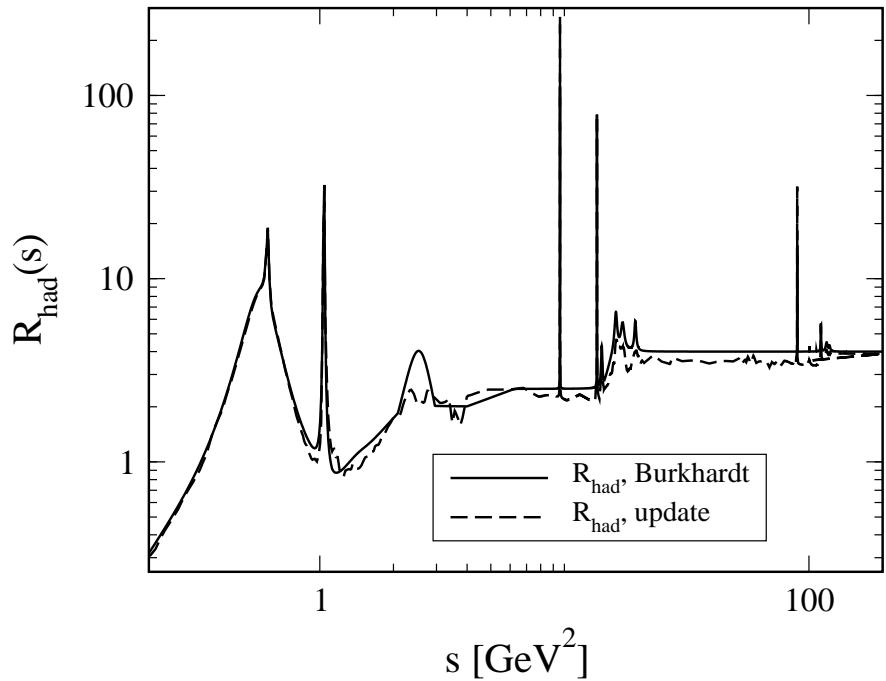}
 \caption[A comparison of \texttt{repi.f} (H. Burkhardt, 1986)  and \texttt{r\_intpl.f} (T. Teubner, private communication, 26 April 2008).]
{\em A comparison of the parametrizations from ~\cite{Burkhardt:1981jk}  and \cite{rintpl:2008AA}.}
 \label{fig:burk_teub}
\end{figure}

In Figure (\ref{fig:burk_teub}) we compare the implementation of  $R_{\rm had}(s)$ taken from 
Burkhardt~\cite{Burkhardt:1981jk} ($R_{\rm had,I}$)  and  our parametrization based on 
\cite{Davier:2002dy}\cite{rintpl:2008AA} \cite{Harlander:2002ur}($R_{\rm had,II}$).
As already stated, the deviations are evidently much smaller than one might expect and may be considered to be irrelevant here.

We close this section with a brief discussion of narrow resonances.
Narrow resonances are implemented replacing the rapidly varying
cross section ratio with the parametrization
\begin{equation}
R_{\rm res}(z)= \frac{9 \pi}{\alpha^2} M_{\rm res} \Gamma^{e^+ e^-}_{\rm res}
\delta(z-M^2_{\rm res}).
\end{equation}
The integration over $z$ is then carried on analytically leading
to the following result for the IR-finite remainder (including
the irreducible box diagrams) of Eq.~\eqref{eqrest}:
\begin{equation}
\frac{d \overline{\sigma}_{\rm rest}}{d \Omega}=
\frac{9 \pi}{\alpha^2} \frac{\Gamma^{e^+ e^-}_{\rm res}}{M_{\rm res}}
\left\{
\frac{F_1(M^2_{\rm res})}{t-M^2_{\rm res}}
+ \frac{1}{s-M^2_{\rm res}} \left[ F_2(M_{\rm res}^2)
+ F_3(M_{\rm res}^2) \ln \left| 1-\frac{M_{\rm res}^2 }{s}\right| \right]
\right\}.
\end{equation}
For the numerical evaluation of the contribution due to the
narrow resonances, we use the values listed in the Burkhardt's
routine~\cite{Burkhardt:1981jk}, collected in Table~\ref{table:BUR}.

\begin{table}[ht]\centering
\setlength{\arraycolsep}{\tabcolsep}
\renewcommand\arraystretch{1.1}
\begin{tabular}{|r|r|r|}
\hline 
resonance & $M_{\rm res}$ [GeV] & $\Gamma^{e^+ e^-}_{\rm res}$ [keV] \\
\hline 
\hline 
$\omega$(782) & 0.7826 & 0.66 \\
$\phi$(1020) & 1.0195 & 1.31\\
J$\slash \psi$(1S) & 3.0969 & 4.7\\
$\psi$(2S) & 3.6860 & 2.1\\
$\psi$(3770) & 3.7699 & 0.26\\
$\psi$(4040) & 4.0300 & 0.75\\
$\psi$(4160)& 4.1590 & 0.77\\
$\psi$(4415) & 4.4150 & 0.47\\
$\Upsilon$(1S) & 9.4600 & 1.22\\
$\Upsilon$(2S)& 10.0234 & 0.54\\
$\Upsilon$(3S)& 10.3555 & 0.40\\
$\Upsilon$(4S)& 10.577 &  0.24\\
$\Upsilon$(10860)& 10.865 &  0.31\\
$\Upsilon$(11020)& 11.019 &  0.13\\
\hline
\end{tabular}
\caption[]{Numerical values for the treatment of narrow resonances,
 taken directly from~\cite{Burkhardt:1981jk}.}
\label{table:BUR}
\end{table}



\section{\label{app-polylog}Evaluation of Polylogarithms}
At several instances, dilogarithms $\litwo(z)$ and trilogarithms $\litri(z)$ of complex argument are needed.
A definition of polylogarithms is:
\begin{eqnarray}
 \lin(z) &=& S_{n-1,1}(z) ~=~ \frac{(-1)^{n}}{(n-2)!}~\int_{0}^{1} \frac{dt}{t}\ln^{n-2}(t)\ln(1-zt).
\end{eqnarray}
They have the special values $\lin(0)=0 $ and $\lin(1)=\zeta(n)$,
where $\zeta(s)$ is the Riemann $\zeta$-function, $\zeta(2)=\pi^2/6, \zeta(3)=1.2020569031595942854\ldots$
An efficient evaluation transforms the arguments to the region where modulus and real part are bound: 
$|z|\leq 1$ and $\Re e(z)<\frac{1}{2}$, using:
\begin{eqnarray}
 \litwo(z) &=& - \litwo\left(\frac{1}{z}\right) - \frac{1}{2}\ln^2(-z) - \zeta(2),
\\
\litwo(z) &=&  -\litwo(1-z) +\zeta(2) - \ln(z)\ln(1-z),
\end{eqnarray}
and
\begin{eqnarray}
 \litri(z) &=&  \litri\left(\frac{1}{z}\right) - \frac{1}{6}\ln^3(-z) - \zeta(2)\ln(-z),
\\
\litri(z) &=& -\litri\left(1-\frac{1}{z}\right) -  \litri(1-z) +\zeta(3) + \frac{1}{6}\ln^3(z) + \zeta(2)\ln(z) -\frac{1}{2}\ln^2(z)\ln(1-z).
\end{eqnarray}
Then,  series expansions with Bernoulli numbers ensure rapid convergence.
For $\litwo(z)$ we follow  Appendix A of \cite{'tHooft:1979xw}:
\begin{eqnarray}
\litwo(z) &=& \sum_{j=0}^{\infty} \frac{B_{j}}{(j+1)!}\left[-\ln(1 - z)\right]^{j+1}
\nonumber \\&=&
-\ln(1 - z) - \frac{1}{4}\ln^2(1 - z)
 + 4\pi \sum_{j=1}^{\infty} \zeta(2j)\frac{(-1)^j}{2j+1}\left[\frac{\ln(1 - z)}{2\pi}\right]^{2j+1}.
\end{eqnarray}
The $B_j$ are Bernoulli numbers, $B_0=1$, etc.
Useful series expansions for $\lin(z)$ are given in Eqns. (48) and (49) of \cite{Vollinga:2004sn}, which we reproduce here for the special case  $n=3$: 
\begin{eqnarray}
  \litri(z) &=& \sum_{j=0}^{\infty} \frac{C_3(j)}{(j+1)!}\left[-\ln(1 - z)\right]^{j+1},
\\
C_3(j)&=& \sum_{k=0}^j 
\begin{pmatrix}
 j \\
 k
\end{pmatrix}
   \frac{B_{j-k} B_k}{1+k}  ,
\end{eqnarray}
with $C_3(0)=1$ etc.
For $\litwo(z)$ and $\litri(z)$ we observe typically that $n$ summation terms give an $n\pm 1$ digits accuracy.
We just mention that we do not allow to evaluate the logarithms and polylogarithms at their cuts (negative real axis beginning at $z=0$ and positive real axis beginning at $z=1$, respectively).
For other conventions we refer to the corresponding remark at p. 19 of  \cite{Vollinga:2004sn}.
Our Fortran code is available as  file  \texttt{cpolylog.f} at the website \cite{webPage:2007xx}.

An alternative, efficient  algorithm for the evaluation of polylogarithms is described in \cite{Crandall:2006} \footnote{U. Langenfeld, private information.}.
 

\clearpage

\providecommand{\href}[2]{#2}
\end{document}